\documentclass[pdflatex,sn-mathphys-num]{sn-jnl}

\usepackage{graphicx}%
\usepackage{multirow}%
\usepackage{amsmath,amssymb,amsfonts}%
\usepackage{amsthm}%
\usepackage{mathrsfs}%
\usepackage[title]{appendix}%
\usepackage{textcomp}%
\usepackage{manyfoot}%
\usepackage{booktabs}%
\usepackage{algorithm}%
\usepackage{algorithmicx}%
\usepackage{algpseudocode}%
\usepackage{listings}%
\usepackage{dsfont}
\usepackage[overload]{empheq}
 \usepackage{caption}
\usepackage{subcaption}
\usepackage{textgreek}
\usepackage[monochrome]{color}

\usepackage{tikz}
\usepackage{pgfplots}
\usepackage{floatpag}
\usepackage{pgfplotstable}
\usetikzlibrary{arrows.meta, decorations.pathreplacing, patterns}

\theoremstyle{thmstyleone}%

\theoremstyle{thmstyletwo}%
\theoremstyle{plain}
\newtheorem{definition}{Definition}%
\newtheorem{lemma}{Lemma}%
\newtheorem{remark}{Remark}%
\newtheorem{corollary}{Corollary}%

\usepackage[numbers,sort&compress]{natbib} 

\usepackage{amsmath}
\usepackage{numprint} 
\usepackage{amsthm} 
\usepackage{url} 
\usepackage{textcomp} 

\usepackage{accents} 
\usepackage{amssymb}
\usepackage{amsfonts}
\usepackage{dsfont}
\usepackage[figuresright]{rotating} 
\usepackage{floatpag} 
	\rotfloatpagestyle{empty} 

\usepackage{longtable}
\usepackage{epstopdf} 
\usepackage{makeidx} 
\usepackage{upgreek} 
\makeindex 

\usepackage{times}
\usepackage{mathdots} 

\usepackage{bm}
\usepackage{bbm}

\theoremstyle{plain}

\def\bb0{{\mathbb{0}}}

\def\ba{{\boldsymbol{a}}}
\def\bb{{\boldsymbol{b}}}

\def\bee{{\boldsymbol{e}}}

\def\bi{{\boldsymbol{i}}}
\def\bj{{\boldsymbol{j}}}

\def\bu{{\boldsymbol{u}}}
\def\bv{{\boldsymbol{v}}}

\def\b0{{\boldsymbol{0}}}

\def\bA{{\boldsymbol{A}}}

\def\bC{{\boldsymbol{C}}}

\def\bI{{\boldsymbol{I}}}

\def\bU{{\boldsymbol{U}}}

\def\b{{\mathrm{b}}}

\def\r0{{\mathbf{0}}}

\def\bbN{{\mathbb{N}}}

\def\bbR{{\mathbb{R}}}

\def\bbZ{{\mathbb{Z}}}

\def\cE{\mathcal{E}}

\def\cK{\mathcal{K}}
\def\cL{\mathcal{L}}

\def\cO{\mathcal{O}}
\def\cP{\mathcal{P}}

\def\cX{\mathcal{X}}

\def\sfL{\mathsf{L}}

\def\bnu{\bm \nu}

\def\bLambda{\bm \Lambda}

\def\sfc{{\mathsf{c}}}
\def\sfd{{\mathsf{d}}}

\def\sfff{{\mathsf{f}}}
\def\sfg{{\mathsf{g}}}

\def\sfu{{\mathsf{u}}}
\def\sfv{{\mathsf{v}}}

\def\sfz{{\mathsf{z}}}

\def\bsf0{{\bm{\mathsf{0}}}}

\def\N0{{N_{\mathrm{0}}}}
\def\sinc{\mathrm{sinc}}

\def\j{\mathrm{j}}

\def\bsf{{\boldsymbol{s}_\mathrm{f}}}

\newcommand{\Trans}{\rm \scriptscriptstyle T}

\newcommand{\Herm}{\rm \scriptscriptstyle H}
\def\diag   {\mbox{\rm diag}}

\def\bCC{{\boldsymbol{\mathsf{C}}}}
\newcommand{\sflambda}{\text{\textsf{\textlambda}}}
\newcommand{\freq}{\text{\textsf{f}}}
\def\sfbLambda{\mathsf{\bm{\Lambda}}}

\def\tr{\mathrm{tr}}

\def\endf{\parallel}

\graphicspath{{./images/}}
 
\begin{document}

\title[Article Title]{Super-Beamforming in Holographic MIMO}


\author*[1]{\fnm{Andrea} \sur{Pizzo}}\email{andrea.pizzo@upf.edu}

\author[1]{\fnm{Angel} \sur{Lozano}}\email{angel.lozano@upf.edu}


\affil[1]{\orgdiv{Department of Engineering}, \orgname{Univ. Pompeu Fabra}, 
\city{Barcelona} \postcode{08005}, \country{Spain}}




\abstract{  
The conventional linear scaling of beamforming gain with the number of antennas $N$ is not a fundamental physical limitation, but rather a consequence of the half-wavelength spacing that minimizes mutual coupling.  
Relaxing this constraint 
enables gains that, 
{\color{blue}in the vicinity of the array axis, exceed those of uncoupled arrays.}
{\color{blue}This paper shows that mutual coupling facilitates the synthesis of super-beams whose widths scale as $1/N$, in contrast to the conventional $1/\sqrt{N}$. 
These narrower beams achieve a quadratic gain scaling with $N$ provided antenna losses remain sufficiently small, decaying at least exponentially with growing $N$ and polynomially with decreasing spacing.}
Notably, this {\color{blue}gain enhancement} does not necessarily require vanishing spacings; it may also emerge for spacings slightly below half wavelength as the array aperture increases.}

\keywords{Holographic MIMO, supergain, superdirectivity, mutual coupling, 
spatial sampling theorem, spectral concentration.}



\maketitle

\section{Introduction}


Arranging $N$ isotropic antennas into a uniform linear array (ULA) can increase the gain by a factor of $N$ relative to a single antenna \cite[Sec.~6]{BalanisBook}.
This linear scaling, however, is not a fundamental limit, but rather a consequence of the half-wavelength antenna spacing typically adopted to ensure that the excitation currents are as they would be for the antennas in isolation (without mutual coupling).
Reducing the spacing below this threshold leads to a holographic array, where currents become inherently coupled and the assumption of independence is superseded \cite{PizzoJSAIT25,dardari2024dynamic}. Far from being a mere parasitic effect, coupling can be harnessed to synthesize 
beams whose gain exceeds that of
half-wavelength-spaced ULAs for the same number of antennas, a phenomenon aptly named supergain (or superdirectivity) \cite{Schelkunoff1943,Uzkov,Hansen1981}.

Supergain is classically associated with a vanishing antenna spacing, with correspondingly small apertures and maximal mutual coupling. 
In this {\color{blue}limit}, the endfire (aligned to the array axis) gain approaches a quadratic scaling with $N$
\cite{Yaghjian2005}.
Such behavior is often dismissed as impractical; it requires precise current distributions that are hypersensitive to structural perturbations and are accompanied by surges in the reactive-to-radiated power ratio, characterized by a high array quality ($Q$) factor \cite{Morgan55,Hansen1981}. 
High-$Q$ arrays are inherently constrained to a narrow operational bandwidth near resonance, and their resonant frequencies cannot be adjusted without structural re-tuning \cite{Gustafsson2006,yaghjian2025}. Furthermore, achieving the requisite $Q$-factors demands delicate engineering and calibration, often at the expense of other design objectives.

Despite these intrinsic challenges, there has been a significant resurgence of interest in supergain \cite{Nossek2010conf,Nossek2010,Marzetta2019,wang2025mutual,McMinn2025} and related effects \cite{Wonseok2015,Heath2023}, driven by the quest for extreme array gains with compact apertures. 
These efforts are driven by recent developments in reconfigurable metasurfaces that may enable ultra-high $Q$-factors at arbitrary, easily tunable resonant frequencies \cite{Huang2023NatCo,fang2024millionq}.
Realizability is further enhanced by intrinsic antenna losses, which give rise to a trade-off between achievable gain and physical sensitivity of the array \cite{Bikhazi2007,Nossek2010}.

Building on these advances, this paper establishes a comprehensive framework for the analysis and design of supergain arrays. This framework generalizes the classical {\color{blue}limit} of electrically small apertures to account for antenna losses. In addition, the framework reveals another supergain {\color{blue}mechanism} in which coupling builds up from the accrual of contributions from a large number of antennas that, having non-vanishing spacings, are only partially coupled.
The specific contributions are as follows:
\begin{itemize}
    \item 
    The physical limitations imposed by dissipative losses on supergain are characterized, extending classical lossless analyses for electrically small apertures \cite{Yaghjian2005}. 
    A rigorous asymptotic analysis as the antenna spacing vanishes reveals the interplay between the number of antennas and the loss factor, formalizing previous numerical observations  \cite{Nossek2010,Sanguinetti2024}.
    \item 
    A novel supergain {\color{blue}mechanism} is identified that emerges as the array aperture widens
    while the antenna spacing remains fixed somewhere below the half-wavelength mark. This rigorously establishes the qualitative behavior predicted in \cite{PizzoSPAWC24}, demonstrating its physical accessibility in holographic arrays.
     \end{itemize}

The manuscript is organized as follows.
Section~\ref{sec:ULA} reviews the discrete-space Fourier transform for ULAs, establishing the array gain and quality factor.
Section~\ref{sec:mutual_coupling} characterizes how mutual coupling impacts array performance measures, and Section~\ref{sec:supergain} examines the physical limits under strong coupling. 
Section~\ref{sec:spectral_concentration} provides a rigorous framework 
based on spectral concentration theory, and Section~\ref{sec:spectral_conc_ULA} specializes this theory to the spatial domain for ULAs. 
Building on this, Sections~\ref{sec:spatial_concentration} and \ref{sec:spectral_conc_Nlarge} formalize the array behavior in limiting {\color{blue}cases}.
Section~\ref{sec:conclusion} concludes with a discussion on potential extensions.

\emph{Notation:} 
%
The Hilbert space of square-integrable complex functions over $\cX$ is denoted by $\mathscr{L}^2(\cX)$, with inner product $\langle f,g \rangle = \int_{\cX} f(x) g^*(x) \, dx$ and norm $\|f\| = \sqrt{\langle f,f \rangle}$. For periodic functions, the $\mathscr{L}^2$-norm is computed over the main period unless otherwise stated. {\color{blue}Similarly, $\ell^2(X)$ denotes the space of square-integrable complex sequences over $X$.}
Landau symbols are employed, namely $f(x)=\cO(g(x))$ indicates that $| f(x)/g(x)|$ is bounded above by a constant, $f(x)=o(g(x))$ indicates that $f(x)$ grows strictly more slowly than $g(x)$, and  $f(x)\sim g(x)$ indicates that $f(x)/g(x)\to 1$. In turn,
$\sinc(x)=\sin(\pi x)/\pi x$ and
$[x]_{\text{rem} \, 2\pi} = x- 2\pi \lfloor x/2\pi \rfloor$ is the remainder of $x$ modulo $2\pi$.

\section{Uniform Linear Arrays} \label{sec:ULA}

All spatial variables are normalized by the wavelength, 
and they are therefore dimensionless.
A time-harmonic and single polarization setting is considered, whereby electromagnetic quantities can be expressed as complex scalars.

\subsection{Space-Time Isomorphism} \label{sec:space-time}

\begin{figure}
     \centering
     \begin{subfigure}[b]{0.49\textwidth}
         \centering
         \includegraphics[width=0.9\textwidth]{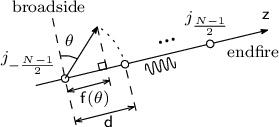}
         \caption{ULA configuration}
         \label{fig:ULA_model}
     \end{subfigure}
     \hfill
     \begin{subfigure}[b]{0.49\textwidth}
         \centering
         \includegraphics[width=\textwidth]{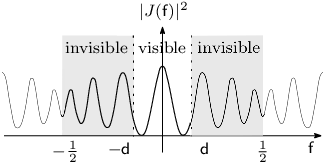}
         \caption{Current power density spectrum}
         \label{fig:ULA_spectrum}
     \end{subfigure}
        \caption{(a) Geometry of a linear array of $N$ punctiform antennas spaced by a normalized distance $\sfd$ and aligned with the $z$-axis. 
        The direction of propagation forms an angle $\theta$ with respect to the array's normal. The $n$th antenna is excited by a complex-valued current weight $j_n$. (b) Periodic power density spectrum of the current vs the  discrete spatial frequency, $\sfff$. The main period splits into a visible and an invisible region.}
        \label{fig:ULA}
\end{figure}

Consider a $z$-aligned uniform array of $N$ punctiform (isotropic) antennas spaced by a normalized distance $\sfd >0$ 
(see Fig.~\ref{fig:ULA}).  
The array is referenced to its midpoint, and the antenna indices are given by the index set
\begin{equation} \label{set_N}
[N] = \left\{ {\textstyle -\frac{N-1}{2}, \ldots, \frac{N-1}{2} } \right\} 
\end{equation}
for odd $N$.
The array is driven by a current vector $\bj = \{j_n\} \in \ell^2([N])$, where each entry $j_n$ is the phasor of the current driving the $n$th antenna.
The discrete-space Fourier transform of $\bj$ yields the current spectrum
\begin{align} \label{current_spectrum}
J(\sfff) & = F \, \bj 
 = \sum_{n\in [N]}  j_n \, e^{-\j 2 \pi \sfff n},
\end{align}
with corresponding inverse Fourier transform (in the energy sense)
\begin{align} \label{current_spectrum_inverse}
j_n & = F^{-1}  J(\sfff)  = \int_{-1/2}^{1/2}  J(\sfff) \, e^{\j 2 \pi \sfff n} \, d\sfff \qquad n\in [N],
\end{align}
where $F$ and $F^{-1}$ are the unitary forward and inverse Fourier transform operators, respectively.
The current spectrum $J(\sfff)$, periodic with period $1$, belongs to the space of square-integrable functions $\mathscr{L}^2[-\tfrac{1}{2}, \tfrac{1}{2}]$.
It corresponds to the far-field radiation pattern as a function of direction, which is encoded through the discrete spatial frequency
\begin{align} \label{f_discrete}
\sfff(\theta) =
\sfd \, \sin \theta = \sfd \, f(\theta),
\end{align}
where $\theta \in [-\frac{\pi}{2},\frac{\pi}{2}]$ is the angle relative to broadside, and $f(\theta) = \sin \theta$ is the  spatial frequency (see Fig.~\ref{fig:ULA}).
Physically, $f(\theta)$ is the frequency along an asymptotically long continuous aperture of the current that must be excited to radiate a plane wave $e^{\j 2 \pi f(\theta) \sfz}$ on direction $\theta$. 
Then, $\sfff(\theta)$
results from sampling that aperture with isotropic antennas spaced by $\sfd$.
Low spatial frequencies correspond to near-broadside propagation ($\theta \approx 0$) whereas high spatial frequencies map to near-endfire directions ($|\theta| \approx \pi/2$).

As $\theta$ varies over its domain, the mapping $\theta \mapsto \sfff$ confines $\sfff$ to the so-called \emph{visible region}, $[-\sfd,\sfd]$, corresponding to directions associated with propagating waves (see Fig.~\ref{fig:ULA}b). Plane waves with spatial frequencies within this region carry real power, contributing to the radiation. 
In contrast, plane waves with spatial frequencies obtained by analytic continuation into the invisible region $|\sfff| > \sfd$ are evanescent and store reactive (imaginary) power, which oscillates without producing net radiation \cite{Rhodes1963, PlaneWaveBook}.
To prevent spatial aliasing, it must hold that $\sfd  \le 1/2$; otherwise, grating lobes would appear within the visible region.

\subsection{Performance Measures}

The \emph{directivity function} in a certain direction specified by $\sfff$ via \eqref{f_discrete} is defined as the ratio of the power density radiated at $\sfff$ to the average radiated power across all visible directions 
\begin{equation}     \label{array_directivity_def} 
D(N,\sfd,\bj,\sfff) = G(N,\sfd,0,\bj,\sfff) = \frac{ |J(\sfff)|^2}{\displaystyle \frac{1}{2 \sfd} \int_{-\sfd}^{\sfd} |J(\sfg)|^2 \, d\sfg}.
\end{equation}
By construction, averaging \eqref{array_directivity_def} over the visible region returns $1$. Thus, the directivity quantifies how effectively the array concentrates power along a specific direction relative to an isotropic radiator. 
This isotropic reference corresponds to $N = 1$, where the array collapses into a single punctiform antenna producing spherical wavefronts.
In practice, current weights are applied to the array by signal generators through an impedance network, but the array gain is independent of such circuitry details \cite{Nossek2010} and the impedance network is therefore omitted.

The introduction of losses via the factor $\rho = R_\text{L}/ R_\text{R} \ge 0$, with $R_\text{R}$ and $R_\text{L}$ the radiation resistance and loss resistance, yields the \emph{array gain function} at $\sfff$ \cite[Sec.~2.9]{BalanisBook}, \cite{Taylor1955}
\begin{equation}     \label{array_gain_def} 
G(N,\sfd,\rho,\bj,\sfff) = \frac{ |J(\sfff)|^2}{\displaystyle  \frac{1}{2 \sfd} \int_{-\sfd}^{\sfd} |J(\sfg)|^2 \, d\sfg +   \rho \int_{-1/2}^{1/2} |J(\sfg)|^2 \, d\sfg} 
\end{equation}
which subsumes \eqref{array_directivity_def} for $\rho=0$, corresponding to lossless antennas. 
The loss factor $\rho$ relates to the antenna radiation efficiency $0 \le \eta \le 1$ via
\begin{equation} \label{antenna_efficiency}
   \eta(\rho) = \frac{1}{1+\rho} = \frac{R_\text{R}}{R_\text{R} + R_\text{L}}.
\end{equation}

Let us also formalize the definition of the \emph{Q factor} 
as the ratio of the power radiated and stored in the fields excited by the antennas to the average power accepted by them
\cite{Taylor1955,Lee1966,Hansen1981,Wallace2005} 
\begin{align}     \label{Q_factor}
Q(N,\sfd,\rho,\bj) & = \frac{\displaystyle \int_{-1/2}^{1/2} |J(\sfg)|^2 \, d\sfg}{\displaystyle \frac{1}{2 \sfd} \int_{-\sfd}^{\sfd} |J(\sfg)|^2 \, d\sfg + \rho \int_{-1/2}^{1/2} |J(\sfg)|^2 \, d\sfg} .
\end{align}
The $Q$ factor plays a central role in the design of arrays. 
Each array can be modeled as a resonant circuit that supports currents at temporal frequencies (in Hz) close to resonance \cite{Hansen1981,Gustafsson2006,yaghjian2025}.
The reciprocal of $Q$ 
is proportional to the ratio of the resonance bandwidth $\Delta_\freq$ to the resonant frequency $\freq_0$,
\begin{equation}
\frac{1}{Q} \propto \frac{\Delta_\freq}{\freq_0},
\end{equation}
which measures the sharpness of the resonance. The lower $Q$, the wider the bandwidth.
(The circuit-based model is accurate only for moderately high $Q$, becoming unreliable for  $Q<2$.)

\subsection{Normalization}

Taking advantage of the scale invariances of \eqref{array_gain_def} and \eqref{Q_factor}, both the current vector $\bj$ and its transform $J(\sfff)$ can be normalized to have unit norms irrespective of $N$, namely
\begin{align} \label{normalization}
   \|\bj \|^2 = \| J\|^2 = 1,
\end{align} 
as per Parseval's theorem.
This ensures that the total (radiated and stored) power is fixed across all current distributions, isolating the effect of beamforming through changes in that distribution. 
Under this normalization, the array gain in \eqref{array_gain_def} becomes
\begin{align}     \label{directivity_current_general_norm} \textstyle
G(N,\sfd,\rho,\bj,\sfff) & = \frac{|J(\sfff)|^2}{\displaystyle \frac{1}{2 \sfd} \int_{-\sfd}^{\sfd} |J(\sfg)|^2 \, d\sfg + \rho} 
\end{align}
and the $Q$ factor in \eqref{Q_factor} satisfies
\begin{align}   \label{Q_factor_norm} 
Q^{-1}(N,\sfd,\rho,\bj) & = \frac{1}{2 \sfd} \int_{-\sfd}^{\sfd} |J(\sfg)|^2 \, d\sfg + \rho.
\end{align}

\subsection{Optimality Criterion}

The optimality of the introduced performance measures is governed by their common term, the average radiated power
\begin{equation} \label{avg_radiated_power}
\frac{1}{2 \sfd} \int_{-\sfd}^{\sfd} |J(\sfg)|^2  d\sfg.
\end{equation}
There is a tension in that maximizing the array gain entails decreasing the power within the visible region, whereas minimizing the $Q$ factor entails the opposite. The ensuing tradeoff embodies the fundamental tension between array gain and usable bandwidth \cite{Hansen1981,Gustafsson2006}.
Additionally, antenna losses act through $\rho$ as a Tikhonov-type physical regularization. Increasing $\rho$ diminishes the impact of \eqref{avg_radiated_power} on performance measures, expanding the usable bandwidth but taking a toll on  gain. Another tradeoff thus arises between maximum gain, for $\rho = 0$, and physical realizability, corresponding to $\rho > 0$ and a surging dissipated power.

In the sequel, the optimality criterion is set to be the maximization of the array gain, while the $Q$ factor is evaluated to quantify the corresponding bandwidth cost. 
Optimizations for minimal $Q$ factor or mixed strategies are also possible \cite{Lee1966,Wallace2005,Gustafsson2013}.
The superscript $^\star$ 
denotes quantities evaluated for the current vector that maximizes the array gain.
For instance, the current $\bj^\star(N,\sfd,\rho,\sfff^\prime)$ that maximizes the gain at a given beamforming direction $\sfff^\prime$ satisfies
\begin{align}     \label{max_current_discrete}
\underset{\bj : \|\bj\|=1}{\text{maximize}} \; G(N,\sfd,\rho,\bj,\sfff^\prime),
\end{align}
with spectrum $J^\star(N,\sfd,\rho,\sfff^\prime,\sfff)$ defined according to \eqref{current_spectrum}.
For the above current, the array gain in the beamforming direction $\sfff^\prime$ is $G^\star(N,\sfd,\rho,\sfff^\prime)  = G(N,\sfd,\rho,\bj^\star,\sfff^\prime)$
while the $Q$ factor equals $Q^\star(N,\sfd,\rho,\sfff^\prime) = Q(N,\sfd,\rho,\bj^\star)$.

\section{Spatial Coupling} \label{sec:mutual_coupling}

\subsection{Uncoupled Arrays}

Consider a fixed spacing $\sfd=1/2$. Isotropic antennas then act as independent transducers, with an overall array response equal to the sum of the individual antenna contributions \cite{OrfanidisBook}. 
From \eqref{Q_factor_norm}, the $Q$ factor with such half-wavelength spacing reduces to the antenna efficiency in \eqref{antenna_efficiency}, regardless of the current vector,
\begin{equation}   \label{Q_factor_norm_lambda2} 
Q(N,\tfrac{1}{2},\rho,\bj) = \eta(\rho).
\end{equation}
As for the array gain, by expanding $J(\sfff)$ in the numerator of \eqref{directivity_current_general_norm} according to \eqref{current_spectrum}, and in light of the normalization in \eqref{normalization}, it reduces to 
\begin{equation}   \label{directivity_lambda2_uniform_main_coupling_Rayleigh}
 G(N,\tfrac{1}{2},\rho,\bj,\sfff)  = N \, \eta(\rho) \, |\ba(\sfff)^{\Herm} \, \bj|^2 
  \end{equation}
where 
\begin{equation} \label{a_N}
\ba(\sfff) = \frac{1}{\sqrt{N}} \, \left(e^{-\j \pi \sfff (N-1)}, \ldots, e^{\j \pi \sfff (N-1)}\right)^{\!\Trans}
\end{equation}
is the array response vector, satisfying $\|\ba\|=1$.
In this case, beamforming via maximum-ratio transmission is optimal for any direction defined by $\sfff^\prime$. 
Indeed, from the Cauchy–Schwarz inequality applied to  \eqref{directivity_lambda2_uniform_main_coupling_Rayleigh},
\begin{equation} \label{u_def}
\bj^\star(N,\tfrac{1}{2},\rho,\sfff^\prime) = \ba(\sfff^\prime),
\end{equation}
yielding 
\begin{align}   \label{directivity_current_halfwavelength_max}
G^\star(N,\tfrac{1}{2},\rho,\sfff^\prime) = N \, \eta(\rho). 
\end{align}
From \eqref{current_spectrum}, the spectrum corresponding to \eqref{u_def} is 
\begin{align}  \label{Dirichlet_kernel_current}
J^\star(N,\tfrac{1}{2},\rho,\sfff^\prime,\sfff) & = \frac{D_N(\sfff-\sfff^\prime)}{\sqrt{N}},
 \end{align}
 where $D_N(\sfff)$ is the $N$th order Dirichlet kernel
\begin{align} \label{Dirichlet_kernel}
D_N(\sfff) = \!\! \sum_{n \in [N]} \! e^{-\j 2 \pi \sfff n} = \frac{\sin(\pi N \sfff)}{\sin(\pi \sfff)}.
\end{align}
This kernel is real-symmetric and positive-definite, attaining its maximum at the origin, where it equals $D_N(0) = N$ \cite[Ch.~6]{BalanisBook}.
From Parseval's theorem, and consistently with \eqref{normalization}, its energy is
\begin{equation}
\int_{-1/2}^{1/2} D_N^2(\sfff) \, d\sfff = N .    
\end{equation}

\subsection{Impact of Coupling}
 
The uniformity in array gain breaks down for $\sfd < 1/2$ due to a surge in antenna interactions.
Expanding $J(\sfff)$ in \eqref{array_gain_def} according to \eqref{current_spectrum} while evaluating the integral
\begin{equation} \label{integral_aux}
\int_{-\sfd}^\sfd e^{\j 2 \pi \sfff n} \, d\sfff = \frac{\sin(2 \pi \sfd n)}{\pi n},
\end{equation}
what emerges is the generalized Rayleigh quotient \cite{Lee1966,Hansen1981}
\begin{align}   \label{directivity_lambda2_uniform_main_coupling}
 G(N,\sfd,\rho,\bj,\sfff) & = N \, \frac{|\ba(\sfff)^{\Herm} \, \bj|^2}{\bj^{\Herm} \bCC( \sfd,\rho) \bj} 
  \end{align}
with $\ba(\sfff)$ as defined in \eqref{a_N} 
while $\bCC \in \bbR^{N \times N}$ denotes the lossy coupling matrix  \cite{Bikhazi2007,Nossek2010}
\begin{equation} \label{C_lossy}
\bCC( \sfd,\rho) =  \bC(\sfd) + \rho \bI_N.
\end{equation}
The matrix $\bC( \sfd)$ represents the coupling in the lossless case. It is real, symmetric, Toeplitz, and positive-definite, its entries given by
\begin{align} \label{prolate_coupling_matrix}
[\bC(\sfd)]_{n,m} 
  = \frac{\sin(2 \pi \sfd (n-m))}{2 \pi \sfd(n-m)} \qquad n,m \in [N] 
\end{align}
specifying, up to a scaling factor, the power radiated by the $n$th antenna and captured by its $m$th counterpart \cite{Nossek2010}. The off-diagonal entries thus quantify the mutual coupling between distinct array antennas.
Let the ordered eigenvalues of $\bC$ be $\lambda_0(N,\sfd) > \cdots > \lambda_{N-1}(N,\sfd) > 0$ with eigenvectors $\bv_0(N,\sfd), \ldots, \bv_{N-1}(N,\sfd)$. 
The affine transformation in \eqref{C_lossy} does not alter the eigenvectors, while the eigenvalues of the lossy coupling matrix $\bCC$ become
\begin{align} \label{eig_lossy}
{\color{blue}\sflambda_k(N,\sfd,\rho)} = \lambda_k(N,\sfd) + \rho.
\end{align}
The lossy coupling matrix $\bCC$ retains the structure of its lossless counterpart and satisfies $\tr(\bCC) = N (1+ \rho)$.
For $\sfd={1}/{2}$, the coupling matrix reduces to identity, confirming that mutual coupling vanishes in uniform linear arrays under this configuration \cite{Nossek2010}.

The $Q$ factor in \eqref{Q_factor_norm}, in turn, can be obtained as \cite{Lee1966,Hansen1981} 
\begin{equation} \label{Q_factor_coupling}
Q(N,\sfd,\rho,\bj) =  \left(\bj^{\Herm} \, \bCC(\sfd,\rho) \,  \bj\right)^{-1} .
\end{equation}

\section{Supergain} \label{sec:supergain}

\subsection{Supergain Characterization}

To isolate the effect of coupling, arising from having $\sfd < 1/2$, from the effects of changing the current distribution, the maximum array gain $G^\star(N,\sfd,\rho,\sfff^\prime)$ in \eqref{directivity_lambda2_uniform_main_coupling} is normalized by its uncoupled,  lossless counterpart $G^\star(N,\tfrac{1}{2},0,\sfff^\prime)$ in \eqref{directivity_current_halfwavelength_max}. This ratio defines the \emph{supergain factor} 
\begin{align}   \label{supergain_def_def}
{\color{blue}\gamma(N,\sfd,\rho,\sfff^\prime)} = \frac{G^\star(N,\sfd,\rho,\sfff^\prime)}{G^\star(N,\tfrac{1}{2},0,\sfff^\prime)}. 
\end{align}
In the lossless case (i.e., for $\rho=0$), $\gamma$ specializes to {\color{blue}the \emph{superdirectivity factor} 
\begin{equation} \label{superdirectivity_def_def}
   \gamma_0(N,\sfd,\sfff^\prime) = \gamma(N,\sfd,0,\sfff^\prime) .
\end{equation}}
{\color{blue}We further introduce the endfire supergain, $\gamma_\endf$, and superdirectivity, $\gamma_{0,\endf}$, corresponding to currents optimized for beamforming directions parallel to the linear array axis (see Fig.~\ref{fig:ULA}).}

Substituting \eqref{directivity_lambda2_uniform_main_coupling} into \eqref{supergain_def_def}, the optimal current for any $\sfff^\prime$ is returned by
\begin{equation} \label{supergain_opt_problem}
    \underset{\bj : \|\bj\|=1}{\text{maximize}} \; \frac{\bj^{\Herm} \bA(\sfff^\prime) \bj}{\bj^{\Herm} \bCC(\sfd,\rho) \bj},
\end{equation}
where $\bCC$ is the 
coupling matrix in \eqref{C_lossy} while
$\bA = \ba \ba^{\Herm}$, recalling that $\ba$ is the array response vector in \eqref{a_N}.
Because the problem is unaffected by scalings, the constraint $\|\bj\|=1$ is momentarily dropped and 
$\bj$ is normalized such that the denominator is unity, i.e.,
\begin{equation}   \label{superdirectivity_problem}
\begin{aligned}
& \underset{\bj}{\text{maximize}}
& &  \bj^{\Herm} \bA(\sfff^\prime) \bj  \\
& \text{subject to}
& & \bj^{\Herm} \bCC(\sfd,\rho) \bj = 1.
\end{aligned}
\end{equation}
The KKT conditions yield the generalized eigenvalue problem
\begin{equation} \label{gen_eig_pb}
    \bA(\sfff^\prime) \bj^\star = {\color{blue}\nu^\star} \, \bCC(\sfd,\rho) \bj^\star \qquad\quad (\bj^\star)^{\Herm} \bCC(\sfd,\rho) \bj^\star = 1.
\end{equation}
Since $\bA(\sfff^\prime)$ is of unit rank, the generalized eigenvector $\bj^\star \in \text{span}(\bCC^{-1} \ba)$ is the unique (up to a factor) 
maximizer.
Reintroducing the normalization condition in \eqref{normalization},
\begin{equation} \label{opt_current_coupling} 
\bj^\star(N,\sfd,\rho,\sfff^\prime) = \frac{\bCC(\sfd,\rho)^{-1} \, \ba(\sfff^\prime)}{ \|\bCC(\sfd,\rho)^{-1} \, \ba(\sfff^\prime)\|}. 
\end{equation}
Also, multiplying each side of the eigenvalue equation in \eqref{gen_eig_pb} by $(\bj^\star)^{\Herm}$ while using the constraint gives
\begin{equation} \label{lambda_star}
{\color{blue}\nu^\star}  = (\bj^\star)^{\Herm} \bA(\sfff^\prime) \bj^\star = \gamma(N,\sfd,\rho,\sfff^\prime).
\end{equation}
%
Plugging \eqref{opt_current_coupling} into \eqref{lambda_star} and using the constraint in \eqref{gen_eig_pb} yields \cite{Nossek2010}
\begin{align}   \label{superdirectivity}
\gamma(N,\sfd,\rho,\sfff^\prime) 
 = \ba^{\Herm}(\sfff^\prime) \, \bCC^{-1}(\sfd,\rho) \, \ba(\sfff^\prime)
\end{align}
while
the $Q$ factor, evaluated for $\bj^\star$, is
\begin{align}   \label{Q_factor_coupling_maxdir}
Q^\star(N,\sfd,\rho,\sfff^\prime) & =  \frac{\|\bCC(\sfd,\rho)^{-1} \, \ba(\sfff^\prime)\|^2}{\ba(\sfff^\prime)^{\Herm} \, \bCC(\sfd,\rho)^{-1} \, \ba(\sfff^\prime) }. 
\end{align}

\begin{figure}
    \centering
    \begin{tikzpicture}
        \pgfplotstableread[col sep=comma]{data/supergain_theta_data.csv}{\data}
        
        \begin{axis}[
            width=11cm, height=7.5cm,
            xlabel={$\theta$ (deg)},
            ylabel={$\gamma(N,\sfd,\rho,\sfff^\prime)$},
            xmin=-90, xmax=90,
            xtick={-90, -60, -30, 0, 30, 60, 90},
            grid=both,
            tick label style={font=\small},
            label style={font=\small}
        ]
        
        \addplot [color=black, solid, thick] table [col sep=comma, x=theta, y=dir_025] {\data}
            node[pos=0.98, sloped, fill=white, inner sep=1pt, font=\footnotesize] {$1/4$};
            
        \addplot [color=black, solid, thick] table [col sep=comma, x=theta, y=dir_033] {\data}
            node[pos=0.98, sloped, fill=white, inner sep=1pt, font=\footnotesize] {$1/3$};

        \draw[black, dashed, thick] (axis cs:0, \pgfkeysvalueof{/pgfplots/ymin}) -- (axis cs:0, \pgfkeysvalueof{/pgfplots/ymax});
        

    \draw[gray, thick] (axis cs:-90,1) -- (axis cs:90,1)
            node[pos=0.95, sloped, fill=white, inner sep=1pt, font=\footnotesize] {$1/2$};

        
        \end{axis}
    \end{tikzpicture}
    \caption{Supergain in \eqref{superdirectivity}, {\color{blue}computed via the generalized eigenvalue problem in \eqref{gen_eig_pb},} as a function of $\theta(\sfff^\prime)$ 
for various antenna spacings $\sfd$. Array with {\color{blue}$N=7$} isotropic antennas and loss factor $\rho=10^{-16}$. Letting $\sfd < 1/2$ ignites coupling effects, which redistribute power density towards the endfire directions.}
\label{fig:dir_f}
\end{figure}

Averaging \eqref{superdirectivity} over $\sfff^\prime$ correctly yields 1 for lossless antennas, confirming that any supergain necessarily comes at the expense of reduced gain elsewhere. 
This is consistent with energy conservation and indicates that coupling can only redistribute power across visible directions,  depending on $\sfd$. 
This is exemplified in Fig.~\ref{fig:dir_f}, where $\gamma(N,\sfd,\rho,\sfff^\prime)$ is plotted as a function of $\theta = \arcsin(\sfff^\prime/\sfd)$ over the visible region, for {\color{blue}$N=7$ antennas}, $\rho = 10^{-16}$, and various $\sfd$. As $\sfd$ falls below $1/2$, coupling arises, redistributing power towards the endfire directions. 
{\color{blue}The objective is computed by solving the generalized eigenvalue problem in \eqref{gen_eig_pb}.}
Although numerical results must be interpreted with care, due to the ill-conditioning of $\bCC$ as $\sfd$ shrinks, they show consistent behavior for moderate $N$ values.
This ill-conditioning manifests in \eqref{gen_eig_pb} as an expanding---encompassing more directions $\sfff^\prime$---intersection between the space spanned by the eigenvectors of $\bCC$ corresponding to arbitrarily small eigenvalues and the 
nullspace of $\bA(\sfff^\prime)$,
resulting in numerical instability \cite{Golub1973}. 

\subsection{Geometric Interpretation}
 
\begin{figure}
\centering\vspace{-0.0cm}
\includegraphics[width=.6\linewidth]{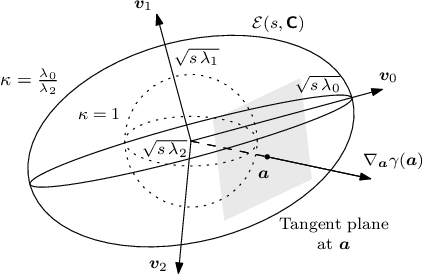}  
\caption{Geometric interpretation 
for $N=3$. The supergain factor defines an ellipsoidal surface, with each point $\ba$ corresponding to a physical direction via the mapping $\sfff \mapsto \ba$ in \eqref{a_N}.
The ellipsoid's maximum-to-minimum elongation 
encodes the dynamic range, 
 while the $Q$ factor quantifies the local rate of change near any point on the 
surface.} 
\label{fig:ellipsoid}
\end{figure}

From the mapping $\sfff^\prime \mapsto \ba$ in \eqref{a_N}, the level set defined by $\gamma(N,\sfd,\rho,\ba) = s$ in \eqref{superdirectivity} describes the surface of an $N$-dimensional ellipsoid, namely (see Fig.~\ref{fig:ellipsoid})
\begin{align}   \label{superdirectivity_geometric}
\cE(s,\bCC) = \{ \ba : \|\ba\|=1 , \ba^{\Herm} \, (s \, \bCC)^{-1} \, \ba = 1 \},
\end{align}
for $\tr(\bCC) = N (1+ \rho)$ and every admissible $s$, meaning $s\in [\sflambda_0^{-1},\sflambda_{N-1}^{-1}]$.
The 
extrema of \eqref{superdirectivity} occur when $\ba$ aligns with either the minimum- or the maximum-eigenvalue eigenvector of $\bCC$, respectively $\bv_{N-1}$ or $\bv_0$, giving 
\begin{equation}   \label{superdirectivity_lambda}
\max_{\ba: \|\ba\|=1} \, \gamma(N,\sfd,\rho,\ba) = \frac{1}{\sflambda_{N-1}(N,\sfd,\rho)} \qquad  \min_{\ba:\|\ba\|=1} \, \gamma(N,\sfd,\rho,\ba) = \frac{1}{\sflambda_0(N,\sfd,\rho)}.
\end{equation}
{\color{blue}(These extrema are achievable only if the corresponding eigenvectors lie on the array manifold traced by $\ba(\sfff)$ for all possible $\sfff$.)}
For $\sfd=1/2$, it holds that $\bCC = (1+\rho) \bI$ and the ellipsoid collapses into a unit-radius hypersphere. 
As $\sfd$ shrinks, coupling deforms this sphere into an ellipsoid introducing an anisotropic scaling specified by $\bCC$ (see again Fig.~\ref{fig:ellipsoid}).
The ratio between the squared major and minor semi-axis lengths 
equals the supergain's dynamic range 
\begin{equation} \label{cond_number}
\kappa(N,\sfd,\rho)  = \frac{\max_{\ba : \|\ba\|=1} \, \gamma(N,\sfd,\rho,\ba)}{\min_{\ba : \|\ba\|=1} \, \gamma(N,\sfd,\rho,\ba)} = \frac{\sflambda_0(N,\sfd,\rho)}{\sflambda_{N-1}(N,\sfd,\rho)} 
\ge 1,
\end{equation}
which coincides with the condition number of $\bCC$, with equality in the uncoupled case.

At any point $\ba = \ba(\sfff^\prime)$ on the level set, from \eqref{Q_factor_coupling_maxdir},
\begin{align}   \label{Q_factor_coupling_maxdir_geometric_ellipsoid}
Q^\star(N,\sfd,\rho,\ba) & =  s^{-1} \, \|\bCC^{-1}(\sfd,\rho) \, \ba\|^2 \\
& = \frac{s^{-1}}{4} \, \| \nabla_\ba \gamma(N,\sfd,\rho,\ba)  \|^2
\end{align}
where $\nabla_\ba \gamma = 2 \bCC^{-1} \ba$ is the gradient vector of $\gamma$ with respect to $\ba$. This vector is normal to the ellipsoid's surface and its norm quantifies the rate of change of $\gamma$ at that point. 
Thus, the $Q$ factor measures the sensitivity of supergain to perturbations in directionality around any $\ba$.

This geometric interpretation clarifies why a high supergain is always accompanied by a high $Q$  factor, while the converse need not hold \cite{Taylor1955,Lee1966}. Anisotropic ellipsoids exhibit steep gradients near their elongated axes, where supergain is highest. This results in both large $\gamma$ and high $Q^\star$.
Conversely, sharp bumps in the ellipsoid's surface can also arise in regions where supergain is moderate, due to local curvature, yielding a high $Q$ for small values of $\gamma$.

\subsection{{\color{blue}Mechanisms} of High Supergain}


\begin{figure}
    \centering 
\begin{tikzpicture}
    \pgfplotstableread[col sep=comma]{data/spectral_concentration_data.csv}{\data}
    
    \tikzset{
    big arrow/.style={<->, >={Stealth[length=7pt, width=5pt]}, thick},
    embedded label/.style={midway, fill=white, inner sep=1.5pt, font=\small}
}

    \begin{axis}[
        width=11cm, height=7.5cm, 
        xlabel={$k/N_0$},
        ylabel={$2\sfd \; \sflambda_k(N,\sfd,\rho)$},
        xmin=0, xmax=2,
        ymin=0, ymax=1,
        grid=both,
        unbounded coords=discard, 
        tick label style={font=\small},
        label style={font=\small}
    ]
    
    \draw[black, thick] (axis cs:1,0) -- (axis cs:1,1);
      
    \addplot [color=black, solid, thick] table [col sep=comma, x=k_5, y=eig_5] {\data}
            node[pos=0.37, sloped, fill=white, inner sep=1pt, font=\footnotesize] {$41$};
            
    \addplot [color=black, solid, thick] table [col sep=comma, x=k_10, y=eig_10] {\data}
            node[pos=0.38, sloped, fill=white, inner sep=1pt, font=\footnotesize] {$81$};
            
    \addplot [color=black, solid, thick] table [col sep=comma, x=k_30, y=eig_30] {\data}
            node[pos=0.39, sloped, fill=white, inner sep=1pt, font=\footnotesize] {$241$};

  \draw[big arrow] (axis cs:0, 0.4) -- (axis cs:1, 0.4) 
    node[embedded label] {$\cK_1$};
\draw[big arrow] (axis cs:0.6, 0.5) -- (axis cs:1.4, 0.5) 
    node[embedded label] {$\cK_2$};
\draw[big arrow] (axis cs:1, 0.4) -- (axis cs:2, 0.4) 
    node[embedded label] {$\cK_3$};

    
    \end{axis}
\end{tikzpicture}
\caption{Normalized sorted eigenvalues of the coupling matrix $\bCC(\sfd,\rho)$ for various $N$ (corresponding to apertures of electrical size $\sfL = (N-1) \sfd = 5, 10, 30$) with $\sfd = 1/8$ and $\rho = 10^{-16}$. 
The eigenvalue indices are normalized to reveal the asymptotic polarization of the curves as the aperture increases relative to the wavelength. This asymptotic behavior naturally partitions the index set into three regions: $\cK_1$ and $\cK_3$, where the eigenvalues respectively concentrate near $1$ and $0$, and $\cK_2$, which defines the transition region.}
\label{fig:eig_C}
\end{figure}

From \eqref{superdirectivity}, supergains much higher than unity require that the coupling matrix $\bCC$ be poorly conditioned. Such high supergains are attained along directions specified by $\sfff^\prime$ for which $\ba(\sfff^\prime)$ 
intersects the essential nullspace of $\bCC$---spanned by eigenvectors associated with eigenvalues that are practically zero \cite{Marzetta2019}. 
In these directions, the optimal current distribution \eqref{opt_current_coupling}
sustains a strong reactive field, as evidenced by $Q^\star\gg1$ in \eqref{Q_factor_coupling_maxdir}. 
This is consistent with supergains relying on the excitation of evanescent waves associated with spatial frequencies in the invisible spectrum \cite{Harrington1958}. Such waves decay exponentially and contribute solely to reactive power, rather than to far-field radiation \cite{PlaneWaveBook,Marzetta2019}.

The optimal currents are inherently unstable and highly sensitive to small perturbations in the antenna excitations or positions \cite{Morgan55}. Antenna losses mitigate this behavior through $\rho$ in \eqref{eig_lossy}, acting as a physical regularization that elevates the small eigenvalues of $\bCC$, improving conditioning while limiting the achievable supergain.

As small eigenvalues indicate strong coupling, the expansion of the essential nullspace of $\bCC$ reflects a surge in coupling. 
Two distinct {\color{blue}mechanisms} by which this occurs are next examined. 

\begin{itemize}
    \item 
For a given $N$, supergain arises as $\sfd$ shrinks, implying a downsizing of the aperture in wavelength units \cite{Schelkunoff1943,Uzkov}. 
As $\sfd\to 0$, the array collapses into a single antenna, 
as reflected by $\bCC$ effectively becoming rank-1.
In this limit, the array \emph{spatially concentrates}, leading to
a quadratic scaling of the endfire gain for lossless antennas \cite{Yaghjian2005}, corresponding to a linear scaling of the endfire supergain factor, namely
\begin{equation} \label{superdirectivity_endfire_lossless}
    \lim_{\sfd\to 0} \gamma_{0,\endf}(N,\sfd) = N.
\end{equation}

Antenna losses via $\rho$ induce a unimodal dependence of $\gamma_\endf$ on $\sfd$, instead of the monotonic behavior observed in the lossless case as $\sfd$ shrinks \cite{Nossek2010conf,Nossek2010}. 
The supergain peak now occurs at a nonzero $\sfd$, which is determined by $N$ and $\rho$, making such gains achievable in practice through realizable antennas.
The asymptotic limits
of the spatial concentration {\color{blue}mechanism} are quantified in Section~\ref{sec:spatial_concentration}.
\item 
A complementary mechanism to spatial concentration is \emph{spectral concentration}. Both mechanisms drive some eigenvalues of $\bCC$ toward zero, but they act in distinct domains.
Spectral concentration yields supergain through an increase in $N$, for fixed $\sfd <1/2$, implying a widening aperture in wavelength units.
Fig.~\ref{fig:eig_C}  illustrates this phenomenon by plotting the ordered eigenvalues of $\bCC$ for growing $N$ and {\color{blue}$\sfd = 1/8$}, $\rho = 10^{-16}$.
As the aperture widens, the eigenvalues polarize into $0$ and $1$, 
the transition between the two levels occurring at the index \cite{Slepian1978}
\begin{equation} \label{DOF_0}
N_0 = 2\sfd N.
\end{equation}
This number represents the spatial degrees of freedom (DOF) of the array, up to an error of order $\cO(\log N)$, which characterizes the width of the transition region.
Hence, normalizing the eigenvalue indices by $N_0$ reveals the concentration as $N\to \infty$ while concealing other effects---such as the impact of inherent losses---asymptotically.

The eigenvalue polarization asymptotically partitions the array into $N_0$ uncoupled antennas and $N-N_0 = N(1-2\sfd)$ fully coupled ones. 
Since full coupling generates a linear endfire supergain factor, this leads to
\begin{equation} \label{gain_growing_N}
\lim_{N\to\infty} \frac{\gamma_{0,\endf}(N,\sfd)}{N} = \tau(\sfd),
\end{equation}
with the impact of $\sfd$ subsumed into the scaling factor $0<\tau(\sfd)<1$. Particularly, $\tau(\sfd)$ approaches $0$ as $\sfd\to 1/2$ and $1$ as $\sfd\to 0$, respectively.
Accounting for losses via $\rho$, the behavior in \eqref{gain_growing_N} persists provided that the essential nullspace of the coupling matrix continues to grow with $N$.
The asymptotic scaling factor $\tau(\sfd)$ and the loss factor $\rho$ compatible with the spectral concentration sustaining this scaling are quantified in Section~\ref{sec:spectral_conc_Nlarge}.
\end{itemize}

\section{Concentration in the Discrete-Space Domain}
\label{sec:spectral_concentration}

This section recalls several results on the concentration of discrete-space sequences, enabling the subsequent quantitative analysis of supergain.
The formulation transposes the classical time-domain analysis \cite{Slepian1961,Slepian1964,Slepian1976,Slepian1978,ProlateBook,FranceschettiBook} to the spatial domain, leveraging the isomorphism between the two domains established in Section~\ref{sec:space-time}. 

\begin{definition}[Indexlimited and Bandlimited Sequences \cite{Donoho1989,FranceschettiBook}] \label{rev:def_bandlimits_indexlimits}
Let $\mathscr{I}_N$ and $\mathscr{B}_\sfd$ be the subspaces spanned by unit-norm sequences supported on the index set $[N]$ defined in~\eqref{set_N} and whose spectra are supported in $[-\sfd,\sfd]$, respectively.
The \emph{indexlimiting operator} $I_N$ truncates a sequence $\{x_n\}_{n\in\bbZ}$ to $[N]$:
\begin{equation}
(I_N x)_n =  \mathds{1}_{[N]}(n) x_n =
\begin{cases}
x_n & n\in [N] \\
0 & \text{otherwise}
\end{cases}
\end{equation}
while the \emph{bandlimiting operator} $B_\sfd$ reconstructs $x_n$ from its spectrum $X(\sfff)$ by retaining only components within the spatial bandwidth $\sfd$,
\begin{equation} \label{Bd_operator}
(B_\sfd x)_n = F^{-1} \mathds{1}_{[-\sfd,\sfd]}(\sfff) F x_n = \int_{-\sfd}^\sfd X(\sfff) \, e^{\j 2 \pi \sfff n} \, d\sfff.
\end{equation}
Both $I_N$ and $B_\sfd$ act as orthogonal projections (in the energy sense) onto $\mathscr{I}_N$ and $\mathscr{B}_\sfd$.
\end{definition}

The asymptotic dimension of the space spanned by sequences that are simultaneously concentrated in both space and discrete spatial frequency is formalized next.

\begin{definition}[Spatial Degrees of Freedom \cite{Slepian1976,Slepian1978,Slepian1983}]
Let $j_n \in \mathscr{I}_N$ with its spectrum $J(\sfff)$ being $\epsilon_\sfd$-concentrated in $\mathscr{B}_\sfd$. Denote by $N_0(N,\sfd,\epsilon_\sfd,\epsilon)$ the minimum number of basis functions 
$U_0(N,\sfd,\sfff),\dots,U_{N_0-1}(N,\sfd,\sfff)$ with indexlimited Fourier inverse transform required to approximate---through  linear combination with some expansion coefficients $\{J_k\}$---any spectrum $J(\sfff)$ over $[-\sfd,\sfd]$ with $\epsilon$-accuracy,
\begin{equation} \label{N_0_formal}
N_0 = \min \left\{M \in \bbN : \| J - \sum_{k=0}^{M-1} J_k \, U_k \|_{[-\sfd,\sfd]} \le \epsilon \right\}  \qquad  0 \le  \epsilon_\sfd < \epsilon <1.
\end{equation}
Asymptotically,
\begin{equation}
\lim_{N\to \infty} \frac{N_0(N,\sfd,\epsilon_\sfd,\epsilon)}{N} = 2 \sfd,
\end{equation}
with the lower-order dependence on $\epsilon_\sfd$ and $\epsilon$ vanishing. For $\sfd = 1/2$, $N_0 = N$.
\end{definition}


Since the norm in \eqref{N_0_formal} represents the residual energy outside the subspace spanned by the basis functions, the optimal subspace consists of the indexlimited sequences whose spectra $U_k(N,\sfd,\sfff)$ are most concentrated within $[-\sfd,\sfd]$ \cite[Ch.~2]{FranceschettiBook}. These are obtained as solutions of the following eigenvalue problem \cite[Ch.~3]{FranceschettiBook}.

\begin{lemma}[Maximally Concentrated Spectra \cite{Slepian1978,FranceschettiBook}]  \label{rev:eig_dpswf}
Let $D_N(\sfff)$ be the Dirichlet kernel in~\eqref{Dirichlet_kernel}. The ordered eigenfunctions $U_0(N,\sfd,\sfff),\ldots,U_{M-1}(N,\sfd,\sfff)$ solving \cite[Eq.~10]{Slepian1978}
\begin{align} \label{int_eq_sin_sin_Dirichlet} 
\int_{-\sfd}^{\sfd} D_N(\sfff-\sfg) \, U_k(N,\sfd,\sfg) \, d \sfg = \mu_k(N,\sfd) \, U_k(N,\sfd,\sfff) \qquad |\sfff|\le \sfd,
\end{align}
span an optimal $M$-dimensional subspace of $\mathscr{I}_N$ for any $M \le N$. The kernel's real symmetry and positive-definiteness ensure that the ordered eigenvalues $1 > \mu_0 > \dots > \mu_{N-1} > 0$ are real and distinct, and that the eigenfunctions are unique and orthogonal over $[-\sfd,\sfd]$.
\end{lemma}

Solving the concentration problem in \eqref{int_eq_sin_sin_Dirichlet} is made possible by establishing a correspondence with a well-known ordinary differential equation. 

\begin{lemma}[Discrete prolate spheroidal wave functions \cite{Slepian1978}]  \label{rev:dpswf_diff_eq}
The eigenfunctions $U_0(N,\sfd,\sfff),\ldots,U_{N-1}(N,\sfd,\sfff)$ are also solution to the Sturm-Liouville eigenvalue problem
\begin{equation} \label{ODE}
    M \, U = \theta \, U,
\end{equation}
induced by the second-order homogeneous differential operator
\begin{align} \label{SL_eig}
M & =  \frac{1}{(2 \pi)^2} \frac{d}{d\sfff} \left\{ \left( \cos(2\pi \sfff) - A\right) \frac{d}{d\sfff} \right\} + \frac{N^2 -1}{4} \cos(2 \pi \sfff), 
\end{align}
where 
\begin{equation} \label{A_cos_def}
    A = \cos(2\pi \sfd).
\end{equation}
These eigenfunctions are associated with a discrete set of real positive values of $\theta$, the eigenvalues of $M$, namely $\theta_0(N,\sfd) > \ldots > \theta_{N-1}(N,\sfd) > 0$. 
\end{lemma}

The correspondence between the concentration problem in \eqref{int_eq_sin_sin_Dirichlet} and ordinary differential equation in \eqref{SL_eig} reveals detailed information about the structure of the $U_k$'s. 

\begin{remark}[Symmetry and Zeros \cite{Slepian1978}] \label{ref:symmetry}
The eigenfunctions $U_0(N,\sfd,\sfff),\ldots,U_{N-1}(N,\sfd,\sfff)$ form a complete set in $\mathscr{L}^2[-\tfrac{1}{2},\tfrac{1}{2}]$.
Each $U_k$ is symmetric for even $k$ and antisymmetric for odd $k$, with period~$1$ when $N$ is odd. It has $N-1$ simple zeros in  $[-\tfrac{1}{2},\tfrac{1}{2}]$, of which $k$ are within the open interval $(-\sfd,\sfd)$. 
\end{remark}


\begin{corollary}[Duality \cite{Slepian1978}] \label{rev:duality}
The ordered eigenvalues $\mu_0, \ldots, \mu_{N-1}$ in \eqref{int_eq_sin_sin_Dirichlet} are also the real solution to the Hermitian counterpart operator of \eqref{int_eq_sin_sin_Dirichlet}, 
\begin{equation} \label{sum_eq_sin}
\boldsymbol{\Omega}(\sfd) \, \bv_k(N,\sfd) = \mu_k(N,\sfd) \, \bv_k(N,\sfd) \qquad \quad k=0, \ldots, N-1,
\end{equation}
where $\boldsymbol{\Omega}\in\bbR^{N\times N}$ is the real, symmetric, and positive-definite \emph{prolate matrix} with entries
\begin{align} \label{prolate_matrix}
[\boldsymbol{\Omega}(\sfd)]_{n,m} = \frac{\sin (2\pi \sfd (n-m))}{\pi (n-m)} \qquad n,m \in [N].
\end{align}
This matrix admits a unique set of real orthonormal eigenvectors $\bv_0(N,\sfd),\ldots,\bv_{N-1}(N,\sfd)$, each equals the discrete prolate spheroidal sequence $v_n^{(k)}(N,\sfd)$
indexlimited to $[N]$. 
\end{corollary}

\begin{lemma}[Fourier relationship \cite{Slepian1978}] \label{rev:analytic}

The discrete prolate spheroidal sequences and discrete prolate spheroidal wave functions are related by 
\begin{equation} \label{Fourier_discrete_pswf} 
U_k(N,\sfd,\sfff) 
= \epsilon_k \sqrt{N} \, \ba(\sfff)^{\Herm} \bv_k(N,\sfd) \qquad \quad |\sfff|\le \frac{1}{2},
\end{equation}
where $\ba(\sfff)$ is defined in \eqref{a_N} and $\epsilon_k$ is $1$ or $\j$ depending on whether $k$ is even or odd.
Conversely,
\begin{equation}  \label{Fourier_inv_discrete_pswf_complete}
\bv_k(N,\sfd)  = \frac{\sqrt{N}}{\epsilon_k}  \int_{-1/2}^{1/2} U_k(N,\sfd,\sfff) \, \ba(\sfff) \, d\sfff,
\end{equation} 
confirming that the sequences, whose spectra are $U_k(N,\sfd,\sfff)$, belong to $\mathscr{I}_N$.
Also, 
\begin{equation}  \label{Fourier_inv_discrete_pswf_complete_bandlimited}
U_k(N,\sfd,\sfff) = \epsilon_k \, \mu_k(N,\sfd) \, \sum_{n\in \bbZ} v_n^{(k)}(N,\sfd) \, e^{-\j 2\pi \sfff n} \qquad \quad |\sfff|\le \sfd,
\end{equation} 
consistent with the spectrum of $v_n^{(k)}(N,\sfd)$ being in $\mathscr{B}_\sfd$.
Also, $\epsilon_k$ ensures real-valued sequences for all $k$, consistent with the symmetry of $U_k(N,\sfd,\sfff)$. 
%
\end{lemma}

\begin{corollary}[Double Orthogonality \cite{Slepian1978,ProlateBook}] \label{rev:double-orthogonality} 
Under the normalization conditions $\|U_k\|=1$ and $\|\bv_k\|=1$, the discrete prolate spheroidal wave functions and corresponding sequences satisfy, for $k,\ell =0, \ldots, N-1$,
\begin{align} \notag
  \int_{-\sfd}^{\sfd} U_k(N,\sfd,\sfff) \, U_\ell(N,\sfd,\sfff) \, d\sfff & = \mu_k(N,\sfd) \int_{-1/2}^{1/2} U_k(N,\sfd,\sfff) \, U_\ell(N,\sfd,\sfff) \, d\sfff  \\ & \hspace{4cm} \label{double_orthogonality_dpswf}
   = \mu_k(N,\sfd) \, \delta_{k \ell} \\ \label{double_orthogonality_dpss}
 \bv_k(N,\sfd)^{\Trans} \bv_\ell(N,\sfd) & = \mu_k(N,\sfd) \sum_{n\in \bbZ} v^{(k)}_n(N,\sfd) v^{(\ell)}_n(N,\sfd) = \delta_{k\ell}.
\end{align}
\end{corollary}


From Corollary~\ref{rev:double-orthogonality}, each 
eigenfunction $U_k(N,\sfd,\sfff)$ contains a fraction $0<\mu_k<1$ of its energy within $[-\sfd,\sfd]$~\cite[Sec.~2.5]{FranceschettiBook}. The principal eigenfunction $U_0$ achieves the highest concentration $\mu_0$, followed by $U_1, U_2, \ldots$ with progressively lower concentration $\mu_1, \mu_2, \ldots$.

  Their truncations

\begin{remark}[Orthogonal Basis \cite{Slepian1978}] \label{ref:completeness}
The 
sequences {\color{blue}$\{v^{(k)}_n\}_{k=0}^\infty$ form a complete orthogonal family in the space $\mathscr{B}_\sfd$. The first $N$ sequences, $v^{(0)}_n,\ldots,v^{(N-1)}_n$, are the discrete prolate spheroidal sequences. Their truncation, $\bv_0(N,\sfd),\ldots,\bv_{N-1}(N,\sfd)$, form an orthonormal basis for $\mathscr{I}_N$. }
Consequently, any $\bj \in \mathscr{I}_N$ admits a unique expansion in terms of $\{\bv_k(N,\sfd)\}_{k=0}^{N-1}$.
\end{remark}

\section{Concentration of Uniform Linear Arrays} \label{sec:spectral_conc_ULA}


\subsection{Spectral Representation} \label{sec:reconciliation}

From Remark~\ref{ref:completeness}, any indexlimited current $\bj \in \mathscr{I}_N$ admits the orthogonal representation
\begin{equation} \label{current_spectrum_pswf_discrete}
\bj(N,\sfd) = \sum_{k=0}^{N-1} \epsilon_k J_k(N,\sfd) \, \bv_k(N,\sfd).
\end{equation}
A discrete-space Fourier transform of both sides as per \eqref{current_spectrum} gives the spectral representation 
\begin{equation} \label{current_spectrum_pswf_discrete}
J(\sfff) = \sum_{k =0}^{N-1} J_k(N,\sfd) \, U_k(N,\sfd,\sfff) \qquad \quad |\sfff|\le \frac{1}{2},
\end{equation}
where $U_k(N,\sfd,\sfff)$ appears because of \eqref{Fourier_discrete_pswf}.
By the double orthogonality of the $U_k$'s (see \eqref{double_orthogonality_dpswf} in Corollary~\ref{rev:double-orthogonality}), the coefficients can be found to be
\begin{align}  \label{coeff_Jk_discrete}
J_k(N,\sfd) &  =   \int_{-1/2}^{1/2} J(\sfff) \, U_k(N,\sfd,\sfff) \, d\sfff  = \frac{1}{\mu_k(N,\sfd)} \int_{-\sfd}^{\sfd} J(\sfff) \, U_k(N,\sfd,\sfff) \, d\sfff  ,
\end{align}
satisfying 
\begin{equation} \label{coeff_Jk_discrete_normalization}
\sum_{k=0}^{N-1} |J_k(N,\sfd)|^2  = 1
\end{equation}
by virtue of \eqref{normalization} and Parseval's theorem.
Building on the link between spatial coupling and spectral concentration, Appendix~\ref{app:optimality} derives the supergain factor at any discrete spatial frequency $\sfff^\prime$ as
\begin{equation}     \label{max_directivity_current_fprime}
\gamma(N,\sfd,\rho,\sfff^\prime)  = \frac{1}{N} \, \sum_{k =0}^{N-1}  \frac{|U_k(N,\sfd,\sfff^\prime)|^2}{\sflambda_k(N,\sfd,\rho)} ,
\end{equation}
where the sum reflects the orthogonality of $\{U_k(N,\sfd,\sfff)\}$.
The corresponding $Q$ factor is
\begin{equation}     \label{Q_factor_norm_prolate}
Q^\star(N,\sfd,\rho,\sfff^\prime)  = \frac{1}{c(N,\sfd,\rho,\sfff^\prime)} \left(\sum_{k =0}^{N-1}  \frac{|U_k(N,\sfd,\sfff^\prime)|^2}{\sflambda_k(N,\sfd,\rho)}\right)^{\!-1},
\end{equation}
where 
\begin{equation} \label{norm_const}
\frac{1}{c(N,\sfd,\rho,\sfff^\prime)} = \sum_{k=0}^{N-1} \frac{|U_k(N,\sfd,\sfff^\prime)|^2}{\sflambda_k^2(N,\sfd,\rho)}
\end{equation}
to enforce \eqref{coeff_Jk_discrete_normalization}.
Such supergain and Q factors are attained by the optimal coefficients 
\begin{equation} \label{coeff_Jk_discrete_optimal_aux}
J_k^\star(N,\sfd,\rho,\sfff^\prime)  = \frac{\sqrt{c(N,\sfd,\rho,\sfff^\prime)} \, U_k(N,\sfd,\sfff^\prime)}{\sflambda_k(N,\sfd,\rho)},
\end{equation}
which yield the maximum-gain spectrum 
\begin{equation} \label{delta_spectrum_discrete_pswf}
J^\star(N,\sfd,\rho,\sfff^\prime,\sfff) = \sqrt{c(N,\sfd,\rho,\sfff^\prime)} \, \sum_{k =0}^{N-1} \frac{U_k(N,\sfd,\sfff^\prime)}{\sflambda_k(N,\sfd,\rho)} \, U_k(N,\sfd,\sfff) .
\end{equation}
Inverting \eqref{delta_spectrum_discrete_pswf} as per \eqref{current_spectrum_inverse} while using \eqref{Fourier_inv_discrete_pswf_complete} gives the corresponding current vector
\begin{equation} \label{optimal_current_discrete_M}
\bj^\star(N,\sfd,\rho,\sfff^\prime) = \sqrt{c(N,\sfd,\rho,\sfff^\prime)} \,  \sum_{k =0}^{N-1}   \frac{\displaystyle \epsilon_k U_k(N,\sfd,\sfff^\prime)}{\displaystyle \sflambda_k(N,\sfd,\rho)} \, \bv_k(N,\sfd) .
\end{equation}

\subsection{Prolate Matrix and Coupling Matrix}


To align with the spatial-domain formulation of Section~\ref{sec:mutual_coupling}, the lossless coupling matrix $\bC(\sfd)$ in \eqref{prolate_coupling_matrix} can be related to the $N$-dimensional prolate matrix $\boldsymbol{\Omega}(\sfd)$ in \eqref{prolate_matrix} via
\begin{equation} \label{Omega_C_relation}
 \bC(\sfd) = (2\sfd)^{-1} \, \boldsymbol{\Omega}(\sfd).
\end{equation}
Note that the coupling matrix is well-defined even for $\sfd=0$.
The two matrices share the same eigenvectors, $\bv_k(N,\sfd)$, while their eigenvalues differ by a constant factor, specifically
 \begin{align} \label{coupling_prolate_identity}
\lambda_k(N,\sfd) & = (2\sfd)^{-1} \,  \mu_k(N,\sfd) \\ \label{avg_spectral_conc}
& = \frac{1}{2 \sfd} \int_{-\sfd}^{\sfd} |U_k(N,\sfd,\sfff)|^2 \, d\sfff = \frac{1}{2 \sfd} \, \| U_k \|^2_{[-\sfd,\sfd]}.
\end{align}
From \eqref{avg_spectral_conc}, each eigenvalue of the coupling matrix represents the \emph{average} spectral concentration of the unit-norm spectrum $U_k(N,\sfd,\sfff)$ 
over $[-\sfd,\sfd]$.

Prolate matrices are classically encountered in the study of statistical correlation, where they capture the covariance in limited domains. Their emergence in the analysis of coupling is not coincidental, rather it reflects a deep isomorphism between correlation and coupling \cite{PizzoJSAIT25}.


For $\sfd=\tfrac{1}{2}$, the prolate and coupling matrices reduce to the identity matrix for lossless antennas. Since there is then only one eigenvalue equal to unity with multiplicity $N$, the orthonormal basis $\big\{ \bv_k(N,\tfrac{1}{2}) \big\}$ is not unique and can be chosen arbitrarily.
Without symmetry constraints, selecting the canonical basis of $\bbR^N$, $\bv_k(N,\tfrac{1}{2}) = \bee_k$, yields the discrete Fourier transform $U_k(N,\tfrac{1}{2},\sfff) = e^{-\j 2\pi \sfff (k-(N-1)/2)}$ via \eqref{Fourier_discrete_pswf} (omitting the normalization factor $\epsilon_k$ enforcing such symmetries).
Plugging this expression into \eqref{norm_const} with $\rho=0$ reveals $c = N^{-1}$.
Applying a change of variable and using \eqref{Dirichlet_kernel}, \eqref{delta_spectrum_discrete_pswf} reduces to the Dirichlet kernel in \eqref{Dirichlet_kernel_current},
\begin{equation} \label{delta_spectrum_discrete_pswf_dhalf}
J^\star(N,\tfrac{1}{2},0,\sfff^\prime,\sfff) = \frac{1}{\sqrt{N}} \, \sum_{k=0}^{N-1} e^{-\j 2\pi (\sfff-\sfff^\prime) \left(k-\tfrac{N-1}{2}\right)} =  \frac{D_N(\sfff-\sfff^\prime)}{\sqrt{N}}.
\end{equation}
For $\sfd < 1/2$, the spectrum of \eqref{opt_current_coupling} follows by substituting \eqref{Fourier_discrete_pswf} 
into \eqref{delta_spectrum_discrete_pswf}, noticing that $\bv_k(N,\sfd)$ are the eigenvectors of the coupling matrix and $|\epsilon_k|=1$ (since $\epsilon_k$ equals $1$ or $\j$). 

\section{Spatial Concentration}

\label{sec:spatial_concentration}

For vanishing $\sfd$, the $U_k$'s converge 
to Legendre polynomials \cite{Slepian1978}. 
This asymptotic correspondence, detailed in Appendix~\ref{app:Legendre_poly}, characterizes the eigenstructure of \eqref{ODE} with eigenvalues satisfying, for $k=0,\ldots,N-1$,
\begin{equation}\label{theta_asymptotic}
\theta_k(N,\sfd)
= \frac{N^2-1}{4} - \frac{k(k+1)}{2} + \cO({\color{blue}\sfd^2}) 
\end{equation}
and corresponding eigenfunctions satisfying, {\color{blue}uniformly for $f=\sfff/\sfd \in [-1,1]$,
\begin{equation}\label{dpswf_legendre}
U_k(N,\sfd,f \sfd) = \sqrt{G_k(N) (2k + 1)} \, (2\pi \sfd)^{k}  \,  P_k(f) \, \left(1 + \cO(\sfd)\right), 
\end{equation}
where
\begin{equation} \label{G_k}
    G_k(N) = \frac{2^{2k} (k!)^6}{(2k+1)^2 [(2k)!]^4} \prod_{j=-k}^k (N-j).
\end{equation}
In turn,} $P_k$ is the Legendre polynomial of degree $k$, {\color{blue}
satisfying
\begin{equation} \label{Pk_bound}
    |P_k(f)| \le 1 \qquad f\in[-1,1],
\end{equation}
with equality attained at the endpoints.}
These polynomials obey the orthogonality relation \cite[Eq.~22.2.1]{AbramowitzStegun}
\begin{equation} \label{norm_Legendre}
\int_{-1}^1 P_k(f) P_\ell(f) \, df = \frac{2}{2k + 1} \delta_{\ell k},
\end{equation}
forming a complete orthogonal basis of $\mathscr{L}^2[-1,1]$. 

Exploiting the Legendre eigenstructure, and that the associated $\sflambda_k$'s are the corresponding average spectral concentrations over $[-\sfd,\sfd]$ asymptotically, Appendix~\ref{app:superdirectivity_vanishing_aperture} analyzes the supergain factor in \eqref{max_directivity_current_fprime} along {\color{blue}the endfire direction as $\sfd$ vanishes.}
Two distinct {\color{blue}asymptotic} regimes emerge, according to $\rho$ relative to 
{\color{blue}$\sfd$ for fixed $N$.
Precisely, a \emph{supergain regime} emerges when the eigenvalues remain larger than $\rho$, allowing high-coupling contributions to sustain a linear growth with $N$.  Conversely, a \emph{loss-dominated regime} occurs when $\rho$ eclipses the smallest eigenvalues, masking their contributions and causing the supergain factor to saturate as $N$ increases.}
  
\begin{itemize}
    \item {\color{blue}\emph{Supergain regime:}
$\rho$ decays at least polynomially with $\sfd$, satisfying $\rho(\sfd) = o(\sfd^{2N-2})$ for $\sfd\to 0$, and the endfire supergain factor admits the expansion
\begin{align} \label{superdirectivity_endfire_lossy_Nlarge_rholl}
    \gamma_\endf(N,\sfd,\rho(\sfd)) 
    = N + o(1),
\end{align}}
which returns the classical {\color{blue}lossless superdirectivity} result in \eqref{superdirectivity_endfire_lossless} for $\rho = 0$.
Unraveling the normalization in \eqref{supergain_def_def}, the maximum endfire gain  grows quadratically with $N$; the second term in \eqref{superdirectivity_endfire_lossy_Nlarge_rholl} represents the correction due to antenna losses.


\item {\color{blue}\emph{Loss-dominated regime:} }
Antenna losses dominate, 
{\color{blue}corresponding to $\rho(\sfd) = \rho_0 > 0$}, and the supergain factor is curtailed and saturates as
{\color{blue}
\begin{align}  \label{endfire_supergain_lossy_regime}
    \gamma_\endf(N,\sfd,\rho(\sfd))  = \frac{1}{N + \rho_0} + \cO(\sfd).
    \end{align}
  (Notably, letting $\sfd\to 0$ in the supergain formula \eqref{superdirectivity} 
  and inverting the resulting coupling matrix by means of the Woodbury identity correctly yields \eqref{endfire_supergain_lossy_regime}, confirming that the spatial and spectral formulations agree.)}
Unraveling the normalization in \eqref{supergain_def_def}, the maximum endfire gain  now grows only linearly with $N$ {\color{blue}for $N \ll \rho_0$}; its slope is approximately $\rho_0^{-1}$, quantifying how more efficient antennas sustain a steeper growth than their less efficient counterparts. 
{\color{blue}However, as the number of antennas increases beyond the loss threshold, $N \gg \rho_0$, the gain saturates to unity. This behavior aligns with the numerical results reported in \cite[Fig.~11]{Nossek2010}, demonstrating that losses suppress supergain enhancement as $\sfd$ shrinks.} 
\end{itemize}

\section{Spectral Concentration} 

\label{sec:spectral_conc_Nlarge}

\subsection{{\color{blue}Qualitative} Behavior}  \label{sec:qualitative}

For widening apertures with fixed antenna spacing ($N\to \infty$), the eigenstructure of the solution to \eqref{ODE} depends on $|\sfff| \le \sfd$ and the index $k$ \cite[Sec.~4.2]{Slepian1978}. 
With {\color{blue}the dependence on the constant $\sfd$ omitted for every function, 
let us apply the substitution }
\begin{equation} \label{U_G_transformation}
   {\color{blue}H(N,\sfff)} =  U({\color{blue}N,} \sfff) \, \sqrt{\cos (2 \pi \sfff) - A} 
\end{equation} 
along with the eigenvalue expansion 
\begin{equation}\label{theta_asymptotic_Nlarge_aux}
\theta({\color{blue}N,} B) = \frac{1}{4} B \, N^2 + \frac{1}{4} C(N) \, N + \cO(1) {\color{blue}\qquad N\to \infty,}
\end{equation}
where $A$ is defined in \eqref{A_cos_def} 
{\color{blue}while $B$ and $C(N)$} are specified later (with $-1 < A \le B \le 1$, and with $A$ and $B$ both independent of $N$).
{\color{blue}Casting} the Sturm-Liouville eigenvalue problem in \eqref{ODE} into the Schr\"{o}dinger form \cite[Eq.~135]{Slepian1978}
\begin{align} \label{ODE_w_G_epsilon}
\frac{d^2 {\color{blue}H(N, B, \sfff)}}{d\sfff^2} = N^2 \, P({\color{blue}N,} B,\sfff) \, {\color{blue}H(N, B, \sfff)},
\end{align}
{\color{blue}where $H$, and in turn $U$, inherit a dependence on $B$ through \eqref{theta_asymptotic_Nlarge_aux}.}
For the problem at hand, 
\begin{equation} \label{Q_potential}
    P({\color{blue}N,} B,\sfff) =  \left(\frac{\pi}{\cos(2\pi \sfff) -A}\right)^{\! 2} \, \big( \! \cos(2\pi \sfff) -y_0 \big) \big(y_1 - \cos(2\pi \sfff) \big),
\end{equation}
which captures the dependence of \eqref{ODE_w_G_epsilon} on $\theta$, subsumed into the roots $y_0 = B + \cO(N^{-1})$ and $y_1 = A + \cO(N^{-1})$. 
(The inequality $A > -1$ applied to \eqref{A_cos_def} implies $\sfd < 1/2$ and ensures that \eqref{Q_potential} is well behaved.)

The WKB method constructs solutions to \eqref{ODE_w_G_epsilon} for large $N$ \cite[Sec.~10]{ODEBook}, 
yielding
\begin{equation}
    {\color{blue}H(N, B, \sfff)} = \exp \! \left(-N \int_\sfff \sqrt{P({\color{blue}N,} B,\sfg)} \, d\sfg + \cO(1)\right) \qquad N\to \infty.
\end{equation}
The sign of $P$ reveals the qualitative behavior of $H$, and in turn of $U$: both functions oscillate {\color{blue}in $|\sfff| \le \sfd$} for $P \le 0$ while they are evanescent (i.e., exponentially decaying in $\sfff$) for $P > 0$. 
Specializing it to the problem at hand via \eqref{Q_potential}, we have that, asymptotically in $N$,
\begin{equation} \label{qualitative_Uk}
    U({\color{blue}N,} B,\sfff) \sim 
    \begin{cases}
        \text{oscillatory in } \sfff  & \quad  {\color{blue}|\sfff|}  \in  \left(0,\frac{\arccos(B)}{2\pi}\right) \, \bigcup \, (\sfd,\tfrac{1}{2}) \\
        \text{evanescent in } \sfff  & \quad  {\color{blue}|\sfff|} \in \left(\frac{\arccos(B)}{2\pi},\sfd \right).
    \end{cases}
\end{equation}
From Lemma~\ref{rev:dpswf_diff_eq}, the Sturm-Liouville problem has a finite number of nontrivial solutions; these are the eigenfunctions {\color{blue}$H_k(N, \sfff) = H(N, B_k,\sfff)$, equivalently $U_k(N,  \sfff) = U(N, B_k,\sfff)$} via \eqref{U_G_transformation}, 
and corresponding eigenvalues {\color{blue}$\theta_k(N) = \theta(N,B_k)$
with $B_k$ and $C_k(N) = C(N,B_k)$} inheriting a dependence on the index $k$ through  
the constraint \cite[Sec.~10.5]{ODEBook} 
\begin{align}  \label{con1_general}
 B_k & = \left\{ A<B<1 : {\color{blue}2} N \! \int_0^{\frac{\arccos(B)}{2\pi}} \!\!\! \sqrt{- P({\color{blue}N}, B,\sfff)} \, d\sfff  = \left(k + \frac{1}{2}\right) \pi + \cO\left(\frac{1}{N} \right)\right\},
\end{align}
{\color{blue}where the factor of $2$ accounts for the symmetry of the integrand.}
For the problem at hand, replacing $P({\color{blue}N,B}, \sfff)$ by \eqref{Q_potential} and applying the substitution $t=\cos(2\pi \sfff)$, 
\begin{align} \label{con1}
    B_k & = \left\{ A<B<1 : \int_B^1 \sqrt{\frac{t-B}{(t-A)(1-t^2)}} \, dt \sim  \frac{k \pi}{N} \right\},
\end{align}
{\color{blue}where the relation was evaluated asymptotically using $k = \cO(N)$, a scaling that is detailed below.  
(Although $N$ appears explicitly on the right-hand side of \eqref{con1}, the ratio $k/N = \cO(1)$ ensures that this dependence on $N$ is only apparent, consistent with our initial assumption.)}

In light of \eqref{qualitative_Uk}, the position of $B_k$ relative to the fixed endpoints $A$ and $1$ determines the spectral behavior of the $U_k$'s by partitioning the index set $\{0, \ldots, N-1\}$ into three subsets, $\cK_1, \cK_2$, and $\cK_3$.
This partition mirrors the distinct asymptotic behaviors of the eigenvalues $\mu_k$ in Fig.~\ref{fig:eig_C}, each representing the average concentration of the associated $U_k$ within the visible region as per \eqref{avg_spectral_conc}.
The dependence of $B_k$ and {\color{blue}$C_k(N)$} on $k$ within each $\cK_i$
is mapped to an asymptotic parameter {\color{blue}$\epsilon >0$, arbitrarily small}, such that each subset is characterized by a common behavior defined by continuous variables $B$ and {\color{blue}$C(N)$}. {\color{blue}These variables are uniquely specified within each $\cK_i$, so that any notational ambiguity is naturally resolved by the context of the subset being analyzed.}

\begin{itemize}
\item
$A < B < 1$ or {\color{blue}$\cK_1  = \left\{ \lfloor N_0 \epsilon \rfloor, \ldots, \lfloor N_0 (1 - \epsilon) \rfloor  \right\}$} \\
Within the visible region, the $U_k$'s exhibit $k = \cO(N)$ oscillations, consistently with Remark~\ref{ref:symmetry}; this corresponds to a \emph{low-pass} behavior.
\item
$A = B$ or {\color{blue}$\cK_2 = \left\{ \lfloor \epsilon \log(N)/\pi  \right\rfloor, \ldots, \left\lfloor N - \epsilon \log(N)/\pi \rfloor \right\}$} \\
The roots collapse at $\sfff=\sfd$, whereby oscillations extend across the entire visible region, corresponding to an \emph{all-pass} behavior.
\item
 $ -A < B$  or {\color{blue}$\cK_3  = \left\{ \lfloor N_0 (1+ \epsilon) \rfloor, \ldots,  \lfloor N - \epsilon N_{0} \rfloor \right\}$} \\
By means of the spatial-frequency symmetry in Remark~\ref{ref:symmetry}, this case reduces to $\cK_1$ with transformed variables $\sfff \mapsto \tfrac{1}{2} - \sfff$, implying $A \mapsto -A$ and $k \mapsto N-k-1$.
The visible region becomes 
$\sfd \le \sfff \le \tfrac{1}{2}$, 
revealing an intrinsic \emph{high-pass} behavior. 
\end{itemize}


{\color{blue}The eigenfunctions $U_k$ and corresponding eigenvalues $\mu_k$}
 define the asymptotic behavior of the supergain factor in~\eqref{max_directivity_current_fprime}  through three contributions from the subsets $\cK_1$, $\cK_2$, and $\cK_3$. 
{\color{blue}Particularly, along the endfire direction,}
the exponential decay of the eigenfunctions {\color{blue}in $\cK_1$}
paired with eigenvalues near unity (see Fig.~\ref{fig:eig_C}), renders its contribution  negligible for large $N$, of order $\cO(e^{-L N})$ for $L>0$.
Consequently, the endfire supergain factor is dominated by contributions from $\cK_2$ and $\cK_3$, 
governed by antenna losses via $\rho$ relative to $N$ and $\sfd$. 
{\color{blue}The array achieves linear growth with $N$ in the {supergain regime} (all eigenvalues are above $\rho$), but saturates in the {loss-dominated regime} as $\rho$ eclipses the smallest eigenvalues in $\cK_3$.}

\subsection{{\color{blue}Asymptotic Analysis at Endfire}}

\begin{figure}
    \centering
    \pgfplotstableread[col sep=comma]{data/B_d_data.csv}{\data}
    
    \begin{tikzpicture}
        \definecolor{forestGreen}{rgb}{0.00, 0.27, 0.11}
        \definecolor{mossGreen}{rgb}{0.05, 0.45, 0.20}
        \definecolor{leafGreen}{rgb}{0.40, 0.65, 0.30}
        
        \begin{axis}[
            width=11cm, height=7cm,
            xlabel={$\sfd$},
            ylabel={$B(\sfd,x)$},
            xmin=0, xmax=0.5,
            ymin=-1.01, ymax=1.01,
            grid=both,
            tick label style={font=\footnotesize},
            label style={font=\small},
        ]
        
        \addplot [color=black, solid, thick] table [col sep=comma, x=d, y=limit_B] {\data};

        \addplot [color=black, solid, thick] table [col sep=comma, x=d, y=B_x_1] {\data}
            node[pos=0.15, sloped, above, font=\footnotesize] {$0$};

        \addplot [color=black, solid, thick] table [col sep=comma, x=d, y=B_x_2] {\data}
            node[pos=0.25, sloped, above, font=\footnotesize] {$0.25$};

        \addplot [color=black, solid, thick] table [col sep=comma, x=d, y=B_x_3] {\data}
            node[pos=0.35, sloped, above, font=\footnotesize] {$0.5$};

        \addplot [color=black, solid, thick] table [col sep=comma, x=d, y=B_x_4] {\data}
            node[pos=0.45, sloped, above, font=\footnotesize] {$0.75$};

        \addplot [color=black, solid, thick] table [col sep=comma, x=d, y=B_x_5] {\data}
            node[pos=0.55, sloped, below, font=\footnotesize] {$1$};
            
        \node[anchor=north west, font=\small, fill=white, fill opacity=0.8, text opacity=1, inner sep=1.5pt] 
            at (axis cs:0.3, 0.2) {$-A = -\cos(2\pi \sfd)$};

        \node[anchor=north west, font=\small] at (axis cs:0.02, 0.8) 
            {Feasibility region};

        \addplot[fill=forestGreen, fill opacity=0.2, draw=none] table [x=d, y=limit_B] {\data} 
            -- (axis cs:0.5, 1) -- (axis cs:0, 1) -- cycle;
        
        \end{axis}
    \end{tikzpicture}
\caption{$B(\sfd,x)$ from \eqref{B_d_main} versus $\sfd$ for various $x \in (0,1)$.
The mapping $x \mapsto B$ is invertible.
The curves show 
that $B\to 1$ as $x \to 1$ and $B\to-A$ as $x \to 0$.}
\label{fig:B_d}
\end{figure}

{\color{blue}Following the framework in \cite{Slepian1978}, the asymptotic {\color{blue}behavior} at endfire of the $U_k$'s, and of their associated $\mu_k$'s, is formalized in Appendix~\ref{app:dpswf_Nlarge}.
(Broadside behavior is recovered setting $\sfff=0$ in lieu of $|\sfff|=\sfd$.)}
{\color{blue}Within the dominant subset $\cK_3$, this behavior} is governed by the   integrals
{\color{blue}
     \begin{subequations}
\label{Li_x3prime} 
  \begin{align}[left ={L_n(\sfd,x) =  \empheqlbrace}]
    & 2(1-B) \int_{0}^{\pi/2} \frac{\cos^2(\phi)}{\left| \left(A - \psi_1(B,\phi)  \right) \left(1 - \psi_1(B,\phi)  \right) \right|^{1/2}} \, d\phi & n=1 \label{L1_prime}  \\
    & 2 \int_{0}^{\pi/2} \left| \left(A - \psi_1(B,\phi)  \right) \left(1 - \psi_1(B,\phi)  \right) \right|^{-1/2} \, d\phi & n=2 \label{L2_prime} \\
    & 2 \int_{0}^{\pi/2} \left(B + \psi_2(B,\phi) \right)  \left| 1 - \psi_2^{2}(B,\phi) \right|^{-1/2} \, d\phi & n=3  \label{L3_prime} \\
   & 2 \int_{0}^{\pi/2} \left| 1 - \psi_2^2(B,\phi) \right|^{-1/2} \, d\phi & n=4, \label{L4_prime}
   \end{align}
  \end{subequations}
where $A = \cos(2\pi\sfd)$ as per \eqref{A_cos_def} and $\psi_1, \psi_2$ are auxiliary functions denoted, for $\phi \in [0, \pi/2]$, by
\begin{align} \label{psi_fun}
\psi_1(B,\phi) = - 1 + (1-B) \sin^2(\phi) \qquad \quad 
\psi_2(B,\phi)  = - B + (A+B) \sin^2(\phi).
\end{align}}
For any $x \in (0,1)$, the parameter $B \in (-A,1)$ is determined {\color{blue}by specializing \eqref{con1} to $k \in \cK_3$ and} applying a change of variables {\color{blue}(Appendix~\ref{app:dpswf_Nlarge})}, yielding the implicit constraint
{\color{blue}
\begin{equation} \label{B_d_main}
2(1-B) \int_{0}^{\pi/2} \frac{\cos^2(\phi) \, d\phi}{\left| \left(A - \psi_1(B,\phi) \right) \left(1 - \psi_1(B,\phi)\right) \right|^{1/2}}  \sim  \pi (1-2\sfd)  (1- x).
\end{equation}}
The above constraint defines an invertible mapping {\color{blue}$x \mapsto B(\sfd,x)$ for any given $\sfd$}, as illustrated in Fig.~\ref{fig:B_d}.
In turn, define
\begin{equation} \label{phase_discont_main}
    C(N,\sfd,x) = \frac{4}{L_2(\sfd,x)} \left[\frac{N}{2} L_1(\sfd,x) + \left(2 + (-1)^{\lfloor (1-2 d) N (1-x)\rfloor}\right) \frac{\pi}{4} \right]_{\text{rem} \, 2\pi},
\end{equation}
which satisfies $0 \le C \le 8\pi/L_2$ as per the modulo operation. 
{\color{blue}Notably, all four integrals in \eqref{Li_x3prime} are strictly positive for $x\in(0,1)$ because their integrands are non-negative on $[0,\pi/2]$ and positive almost everywhere (except at endpoints where they could be zero).}

With the asymptotic foundations established, {\color{blue}bounds are constructed in Appendix~\ref{app:endfire_supergain_wide_apertures}} for the endfire supergain factor $\gamma_\endf(N,\sfd,\rho)$ as $N \to \infty$. {\color{blue}Because the subsets $\cK_2$ and $\cK_3$ are inherently non-disjoint (see Fig.~\ref{fig:eig_C} and their definitions in Sec.~\ref{sec:qualitative}), combining their contributions yields non-achievable limits:} a lower bound is derived from $\cK_3$ alone, while an upper bound sums both subsets, 
\begin{equation} \label{bound_supergain_largeN}
\underline{\gamma}_\endf(N,\sfd,\rho)  < \gamma_\endf(N,\sfd,\rho)  <  
{\color{blue}\overline{\gamma}_\endf(N,\sfd,\rho).}
\end{equation}
The lower bound captures the linear scaling of the supergain factor via
\begin{align}  \label{supergain_endfire_lossy_KK3}
  \underline{\gamma}_\endf(N,\sfd,\rho) & \sim  \frac{\pi N (1-2\sfd)}{\sinc(2 \sfd)} \int_{0}^{1}  
         \frac{1}{L_2(\sfd,x)} 
            \left(1 + 2 \sfd \rho \, e^{C(N,\sfd,x) L_4(\sfd,x)/2} \, e^{ L_3(\sfd,x) N}\right)^{-1}
             \, dx,
\end{align}
{\color{blue}while the upper bound, 
\begin{equation} \label{bound_supergain_largeN}
 \overline{\gamma}_\endf(N,\sfd,\rho) = \underline{\gamma}_\endf(N,\sfd,\rho) + \delta_\endf(\sfd,\rho),
\end{equation}
also includes the $\cO(1)$ contribution, independent of $N$
and originating from $\cK_2$}, 
\begin{align} \label{Delta_rho_gap}
\delta_\endf(\sfd,\rho) & 
\sim 
12 \sfd \ln \! \left(1+\frac{1}{2\sfd \rho} \right).
\end{align}
{\color{blue}
As established in Appendix~\ref{app:superdirectivity_largeN_subsetK3}, in the lossless case ($\rho = 0$) 
the contribution from $\cK_3$ grows linearly with $N$, whereas those from $\cK_1$ and $\cK_2$ are sublinear. Consequently, the difference between the upper and lower bounds is $o(N)$, and the bounds are asymptotically tight.
In particular,
\begin{equation} \label{tau_d_lossless_factor}
\tau(\sfd)
=
\frac{\pi (1-2\sfd)}{\sinc(2\sfd)}
\int_0^1
\frac{1}{L_2(\sfd,x)}\,dx,
\end{equation}
quantifies the asymptotic linear growth with $N$ predicted  in \eqref{gain_growing_N}. Here, recall, the integrand remains bounded as $L_2$ is positive according to \eqref{L2_prime}. }

{\color{blue}Also relevant} is that the gap $\delta_\endf(\sfd,\rho)$ diverges logarithmically as $\rho\to0$ resulting in an unbounded superdirectivity.
However, this divergence is physically inconsequential; this term is driven by contributions from the subset $\cK_2$ that are progressively far from the transition index $N_0$ as $N$ increases (see Appendix~\ref{app:endfire_supergain_wide_apertures}).
{\color{blue}This extreme sensitivity to losses renders} such components incompatible with 
the limit $\rho \to 0$ {\color{blue}(see Fig.~\ref{fig:eig_C}), which must therefore be treated separately, as above.} 
 The inherent physical constraints and finite apertures further regularize this behavior. 
{\color{blue}Then, $\gamma_\endf(N,\sfd,\rho)$ is expected to scale linearly as $\cO(N)$  according to \eqref{supergain_endfire_lossy_KK3}.}

{\color{blue}\subsection{Impact of Array Aperture on Supergain}}

\begin{figure}
    \centering
    \pgfplotstableread[col sep=comma]{data/G_N_data.csv}{\data}
    
    \begin{tikzpicture}
        \begin{axis}[
        width=11cm, height=7cm,
                xlabel={$N$},
                ylabel={$\underline{\gamma}_\endf(N,\sfd,\rho)$},
                xmin=20, xmax=200,
                ymin=0, ymax=45,
                grid=both,
                tick label style={font=\small},
                label style={font=\small},
                xtick={20, 40, 60, 80, 100, 120, 140, 160, 180, 200}, 
                legend pos=north east,
                legend style={font=\footnotesize, cells={anchor=west}},
        ]

       \addplot [color=black, solid, thick] table [col sep=comma, x=N, y=D] {\data}
    node[pos=0.6, sloped, above, fill=white, inner sep=2pt, font=\footnotesize] {Lossless $ = \tau(\sfd) N$};

            \addplot [color=black, solid, thick] table [col sep=comma, x=N, y=G_rho1] {\data}
            node[pos=0.8, fill=white, inner sep=2pt, font=\footnotesize] {$10^{-8}$};

        \addplot [color=black, solid, thick] table [col sep=comma, x=N, y=G_rho2] {\data}
            node[pos=0.8, fill=white, inner sep=2pt, font=\footnotesize] {$10^{-12}$};

            \addplot [color=black, solid, thick] table [col sep=comma, x=N, y=G_rho3] {\data}
            node[pos=0.8, fill=white, inner sep=2pt, font=\footnotesize] {$10^{-16}$};
            
        \end{axis}
    \end{tikzpicture}
    \caption{Lower bound 
    in \eqref{supergain_endfire_lossy_KK3} as a function of $N$ for $\sfd=0.45$ and various $\rho$. The curves are benchmarked against the lossless limit $\tau(\sfd) N$. For fixed $\rho$ and $\sfd$, {\color{blue}increasing $N$ breaks the condition required for supergain in \eqref{supergain_regime_condition_N}, causing losses to saturate the linear scaling}.}
\label{fig:G_N}
\end{figure}

\begin{figure}
    \centering
    \pgfplotstableread[col sep=comma]{data/G_N_loss_data.csv}{\data}
    
    \begin{tikzpicture}
        \begin{axis}[
        width=11cm, height=7cm,
                xlabel={$N$},
                ylabel={{\color{blue}$\overline{\gamma}_\endf(N,\sfd,\rho)$}}, 
                xmin=20, xmax=200,
                ymin=0, ymax=50,
                grid=both,
                tick label style={font=\small},
                label style={font=\small},
                xtick={20, 40, 60, 80, 100, 120, 140, 160, 180, 200}, 
                legend pos=north east,
                legend style={font=\footnotesize, cells={anchor=west}, inner sep=2pt, row sep=-1pt}, 
        ]

       \addplot [color=black, dashed, thick] table [col sep=comma, x=N, y=Delta_rho1] {\data};

    \addlegendentry{$\delta_\endf(\sfd,\rho)$}

    \addplot [color=black, dashed, thick, forget plot] table [col sep=comma, x=N, y=Delta_rho2] {\data};

    \addplot [color=black, dashed, thick, forget plot] table [col sep=comma, x=N, y=Delta_rho3] {\data};

            \addplot [color=black, solid, thick] table [col sep=comma, x=N, y=G_rho1] {\data}
            node[pos=0.4, fill=white, inner sep=2pt, font=\footnotesize] {$10^{-1}$};
            
\addlegendentry{{\color{blue}$\overline{\gamma}_\endf(N,\sfd,\rho)$}};

        \addplot [color=black, solid, thick, forget plot] table [col sep=comma, x=N, y=G_rho2] {\data}
            node[pos=0.4, fill=white, inner sep=2pt, font=\footnotesize] {$10^{-2}$};

            \addplot [color=black, solid, thick, forget plot] table [col sep=comma, x=N, y=G_rho3] {\data}
            node[pos=0.4, fill=white, inner sep=2pt, font=\footnotesize] {$10^{-3}$};
            
        \end{axis}
    \end{tikzpicture}
    \caption{Upper bound in \eqref{bound_supergain_largeN} as a function of $N$ for $\sfd=0.45$ and various $\rho$. 
    {\color{blue}For each fixed $\rho$, the upper bound remains approximately independent of $N$, with only small residual oscillations due to the phase discontinuity in \eqref{phase_discont_main}, indicating that losses  limit the benefit of adding antennas; each curve saturates to the gap term $\delta_\endf(\sfd,\rho)$ in \eqref{Delta_rho_gap}.}}
\label{fig:G_N_loss}
\end{figure}

{\color{blue}As established qualitatively, two operational regimes emerge depending on antenna losses via $\rho$ relative to $N$ with fixed $\sfd\in(0,1/2)$.}

\begin{itemize}
    \item {\color{blue}\emph{Supergain regime:} When $\rho$ decays 
    sufficiently rapidly with growing $N$, the loss term in \eqref{supergain_endfire_lossy_KK3} vanishes uniformly over $x\in(0,1)$. In particular, since Appendix~\ref{app:L3_monotonicity} establishes that
\begin{equation}
\sup_{x\in(0,1)} L_3(\sfd,x) = \ln\left(\frac{1-A}{3+A - 2 \sqrt{ 2(1+A)}} \right) = \sfL_3 >0,
\end{equation}
with $A = \cos(2\pi\sfd)$ as defined in \eqref{A_cos_def}, a sufficient condition is
\begin{equation} \label{supergain_regime_condition_N}
\rho(N) = o\!\left(e^{-\sfL_3 N}\right) \qquad N\to\infty.
\end{equation}
Under this condition, the lower bound in \eqref{supergain_endfire_lossy_KK3}  
   grows linearly with $N$} with slope $\tau(\sfd)$.
This lower bound is plotted in Fig.~\ref{fig:G_N} against $N$ for $\sfd=0.45$ and various $\rho$, alongside the lossless limit $\tau(\sfd) N$. 
(The fluctuations in the numerical results arise from the phase discontinuity in \eqref{phase_discont_main} {\color{blue}with varying $N$}.)
The figure shows that even the slightest deviation from a half-wavelength spacing can produce linear supergains, provided that losses decay according to \eqref{supergain_regime_condition_N}, thereby sustaining this asymptotic scaling.

  \item {\color{blue} \emph{Loss-dominated regime:}} If condition \eqref{supergain_regime_condition_N} is violated, antenna losses override the eigenvalues in $\cK_3$. The linear growth in \eqref{supergain_endfire_lossy_KK3} collapses, and the supergain factor saturates to the constant upper bound $\delta_\endf(\sfd,\rho)$ in \eqref{Delta_rho_gap}, as confirmed in Fig.~\ref{fig:G_N_loss}.
  This behavior emerges 
  {\color{blue}when the loss term does not decrease sufficiently fast with $N$,}
    causing the contributions to progressively concentrate within $\cK_2$ and capturing the transition around $N_0$ (see Fig.~\ref{fig:eig_C}).
\end{itemize}

{\color{blue}While, for fixed $\rho>0$, the gap
$\delta_\endf(\sfd,\rho)$ in \eqref{Delta_rho_gap} is independent of $N$,
it can grow with $N$ in the supergain regime. Indeed, under
\eqref{supergain_regime_condition_N}, for $N\to \infty$
\begin{equation}
    \rho(N) = e^{-\sfL_3 N} \, w(N) \qquad w(N)\to0, 
\end{equation}
and therefore
\begin{align}
\delta_\endf(\sfd,\rho(N))
&\sim
12\sfd
\ln\!\left(
1+
\frac{e^{\sfL_3N}}{2\sfd w(N)}
 \right) 
\qquad N\to\infty.
\end{align}
Hence, the gap is not necessarily sublinear in $N$; its growth depends on
the rate at which $w(N)$ decays with increasing $N$.
Unlike in the lossless case, the difference between the upper and lower bounds need not vanish after normalization by $N$, and the bounds in
\eqref{bound_supergain_largeN} are not necessarily asymptotically tight.}

{\color{blue}\subsection{Impact of Antenna Spacing on Supergain}}

\begin{figure}
    \centering
    \pgfplotstableread[col sep=comma]{data/tau_d_data.csv}{\data}
    
    \begin{tikzpicture}
        \begin{axis}[
        width=11cm, height=7cm,
                xlabel={$\sfd$},
                ylabel={$\underline{\gamma}_\endf(N,\sfd,\rho)/N$},
                xmin=0, xmax=0.5,
                ymin=0, ymax=1,
                grid=both,
                tick label style={font=\small},
                label style={font=\small},
                xtick={0.1, 0.2, 0.3, 0.4, 0.5}, 
                legend pos=north east,
                legend style={font=\small, cells={anchor=west}},
        ]

        \addplot [color=black, solid, thick] table [col sep=comma, x=d, y=tau] {\data}
            node[pos=0.3, sloped, fill=white, inner sep=2pt, font=\footnotesize] {Lossless $ = \tau(\sfd)$ };

            \addplot [color=black, solid, thick] table [col sep=comma, x=d, y=tau_rho1] {\data}
            node[pos=0.4, fill=white, inner sep=2pt, font=\footnotesize] {$10^{-8}$};

        \addplot [color=black, solid, thick] table [col sep=comma, x=d, y=tau_rho2] {\data}
            node[pos=0.4, fill=white, inner sep=2pt, font=\footnotesize] {$10^{-16}$};

            \addplot [color=black, solid, thick] table [col sep=comma, x=d, y=tau_rho3] {\data}
            node[pos=0.4, fill=white, inner sep=2pt, font=\footnotesize] {$10^{-24}$};
            
        \end{axis}
    \end{tikzpicture}
    \caption{Lower bound on the asymptotic linear scaling of the endfire supergain factor as a function of $\sfd$ for various $\rho$ and $N=41$. The curves are benchmarked against the lossless limit $\tau(\sfd)$ in \eqref{tau_d_lossless_factor}; the linear scaling is fully exploited as $\sfd \to 0$ whereas it is suppressed as $\sfd \to 1/2$, reflecting a progressive decrease in mutual coupling. As $\rho$ departs from zero (with fixed $N$ and $\sfd$), the curves deviate significantly from $\tau(\sfd)$. With $N$ and $\rho$ fixed, 
    the impact of losses intensifies as $\sfd$ decreases, violating the scaling requirement in \eqref{supergain_regime_condition_d}.}
\label{fig:tau_d}
\end{figure}

The supergain behavior with decreasing $\sfd$ for fixed, large $N$ is illustrated in Fig.~\ref{fig:tau_d}. 
Expectedly, the lossless scaling factor $\tau(\sfd)$ in \eqref{tau_d_lossless_factor} vanishes as $\sfd \to 1/2$, since half-wavelength spacing suppresses mutual coupling among isotropic antennas. In contrast, $\tau(\sfd)\to 1$ as $\sfd \to 0$ (see Appendix~\ref{app:superdirectivity_largeN_subsetK3}  and the  discussion preceding \eqref{gain_growing_N}).
When losses are included via the lower bound in \eqref{supergain_endfire_lossy_KK3}, however, the behavior deviates significantly from the lossless limit according to $\rho$ relative to $\sfd$.

\begin{itemize}
    \item {\color{blue}\emph{Supergain regime:} For the iterated asymptotic regime in which $N\to\infty$ is considered
first and $\sfd\to0$ subsequently, Appendix~\ref{app:supergain_condition}
shows that a sufficient condition for maintaining the linear scaling is
that $\rho$ decays according to
    \begin{equation} \label{supergain_regime_condition_d}
        \rho(N,\sfd) = o\!\left(\sfd^{N + 7}\right) \qquad \sfd \to 0, \quad N\to \infty.
    \end{equation}}
    For fixed $\rho$, reducing $\sfd$ from $1/2$ initially increases the slope monotonically in Fig.~\ref{fig:tau_d}, 
 confirming the constructive effect of coupling.
 Yet this trend only persists down to a critical spacing, beyond which the curves become unimodal and the supergain factor rapidly deteriorates as $\sfd$ decreases further, consistent with \cite{Nossek2010}.
Since $\rho$ is kept constant in Fig.~\ref{fig:tau_d}, the condition \eqref{supergain_regime_condition_d} is inevitably violated as $\sfd$ becomes small, explaining the breakdown of the supergain behavior.

    \item {\color{blue}\emph{Loss-dominated regime:}} Once \eqref{supergain_regime_condition_d} fails, the behavior is governed by the gap term $\delta_\endf(\sfd,\rho)$ in \eqref{Delta_rho_gap}. 
 Since $\delta_\endf(\sfd,\rho) \sim -12 \sfd \ln (\sfd) $ as  $\sfd\to 0$, both the upper and lower bounds converge to zero, implying by the squeeze theorem that the supergain factor itself vanishes. In this {\color{blue}limit}, each additional antenna introduces an exponential penalty relative to the uncoupled case. 
Notice how this behavior differs from that of  \eqref{endfire_supergain_lossy_regime} for electrically small apertures, where the supergain factor remains nonzero, also observed  in \cite{Sanguinetti2024}.
\end{itemize}

{\color{blue}\subsection{Asymptotic Beamwidth Analysis at Endfire}

\begin{figure}
  \centering
  \begin{tikzpicture}[>=Stealth, scale=1.1]
    \draw[dashed, gray!80!black, thick] (-2.6, 0) -- (8.5, 0);

    \filldraw[fill=white, draw=black, thick] (-2.6, 0) circle (0.12);
    \filldraw[fill=white, draw=black, thick] (-2, 0) circle (0.12);
    \filldraw[fill=white, draw=black, thick] (-1.4, 0) circle (0.12);
    \filldraw[fill=white, draw=black, thick] (0, 0) circle (0.12);

    \draw[decorate, decoration={brace, amplitude=6pt}, thick] 
      (-2.6, 0.25) -- (0, 0.25) 
      node[midway, above=8pt, font=\small] {$N\text{ antennas}$};

    \draw[thick] 
      (0, 0) .. controls (1.5, 2.2) and (4.5, 1.1) .. (4.5, 0)
             .. controls (4.5, -1.1) and (1.5, -2.2) .. (0, 0);

    \node[font=\small\itshape, rotate=10, anchor=south] at (1.9, 0.4) {loss-dominated region};

    \filldraw[thick, fill=gray!15, pattern=north west lines, pattern color=gray!60] 
      (0, 0) .. controls (2.0, 0.9) and (7.5, 0.4) .. (7.5, 0)
             .. controls (7.5, -0.4) and (2.0, -0.9) .. (0, 0);

    \node[font=\small\itshape, fill=white, inner sep=1pt] at (4.5, 0) {supergain region};

    \draw[<->, thick] (19.9:3.4) arc (19.9:-19.9:3.4);
    \node[font=\small, anchor=north west] at (3.3, -1.0) {$o\!\left(\frac{1}{\sqrt{N}}\right)$};

    \draw[<->, thick] (2.8:6.47) arc (2.8:-2.8:6.47);
    \node[font=\small, anchor=north west] at (6.4, -0.3) {$o\!\left(\frac{1}{N}\right)$};

    \node[font=\small, align=center, anchor=west] at (7.3, 0.4) {($\theta = \theta_\endf$) \\ endfire};
  \end{tikzpicture}
  \caption{{\color{blue}Asymptotic endfire beamwidth scaling versus $N$. The loss-dominated regime exhibits an $o(1/\sqrt{N})$ behavior,  
  recovering classical beamwidth scalings for endfire arrays under unit-amplitude excitations (including Hansen-Woodyard configurations). 
Conversely, the supergain regime enables asymptotically narrower beamwidths of order $o(1/N)$.}}
\label{fig:beamwidth_N}
\end{figure}

Physically, an array of finite aperture does not radiate exclusively along an isolated direction; rather, spatial truncation smears power across a finite angular beam.
Recalling that $\theta \in [-\pi/2,\pi/2]$ is the angle relative to the array's normal
and letting $\theta_\endf = \pm \pi/2$ denote the endfire direction,
it is established in Appendix~\ref{app:beamwidth_N} how narrow a beam must be---asymptotically with increasing $N$ and for a fixed $\sfd$---for the supergain and loss-dominated regimes to manifest (refer to Fig.~\ref{fig:beamwidth_N}).
\begin{itemize}
 \item  \emph{Loss-dominated regime:} 
 This regime manifests whenever
\begin{equation} \label{loss_theta}
|\theta(N) - \theta_\endf| = o \! \left(\frac{1}{\sqrt{N}} \right) \qquad N\to\infty,
\end{equation}
which aligns with conventional endfire array theory under phase-only, optimized current excitation (the Hansen-Woodyard endfire configuration retains an identical asymptotic scaling) \cite[Ch.~6]{BalanisBook}. For $\pi \sfd \ll 1$, the one-sided half-power beamwidth at endfire is \cite[Table~6.4]{BalanisBook}
\begin{equation} \label{ThetaHPBW_endfire}
|\theta_\text{HPBW}(N) - \theta_\endf| \approx  \arccos \! \left(1 - \frac{1.39}{\pi N \sfd} \right).
\end{equation}
Using $\arccos(1-y) = \sqrt{2 y} +  \cO(y^{{3/2}})$ with $y = 1.39/(\pi \sfd N) \to 0$ ($N\to \infty$) yields, to leading order,
\begin{equation} \label{ThetaHPBW_endfire_Ninf}
|\theta_\text{HPBW}(N) - \theta_\endf| \approx  \frac{0.94}{\sqrt{\sfd}} \sqrt{\frac{1}{N}} + \cO \! \left(\frac{1}{N^{3/2}}\right) = \cO \! \left(\frac{1}{\sqrt{N}} \right) \qquad N\to \infty.
\end{equation}
The misalignment between \eqref{ThetaHPBW_endfire_Ninf} and our derived $o(1/\sqrt{N})$ limit stems from the expanded search space in the current optimization. Unit-amplitude currents can attain \eqref{ThetaHPBW_endfire_Ninf}, but non-uniform amplitudes enable the synthesis of even narrower beams.
 
    \item \emph{Supergain regime:} This regime requires satisfying a stricter condition than \eqref{loss_theta}:
\begin{equation} \label{supergain_theta}
|\theta(N) - \theta_\endf| = o \! \left(\frac{1}{N} \right) \qquad N\to\infty.
\end{equation}
The supergain regime recovers the concentration typical of broadside arrays 
whose one-sided half-power beamwidth is given, for any $N$ and $\pi \sfd \ll 1$, by  \cite[Table~6.2]{BalanisBook}
\begin{equation} \label{ThetaHPBW_broadside}
|\theta_\text{HPBW}(N)| \approx  \pi - \arccos \! \left( \frac{1.39}{\pi N \sfd} \right).
\end{equation}
Indeed, letting $N\to \infty$ and using $\arccos(y) = \pi/2 - y + \cO(y^3)$ at $y = 0$ (for $N\to \infty$) yields, to leading order,
\begin{equation} \label{ThetaHPBW_broadside_Ninf}
|\theta_\text{HPBW}(N)| \approx  \frac{0.44}{\sfd} \frac{1}{N} +\cO \! \left(\frac{1}{N^3}\right) = \cO \! \left(\frac{1}{N}\right) \qquad N\to \infty.
\end{equation}
While expanding the aperture sharpens the beam in both regimes, in the supergain regime this behavior shifts from $\cO(1/\sqrt{N})$ to $\cO(1/N)$, departing significantly from conventional endfire designs, and consistently with the observed superlinear gain scalings.
\end{itemize}}

\section{Conclusion} \label{sec:conclusion}



Through a spectral analysis, this paper has characterized the fundamental limits of the gain achievable by linear holographic arrays. 
The results confirm that, properly harnessed, mutual coupling arising as antenna spacing falls below a half-wavelength yields a gain that scales quadratically with the number of antennas $N$ along the endfire direction. The price to pay for such enhancement is a gain loss everywhere else as coupling can only redistribute power across directions.

The quadratic scaling is achievable whenever the coupling is strong, provided the array can withstand the reactive behavior and can sustain highly coupled modes against the backdrop of ohmic losses, $\rho$. 
Specifically:
\begin{itemize}
    \item 
    The spatial concentration {\color{blue}mechanism} (vanishing antenna spacing): 
    As antennas collapse into a single one, coupling intensifies to an extreme degree. If losses decay at least {\color{blue}polynomially with $\sfd$}, the gain sustains a quadratic growth. Conversely, {\color{blue}if losses do not vary with $\sfd$,} the gain tends to {\color{blue}$N/(N+\rho)$; 
    it grows linearly with $N$ with slope $\rho^{-1}$ when $N \ll \rho$, whereas it saturates to unity when $N \gg \rho$.}
    \item 
    The spectral concentration {\color{blue}mechanism} (increasing aperture with  fixed spacing below a half wavelength): 
    The quadratic scaling emerges again, not from a singular collapse, but from the cumulative contribution to mutual coupling across progressively more distant antennas.    
    To sustain this growth, 
    the tail contributions from far-apart antennas must not saturate, ensured by $\rho$ decaying {\color{blue}at least} exponentially with $N$ for a fixed $\sfd$. 
    Further gains are achievable through array densification, provided that $\rho$ decays {\color{blue}at least} polynomially with $\sfd$.
\end{itemize}

These findings extend beyond holographic arrays to arrays that seek to  minimize mutual coupling---such as half-wavelength spaced arrays---in wideband scenarios. Because these are designed for a center frequency, variations across the bandwidth perturb their conditioning, activating the aforedescribed spectral concentration mechanisms \cite{PizzoSPAWC24}. As $N$ increases, these deviations accumulate, producing non-negligible tail contributions to mutual coupling whose impact on  beamforming remains unexplored.

In general, mutual coupling intensifies for large apertures, off-broadside beamforming, and wideband operation, which are also the {\color{blue}domains} where beam squint becomes most pronounced \cite{Heedong2024}. Since both phenomena introduce frequency-dependent distortions of the array response, their interplay 
warrants further investigation.

\section*{Acknowledgment}

Work supported by MICIU under the Maria de Maeztu Units of Excellence Programme (CEX2021-001195-M), and by AGAUR (Catalan Government).

\begin{appendices}

\section{} \label{app:optimality}

The dependence of every variable on $\sfd$ and $N$ is henceforth omitted as they are kept fixed throughout.
Let us rewrite \eqref{current_spectrum_pswf_discrete} compactly as
\begin{align} \label{current_spectrum_dpswf}
J(\sfff) 
= \bu^{\Trans}(\sfff) \bi  
\end{align}
with $[\bu(\sfff)]_k = U_k(\sfff)$ and $[\bi]_k = J_k$ the $k$th spectral basis function and corresponding expansion coefficient, for $k = 0, \ldots, N-1$. Under this mapping, the normalization condition in \eqref{coeff_Jk_discrete_normalization} implies $\|\bi\|=1$.
Squaring the absolute value of \eqref{current_spectrum_dpswf} and averaging over $[-\sfd,\sfd]$,
\begin{align} \label{current_spectrum_power}
\frac{1}{2\sfd} \int_{-\sfd}^{\sfd} |J(\sfg)|^2 \, d\sfg  = \bi^{\Herm}  \left(\frac{1}{2\sfd} \int_{-\sfd}^{\sfd} \bu(\sfg) \bu^{\Trans}(\sfg)  \, d\sfg \right) \bi   = \bi^{\Herm} \bLambda \bi ,
\end{align}
where $\bLambda = \diag(\lambda_0, \ldots, \lambda_{N-1})$, with $\lambda_k > 0$ the eigenvalues of $\bC$ in \eqref{prolate_coupling_matrix}.
The above equation follows from the orthogonality of the $U_k$'s in \eqref{double_orthogonality_dpswf} and the coupling-prolate identity in  \eqref{coupling_prolate_identity}.
Substituting the square value of \eqref{current_spectrum_dpswf} and \eqref{current_spectrum_power} respectively into the numerator and denominator of \eqref{directivity_current_general_norm}, the array gain function equals  
\begin{align}     \label{G_N_app} 
G(\rho,\bi,\sfff) 
& =  \frac{|\bu(\sfff)^{\Trans} \bi |^2}{\bi^{\Herm} \sfbLambda(\rho) \bi},
\end{align}
where $\sfbLambda(\rho) =  \bLambda + \rho \bI_N$ is diagonal with its entries being the ordered eigenvalues of $\bCC(\rho)$ in \eqref{eig_lossy}.
Similarly, the $Q$ factor in \eqref{Q_factor_norm} is derived using \eqref{current_spectrum_power} as
\begin{align}   \label{Q_factor_norm_app} 
Q(\rho,\bi)  =  (\bi^{\Herm} \sfbLambda(\rho) \bi)^{-1}.
\end{align}
(The dependence of $G(\rho,\bi,\sfff)$ and $Q(\rho,\bi)$ on the spectrum $J(\sfff)$ is encoded into $\bi$.)
The maximum of \eqref{G_N_app} at any $\sfff^\prime$ is obtained by solving the generalized Rayleigh quotient  
\begin{equation}   \label{max_directivity_app_discrete_aux}
\begin{aligned}
& \underset{\bi : \|\bi\|=1}{\text{maximize}}
& & \frac{\bi^{\Herm} \bU(\sfff^\prime) \bi}{\bi^{\Herm} \sfbLambda(\rho) \bi},
\end{aligned}
\end{equation}
where $\bU = \bu \bu^{\Trans}$ is a unit-rank matrix.
Following the same steps that lead from \eqref{supergain_opt_problem} to \eqref{opt_current_coupling}, the maximizer of \eqref{max_directivity_app_discrete_aux} is 
the unique vector of spectral coefficients
\begin{equation} \label{opt_current_G_N_app} 
\bi^\star(\rho,\sfff^\prime) = \frac{\sfbLambda(\rho)^{-1} \, \bnu(\sfff^\prime)}{\|\sfbLambda(\rho)^{-1} \, \bnu(\sfff^\prime)\|}
\end{equation}
with spectrum, according to \eqref{current_spectrum_dpswf},
\begin{align} \label{current_spectrum_dpswf_optimal}
J^\star(\sfff) 
= \bu^{\Trans}(\sfff) \, \bi^\star(\rho,\sfff^\prime) =  \frac{\bu^{\Trans}(\sfff) \sfbLambda(\rho)^{-1} \, \bu(\sfff^\prime)}{\|\sfbLambda(\rho)^{-1} \, \bu(\sfff^\prime)\|}.
\end{align}
(Equivalently, from \eqref{max_directivity_app_discrete_aux}, changing variables according to $\bi^\prime = \sfbLambda^{1/2}(\rho) \bi$, yields a Rayleigh quotient optimization as in \eqref{directivity_lambda2_uniform_main_coupling_Rayleigh}, solved via Cauchy–Schwarz inequality.)
As per \eqref{supergain_def_def}, the supergain is obtained by substituting \eqref{opt_current_G_N_app} into \eqref{G_N_app} and dividing by $N$, i.e.,
\begin{align}  \label{superdirectivity_app_discrete}
\gamma(\rho,\sfff^\prime) & = \frac{1}{N} \, \bu(\sfff^\prime)^{\Trans} \sfbLambda(\rho)^{-1} \bu(\sfff^\prime).
\end{align}
The $Q$ factor associated with the optimized current \eqref{opt_current_G_N_app} is then
\begin{align}   \label{Q_factor_norm_app_opt} 
Q^\star(\rho,\sfff^\prime)  =  \frac{\|\sfbLambda(\rho)^{-1} \, \bu(\sfff^\prime)\|^2}{\bu(\sfff^\prime)^{\Trans} \sfbLambda(\rho)^{-1} \bu(\sfff^\prime)}.
\end{align}
Expanding \eqref{superdirectivity_app_discrete}, \eqref{Q_factor_norm_app_opt}, and \eqref{current_spectrum_dpswf_optimal} gives the expressions  in \eqref{max_directivity_current_fprime}, \eqref{Q_factor_norm_prolate}, and \eqref{delta_spectrum_discrete_pswf} respectively.

\section{} \label{app:Legendre_poly}
{\color{blue}\subsection{Small-$d$ Expansion of the Eigenfunctions}}

Expanding the derivatives in \eqref{SL_eig} and using \eqref{A_cos_def}, \eqref{ODE} becomes
\begin{align} \label{ODE_app}
& \frac{1}{(2\pi)^2} \left(\cos(2\pi\sfff)-\cos(2\pi\sfd)\right) \frac{d^2 U(\sfff)}{d\sfff^2}
 - \frac{1}{2\pi} \sin(2\pi\sfff) \frac{dU(\sfff)}{d\sfff}
\nonumber \\
&\hspace{6cm} + \left(\frac{N^2-1}{4} \cos(2\pi\sfff) - \theta\right) \, U(\sfff) = 0.
\end{align}
To characterize the asymptotic behavior for small $\sfd$, we restrict attention to the visible region $|\sfff|\le\sfd$ and apply the variable change in \eqref{f_discrete},
{\color{blue}\begin{equation} \label{var_change_y}
 f = \sfff/\sfd \in [-1,1] \qquad \qquad  V(f) = U(f \sfd).
\end{equation}}
By the chain rule, $\frac{d}{d\sfff}  = \frac{1}{\sfd} \frac{d}{df}$ and $\frac{d^2}{d \sfff^2}  = \frac{1}{\sfd^2} \frac{d^2}{df^2}$, which 
substituted into \eqref{ODE_app} yield
\begin{align} \notag
& \frac{1}{(2 \pi \sfd)^2} \left( \cos(2\pi f \sfd) - \cos(2 \pi \sfd)\right) \frac{d^2 V(f)}{df^2} - \frac{1}{2 \pi \sfd} \sin(2\pi f \sfd) \frac{d V(f)}{df}  \\ \label{ODE_y}
& \hspace{5.5cm} + \left( \frac{N^2 -1}{4} \cos(2 \pi f \sfd) - \theta \right) \, V(f) = 0.
\end{align}
The expansions $\cos x = 1-\frac{x^2}{2} + \frac{x^4}{4!} + \cO(x^6)$ and $\sin x  = x - \frac{x^3}{3!} + \cO(x^5)$ for $x\to 0$
yield
\begin{equation}
\begin{aligned}
    \cos(2\pi f \sfd) - \cos(2 \pi \sfd) & = 2 \pi^2 \sfd^2 \, (1-f^2) - \frac{2}{3} \pi^4 \sfd^4 (1-f^4) + \cO(\sfd^6) \\
    \cos(2\pi f \sfd) & = 1 - 2 \pi^2 f^2 \sfd^2 + \cO(\sfd^4) \\
    \sin(2\pi f \sfd) & = 2\pi f \sfd - \frac{4}{3} \pi^3 f^3 \sfd^3 + \cO(\sfd^5) \qquad \sfd\to 0,
\end{aligned}
\end{equation}
uniformly in $|f|\le 1$.
Substituting into \eqref{ODE_y} and multiplying through by $2$ gives
\begin{equation} \label{ODE_y_Taylor}
\begin{aligned}
& \left(1-f^2 - \frac{\pi^2 \sfd^2 (1-f^4)}{3} + \cO(\sfd^4) \right) \frac{d^2 V(f)}{df^2} \\
& \hspace{1.5cm} - \left(2f - \frac{4}{3} \pi^2 f^3 \sfd^2 + \cO(\sfd^4)\right) \frac{d V(f)}{df} \\ 
& \hspace{3cm} + \left( \frac{N^2 -1}{2} (1 - 2\pi^2 f^2 \sfd^2 + \cO(\sfd^4)) - 2\theta \right) \, V(f) = 0.
\end{aligned}
\end{equation}
Letting $\zeta^{\scriptscriptstyle(0)}(\theta) = 2\theta - \frac{N^2 -1}{2}$, {\color{blue}the zeroth-order terms in \eqref{ODE_y_Taylor} formulate the unperturbed eigenvalue problem 
$L_P^{\scriptscriptstyle(0)} V = \zeta^{\scriptscriptstyle(0)} V$, governed by the Legendre operator \cite[Eq.~22.6.1]{AbramowitzStegun}}
\begin{equation} \label{LP_operator}
  L_P^{\scriptscriptstyle(0)} {\color{blue}= (1-f^2) \frac{d^2}{df^2} -2f \frac{d}{df}} = \frac{d}{df} \left\{\big( 1-f^2 \big) \frac{d}{df}\right\}.
 \end{equation}
This operator admits a complete set of real orthogonal eigenfunctions $V_k^{\scriptscriptstyle(0)}(f)$ in $\mathscr{L}^2[-1,1]$, {\color{blue}given by the Legendre polynomials $P_k(f)$,} with associated eigenvalues \cite[Eq.~22.5.35]{AbramowitzStegun}
\begin{equation} \label{xi_0_k}
    \zeta_{k}^{\scriptscriptstyle(0)} = -k(k+1), \qquad  k=0, \ldots, N-1.
\end{equation}
{\color{blue}
Reintroducing the higher-order perturbation terms and defining the operator
\begin{align} 
 L_P^{\scriptscriptstyle(1)} = - \frac{\pi^2 (1-f^4)}{3} \frac{d^2}{df^2} + \frac{4}{3} \pi^2 f^3 \frac{d}{df} - (N^2 -1) \pi^2 f^2,
\end{align}
the differential equation \eqref{ODE_y_Taylor} takes the form of a perturbed eigenvalue problem $L_P V = \zeta V$, where $L_P$ expands as
\begin{equation}
L_P = L_P^{\scriptscriptstyle(0)} + \sfd^2 L_P^{\scriptscriptstyle(1)} + \cO(\sfd^4).
\end{equation}
The perturbed eigenfunctions $V_k(f)$ and associated eigenvalues $\zeta_{k}$ can be expressed as power series in $\sfd^2$, leading to \cite[Sec.~7.3]{ODEBook},
\begin{align} \label{Vkf}
     V_k(f) & = V_k^{\scriptscriptstyle(0)}(f) + \cO(\sfd^2) \\ \label{xik}
    \zeta_{k} & =\zeta_{k}^{\scriptscriptstyle(0)} + \cO(\sfd^2),
\end{align}
where $V_k^{\scriptscriptstyle(0)}(f)$ and $\zeta_{k}^{\scriptscriptstyle(0)}$ denote the unperturbed eigenfunctions and eigenvalues, respectively.
Expressing $\zeta^{\scriptscriptstyle(0)}=\zeta_{k}^{\scriptscriptstyle(0)}$ in terms of $\theta=\theta_k$ using \eqref{xi_0_k} and substituting into \eqref{xik} gives \eqref{theta_asymptotic}.
Similarly, plugging $V_k^{\scriptscriptstyle(0)}(f)$ into \eqref{Vkf} and inverting the transformation in \eqref{var_change_y}, the eigenfunctions converge, uniformly for $|\sfff^\prime|\le \sfd$, to
\begin{equation}\label{dpswf_legendre_app}
U_k(N,\sfd,f \sfd) = P_k(f) + \cO(\sfd^2)  \qquad \sfd\to 0.
\end{equation}}
{\color{blue}Since the $U_k$'s are solutions of the eigenvalue problem in \eqref{ODE}, they are defined only up to a multiplicative constant. 
In the lossless case ($\rho = 0$), this factor is immaterial because both $|U_k|^2$ in the numerator of \eqref{max_directivity_current_fprime} and the corresponding spectral concentration term in \eqref{avg_spectral_conc} in the denominator scale by the same constant. 
However, when $\rho >0$, the eigenvalues contain an additive loss term which is unaffected by $U_k$, making the normalization condition in \eqref{double_orthogonality_dpswf} essential, for $k = 0, \ldots, N-1$,
\begin{align} 
    \int_{-1/2}^{1/2} \left|U_k\left(N,\sfd,\sfff\right)\right|^2 \, d\sfff 
    & = \frac{1}{\mu_k(N,\sfd)} \int_{-\sfd}^{\sfd} \left|U_k\left(N,\sfd,\sfff\right)\right|^2 \, d\sfff \\ \label{norm_double_orthogonality_dpswf}
    & = \frac{\sfd}{\mu_k(N,\sfd)} \int_{-1}^{1} \left|U_k\left(N,\sfd,f\right)\right|^2 \, df = 1.
\end{align}}

{\color{blue}\subsection{Eigenfunction Normalization}

With a slight abuse of notation, let us have $U_k(N,\sfd,f) = U_k(N,\sfd,f \sfd)$ with $f = \sfff/\sfd$.
Squaring the absolute value of the expansion in \eqref{dpswf_legendre_app} and leveraging the boundedness of Legendre polynomials in \eqref{Pk_bound}, implying $P_k(f) = \cO(1)$ uniformly for $|f|\le 1$, returns
\begin{align}
    |U_k(N,\sfd,f)|^2 
    & = \sfc_k(N,\sfd) \,  \left(P_k(f) + \cO(\sfd^2)\right)^2 \\
    & =  \sfc_k(N,\sfd) \,  \left( P_k^2(f) + \left| P_k(f)\right| \cO(\sfd^2) + \cO(\sfd^4)\right) \\  \label{dpswf_legendre_app_squared}
    & =  \sfc_k(N,\sfd) \  \left( P_k^2(f) + \cO(\sfd^2)\right) \qquad \sfd\to 0,
\end{align}
where $\sfc_k(N,\sfd)$ enforces the normalization condition in \eqref{norm_double_orthogonality_dpswf}.
Substituting \eqref{dpswf_legendre_app_squared} into \eqref{norm_double_orthogonality_dpswf} and interchanging the limit and the integral yields 
\begin{align}
\int_{-1}^{1} \left|U_k\left(N,\sfd,f\right)\right|^2 \, df
& = \sfc_k(N,\sfd) \left(\int_{-1}^{1} P_k^2(f) \, df  +  \int_{-1}^{1}  \cO(\sfd^2)  \, df \right)  \\
& =  \sfc_k(N,\sfd) \left(\frac{2}{2k + 1}  + \cO(\sfd^2)\right)  \\ \label{int_U_ka}
& = \frac{2}{2k + 1} \sfc_k(N,\sfd) \left(1 + \cO(\sfd^2)\right),
\end{align}
after using the orthogonality relation \eqref{norm_Legendre}.
The unscaled eigenvalues $\mu_k(N,\sfd)$, sharing the same normalization condition \eqref{double_orthogonality_dpswf}, expand as \cite[Eq.~64]{Slepian1978}
\begin{equation} \label{eigmu_dsmall}
    \mu_k(N,\sfd) 
    =  \frac{G_k(N)}{\pi} (2\pi \sfd)^{2k+1} \, (1 + \cO(\sfd)) 
    \qquad \sfd\to 0,
\end{equation}
where $G_k(N)$ is reported in \eqref{G_k}.
Substituting \eqref{int_U_ka} and \eqref{eigmu_dsmall} into  \eqref{norm_double_orthogonality_dpswf} yields
\begin{align}
   \sfc_k(N,\sfd) & = G_k(N) (2k + 1)  (2\pi \sfd)^{2k}  \, \frac{1 + \cO(\sfd)}{1 + \cO(\sfd^2)} \\
   & =  G_k(N) (2k + 1)  (2\pi \sfd)^{2k}  \, (1 + \cO(\sfd))
    \qquad \sfd\to 0.
\end{align}
Plugged into \eqref{dpswf_legendre_app}, using $\sqrt{1+x} = 1+ \frac{1}{2} x + \cO(x^2)$ with the substitution $x = \cO(\sfd) \to 0$ ($\sfd\to 0$) and expanding the products, the normalized eigenfunction emerges as 
\begin{align}
    U_k(N,\sfd,f) & = \sqrt{\sfc_k(N,\sfd)} \,  (P_k(f) + \cO(\sfd^2)) \\
    & = \sqrt{G_k(N) (2k + 1)} \, (2\pi \sfd)^{k}  \, (1 + \cO(\sfd)) \,  (P_k(f) + \cO(\sfd^2)) \\ 
    & = \sqrt{G_k(N) (2k + 1)} \, (2\pi \sfd)^{k}  \,  (P_k(f) + \cO(\sfd)) \\ \label{Uk_normalized}
    & = \sqrt{G_k(N) (2k + 1)} \, (2\pi \sfd)^{k}  \,  P_k(f) \, (1 + \cO(\sfd)), \qquad \sfd\to 0.
\end{align}
Correctly, inserting \eqref{Uk_normalized} into the spectral concentration formula \eqref{avg_spectral_conc} (unscaled by $2\sfd$ according to \eqref{coupling_prolate_identity}) recovers the asymptotic behavior for $\mu_k(N,\sfd)$ in \eqref{eigmu_dsmall}.}



\section{} \label{app:superdirectivity_vanishing_aperture}

{\color{blue}
Let us henceforth suppress the constant $N$ in the notation. 
Dividing \eqref{eigmu_dsmall} by $2\sfd$ according to \eqref{coupling_prolate_identity} and reinserting losses according to \eqref{eig_lossy} yields
\begin{equation} \label{eiglambda_dsmall}
    \sflambda_k(\sfd,\rho) = \rho + G_k (2\pi \sfd)^{2k} \, (1 + \cO(\sfd)) 
    \qquad \sfd\to 0.
\end{equation}
Substituting \eqref{Uk_normalized} and \eqref{eiglambda_dsmall} into the supergain factor \eqref{max_directivity_current_fprime} gives, for $|f^\prime|\le 1$, 
\begin{align}
\gamma(\sfd,\rho,f^\prime \sfd)
& =
\frac{1}{N} 
\sum_{k=0}^{N-1} 
\frac{G_k (2k + 1) \, (2\pi \sfd)^{2k}  \,  P_k^2(f^\prime) + \cO(\sfd^{2k+1})}
{\rho + G_k (2\pi \sfd)^{2k} \, (1 + \cO(\sfd))} \\
& =
\frac{1}{N}  \sum_{k=0}^{N-1} 
\left(\frac{(2k+1) \, P_k^2(f^\prime)}{1 + \rho \,  m_k \, \sfd^{-2k}} \right) + \cO(\sfd)
\qquad
\sfd\to0,
\label{Sstar_dzero_lambdak}
\end{align}
where
\begin{equation} \label{Mksequence}
 \frac{1}{m_k} = (2\pi)^{2k} G_k,
\end{equation}
with $G_k$ as defined in \eqref{G_k}.
Unscaled by $N$, \eqref{Sstar_dzero_lambdak} subsumes the maximum directivity in \cite[Eq.~A.28]{Yaghjian2005}, also incorporating antenna losses and asymptotic correction terms.
At endfire ($|f^{\prime}|=1$), since $|P_k(1)| = 1$ for every $k$, as given by \eqref{Pk_bound}, \eqref{Sstar_dzero_lambdak} specializes to
\begin{align}
\gamma_{\endf}(\sfd,\rho)
& = \frac{1}{N}  \sum_{k=0}^{N-1} 
\left(\frac{2k+1}{1 + \rho \, m_k\, \sfd^{-2k}} \right) + \cO(\sfd)
\qquad
\sfd\to0.
\label{Sstar_dzero_lambdak_endf}
\end{align}
Allowing $\rho$ depend on $\sfd$, denoted by $\rho(\sfd)$, and defining
\begin{equation}
     \frac{2k+1}{1 + m_k \, \rho(\sfd) \sfd^{-2k}} 
     = a_k(\sfd),
    \label{a_k_definition}
\end{equation}
the endfire supergain factor, $\gamma_\endf(\sfd) = \gamma_\endf(\sfd,\rho(\sfd))$, becomes
\begin{equation}
    \gamma_\endf(\sfd) = \frac{1}{N}  \sum_{k=0}^{N-1} a_k(\sfd) + \cO(\sfd)
\qquad
\sfd\to0.
\label{Sstar_dzero_lambdak_endf_final}
\end{equation}

\begin{itemize}
    \item  Supergain Regime: When the loss factor satisfies $\rho(\sfd) = o(\sfd^{2N-2})$,
\begin{equation} \label{condrhod}
    \rho(\sfd) \sfd^{-2k} = o\!\left( \sfd^{2(N-1-k)} \right) = o(1) \qquad \sfd \to 0,
\end{equation}
uniformly for $k=0,\ldots,N-1$. Expanding $(1 + t_k)^{-1} = 1 - t_k + \cO(t_k^2)$ with $t_k = m_k \rho(\sfd) \sfd^{-2k}$ gives
\begin{equation}
    a_k(\sfd) = (2k+1) \left( 1 - m_k \rho(\sfd) \sfd^{-2k} + \cO\!\left( \rho^2(\sfd) \sfd^{-4k} \right) \right),
\end{equation}
which guarantees uniform convergence of every term in the sum of \eqref{Sstar_dzero_lambdak_endf} as $\sfd\to0$.
From \eqref{Sstar_dzero_lambdak_endf_final}, averaging over $k = 0, \ldots, N-1$ leads to
\begin{equation} \label{Sstar_dzero_lambdak_endf_final_end}
    \gamma_\endf(\sfd) = \frac{1}{N} \sum_{k=0}^{N-1} (2k+1) - \frac{\rho(\sfd)}{N} \sum_{k=0}^{N-1} \left((2k+1) m_k \sfd^{-2k}\right) + \cO\!\left( \rho^2(\sfd) \sfd^{-4N + 4} \right) + \cO(\sfd).
\end{equation}
The leading term evaluates directly using the arithmetic progression formula \cite[Eq.~0.111]{IntegralsBook} to
\begin{equation}
    \frac{1}{N} \sum_{k=0}^{N-1} (2k+1) = N.
\end{equation}
For the leading-order correction, the term corresponding to $k = N-1$ dominates the sum asymptotically as $\sfd\to 0$:
\begin{align}
    \sum_{k=0}^{N-1} (2k+1) m_k \sfd^{-2k} 
    & = (2N-1) m_{N-1} \sfd^{-2N+2} + \sum_{k=0}^{N-2} \left((2k+1) m_k \sfd^{-2k}\right) \\
     &= (2N-1) m_{N-1} \sfd^{-2N+2} + \cO\left( \sfd^{- 2N+4} \right) \\
     &= (2N-1) m_{N-1} \sfd^{-2N+2} (1 + \cO(\sfd^{2})).
\end{align}
Multiplying both sides by $\rho(\sfd)$, by hypothesis,
\begin{align}
    \rho(\sfd) \sum_{k=0}^{N-1} (2k+1) m_k \sfd^{-2k} 
    &= (2N-1) m_{N-1} \rho(\sfd) \sfd^{-2N+2} (1 + \cO(\sfd^{2}))
     = o(1).
\end{align}
As for the quadratic remainder, since $\cO(x^2) = o(1)$ for $x\to 0$, this term can be subsumed into the leading-order correction.
Combining these results into \eqref{Sstar_dzero_lambdak_endf_final_end} gives
\begin{equation} \label{gamma_supergain_general}
    \gamma_\endf(\sfd) = N + o(1) + \cO(\sfd) = N + o(1).
\end{equation}


\item Loss-dominated regime: 
For a constant loss factor $\rho(\sfd) = \rho_0 > 0$, factoring $m_k \rho_0 \sfd^{-2k}$ from the denominator of \eqref{a_k_definition} yields
\begin{align}
    a_k(\sfd) & = \frac{2k+1}{m_k \, \rho(\sfd) \sfd^{-2k} (1 + (m_k \, \rho(\sfd) \sfd^{-2k})^{-1})} \\ \label{ak_definition_loss}
    & = \frac{2k+1}{m_k \, \rho_0 \sfd^{-2k} (1 + m_k^{-1} \, \rho_0^{-1} \sfd^{2k})}.
\end{align}
Since $2k \ge 2 > 0$ for $k = 1, \ldots, N-1$, the expansion parameter
\begin{equation}
   \left( \rho(\sfd) \sfd^{-2k} \right)^{\!-1} = \cO \! \left( \sfd^{2k} \right) = \cO(\sfd^2) \qquad  \sfd \to 0,
\end{equation}
uniformly across these indices.
Expanding \eqref{ak_definition_loss} via $(1 + s_k)^{-1} = 1 - s_k + \cO(s_k^2)$ with $s_k = m_k^{-1} \rho_0^{-1} \sfd^{2k}$ gives
\begin{align} 
    a_k(\sfd) & = \frac{2k+1}{m_k \, \rho_0 \sfd^{-2k}} \left( 1 - \frac{1}{m_k \rho_0} \sfd^{2k} + \cO \! \left( \sfd^{4k} \right) \right) \\ \label{akkkk}
    & =  \frac{2k+1}{m_k \, \rho_0} \left(\sfd^{2k} - \frac{1}{m_k \rho_0} \sfd^{4k} + \cO \! \left( \sfd^{6k} \right) \right).
\end{align}
For $k=0$, setting $m_0 = 1/N$ (equivalently $G_0 = N$) as per \eqref{Mksequence} gives $s_0 = m_0^{-1} \rho_0^{-1} = N/\rho_0 = \cO(1)$, resulting in
\begin{equation} \label{a0000}
a_0 = \frac{1}{1 + \rho_0 m_0} = \frac{N}{N + \rho_0}.
\end{equation}
Splitting the sum \eqref{Sstar_dzero_lambdak_endf_final} into the $k=0$ term and the remaining ones yields
\begin{align} \label{Sstar_dzero_lambdak_endf_final_split}
    \gamma_\endf(\sfd) 
    = \frac{1}{N} \left(a_0 + \sum_{k=1}^{N-1} a_k(\sfd)\right) + \cO(\sfd). 
    \end{align}
Since the sum is dominated by the term corresponding to $k = 1$ asymptotically as $\sfd\to 0$:
\begin{align}
    \sum_{k=1}^{N-1} a_k(\sfd)
    & = \frac{1}{\rho_0} \sum_{k=1}^{N-1} \frac{2k+1}{m_k} \left(\sfd^{2k} - \frac{1}{m_k \rho_0} \sfd^{4k} + \cO \! \left( \sfd^{6k} \right) \right) 
    = \cO(\sfd^{2}).
\end{align}
Combining these gives the expansion
\begin{align}
    \gamma_\endf(\sfd) 
     & = \frac{1}{N + \rho_0} + \cO(\sfd^2) + \cO(\sfd) \\
     & = \frac{1}{N + \rho_0} + \cO(\sfd).
\end{align}

\end{itemize}}

\section{} \label{app:dpswf_Nlarge}

\subsection{{\color{blue}Large-$N$ Analysis}} \label{app:largeN_analysis}

{\color{blue}Since the supergain \eqref{max_directivity_current_fprime} depends on the squared eigenfunctions, which inherit symmetry or antisymmetry from their indices (see Remark~\ref{ref:symmetry}), it suffices to consider $\sfff \in[0, \sfd]$ within the visible region.
 To facilitate the analysis near endfire, $\sfff=\sfd$, 
apply the variable change
\begin{equation} \label{map_u_w}
\sfu = \cos(2\pi\sfff)-A  \qquad  \quad\Xi_k(N,\sfd,\sfu) = U_k \! \left(N,\sfd,\arccos \! \left( \frac{\sfu+A}{2\pi}\right)\right),
\end{equation}
where $A = \cos(2\pi \sfd)$ is defined in \eqref{A_cos_def} 
and $\sfu \ge 0$ represents the distance from endfire.}
{\color{blue}In turn,  $\sfd$ is omitted from the notation and $\sfu(N)$ is treated as an implicit function of $N$. 
(The specific behavior of $\sfu$ in each subset $\cK_i$ is detailed next.) 
Accordingly, let us define $\Xi_k(N) = \Xi_k(N,\sfd,\sfu)$ and $\mu_k(N) = \mu_k(N,\sfd)$.}

\begin{itemize}
\item Case {\unboldmath $-1 < A < B < 1 \; \mathrm{or} \; \cK_1 = \{k = \lfloor N_0 (1-x)\rfloor, 0 < x < 1\} $ }

Specializing the asymptotic constraint in \eqref{con1} to $k \in \cK_1$ and omitting rounding errors, which only produce {\color{blue}$\cO(1)$ corrections while $k=\cO(N)$},
\begin{equation} \label{L1_B_I2}
    B(x) = \left\{A < B< 1 : \int_B^1 \sqrt{\frac{t-B}{(t-A)(1-t^2)}} \, dt
    \sim 2 \pi \sfd (1-x) \right\},
\end{equation}
where $N_0 = 2\sfd N$ was used {\color{blue}from \eqref{DOF_0}. }
Let us introduce the positive integrals \cite[Eq.~47]{Slepian1978}
\begin{equation}
\begin{aligned} \label{Li_Slepian}
       L_1(B) & = \int_{B}^1 R(B,t) \, dt \hspace{1.5cm} 
       L_2(B) = \int_{B}^1 S(B,t) \, dt  \\
   L_3(B) & = \int_{A}^B R(B,t) \, dt \hspace{1.5cm} 
   L_4(B) = \int_{A}^B S(B,t) \, dt,
\end{aligned}
\end{equation}
where
\begin{equation}
\begin{aligned}
    R(B,t)   = \left|\frac{t-B}{(t-A)(1-t^2)}\right|^{1/2} \qquad S(B,t)  = \left|(t-B)(t-A)(1-t^2)\right|^{-1/2}.
    \end{aligned}
\end{equation}
Each integral is an implicit function of the asymptotic parameter $x$ via $B$ through the constraint  \eqref{L1_B_I2}.
This dependence is henceforth subsumed into that of $x$ by defining $L_n(x) = L_n(B(x))$ for $n=1,2,3,$ and $4$.
For any admissible $x$, further define \cite[Eqs.~43, 45]{Slepian1978}
\begin{equation} \label{C_eq_k}
    C(N,x)  = \frac{4}{L_2(x)} \left[\frac{N}{2} L_1(x) + \left( 2 + (-1)^{\lfloor 2\sfd N (1-x)\rfloor} \right) \frac{\pi}{4} \right]_{\text{rem} \, 2\pi}.
\end{equation}
{\color{blue}Letting $\sfu = \sfv/ N^2$ with $\sfu$ as defined in \eqref{map_u_w}, the eigenvalue problem \eqref{ODE} with $\theta$ replaced by its large-$N$ expansion in \eqref{theta_asymptotic_Nlarge_aux}, becomes, with a slight abuse of notation,
\begin{equation}
\sfv \frac{d^{2} \Xi}{d\sfv^{2}} + \frac{d \Xi}{d\sfv} - \frac{B-A}{4(1-A^2)} \Xi + \cO \! \left(\frac{1}{N}\right) = 0.
\end{equation}
Then, the corresponding eigenfunctions asymptotically behave} as \cite[Eq.~42]{Slepian1978}
\begin{equation} \label{Xikd1}
    \Xi_k(N) \sim  c_{1}(N,x) \, I_0\!\left(N \sqrt{\frac{B-A}{1-A^2} \sfu}\right) {\color{blue}= \Xi^{(1)}_\endf(N,x)} \qquad N\to \infty,
\end{equation}
where $I_0(\cdot)$ is the modified Bessel function of the first kind, and 
\begin{align} \label{d4k}
c_{1}(N,x)  = \pi \sqrt{\frac{ N e^{-C(N,x) L_4(x)/2} e^{-N L_3(x)} }{L_2(x) \sqrt{1-A^2}}}.
\end{align}
{\color{blue}Let $z(N)=N \sqrt{a \sfu(N)} = \sqrt{a N^2 \sfu(N)}$ with $a = (B-A)/(1-A^2)$. Setting $\sfu(N) =  o(1/N^2)$ implies that $z(N) = o(1)$ as $N\to \infty$ whereby the small-argument expansion applies to the Bessel function in \eqref{Xikd1},
\begin{align}
I_0(z(N)) & = 1 + \frac{z^{2}(N)}{4} + o(z^2(N)) \\
& = 1 +  \frac{a N^2 \sfu(N)}{4} + o(N^2 \sfu(N)) = 1 + o(1) \qquad N\to \infty.
\end{align}
Then, using \eqref{Xikd1} and taking the squared absolute value, asymptotically gives}
\begin{align}   \label{U_lowpass_wideband}
   |\Xi^{(1)}_\endf(N,x)|^{2}   \sim
      \frac{\pi^2 e^{-C(N,x) L_4(x)/2}}{L_2(x) \sqrt{1-A^2}} \,  N e^{-L_3(x) N} \qquad N\to \infty.
\end{align}
{\color{blue}Since the mapping $z\mapsto |z|^2$ is everywhere continuous, asymptotic equivalence is preserved under taking the squared modulus. 
This fact is used repeatedly without further mention.}
The eigenvalues converge to unity exponentially fast with increasing $N$. {\color{blue}From \cite[Eq.~59]{Slepian1978},}
\begin{equation} \label{mu_1}
   1- \mu_k(N) \sim  e^{- C(N,x) L_4(x)/2} \, e^{- L_3(x) N} {\color{blue}= 1- \mu^{(1)}(N,x)} \qquad N\to \infty.
\end{equation}
    
\item Case {\unboldmath $-1 < A = B <1 \; \mathrm{or} \; \cK_2 = \{k = \lfloor N_0 + \frac{x}{\pi} \log(N)\rfloor, -\frac{\pi N_0}{\log(N)} < x < \frac{\pi (N - N_0)}{\log(N)}  \}$}

{\color{blue}Letting $ \sfu = \j \xi/ (N \beta)$ and $\tilde{\Xi} = \Xi e^{\xi/2}$, the eigenvalue problem \eqref{ODE}, with $\theta$ replaced by its large-$N$ expansion in \eqref{theta_asymptotic_Nlarge_aux} as well as $B=A$, morphs into \cite[Eq.~170]{Slepian1978} 
\begin{equation}
\xi \frac{d^{2} \tilde{\Xi}}{d\xi^{2}} + (1-\xi) \frac{d \tilde{\Xi}}{d\xi} - \left(\frac{1}{2} - \frac{\j}{2} E \beta \right)  \tilde{\Xi} + \cO \! \left(\frac{1}{N}\right) = 0.
\end{equation}
Then, the corresponding eigenfunctions asymptotically behave} as  \cite[Eq.~50]{Slepian1978}
\begin{align} \label{Xitildek_logN} 
    \tilde{\Xi}_k(N)  \sim  c_2(N,x) \, \Phi\!\left(\frac{1}{2}- \frac{\j}{2} E(N,x) \beta, 1; \xi \right) = \tilde{\Xi}^{(2)}(N,x)  \qquad N\to \infty,
\end{align}
where $\beta = (1-A^2)^{-1/2}$, $\Phi(\cdot)$ is the confluent hypergeometric function, and $c_2(N,x)$ is a normalization constant, given by
\begin{align}  \label{e2}
c_2(N,x) & = (-1)^{\left\lfloor\frac{N_0}{2} + \frac{x}{2 \pi} \log(N) \right\rfloor}  \left|\Gamma\left(\frac{1}{2}-\frac{\j E(N,x) \beta}{2} \right)\right| \, e^{\frac{\pi}{4} E(N,x) \beta} \\ \nonumber 
& \hspace{4cm} \cdot \sqrt{\frac{3 \pi}{(1+e^{\pi E(N,x) \beta})}} \sqrt{\frac{N}{\log N}},
\end{align}
with $\Gamma(\cdot)$ the Gamma function. 
In turn, $E(N,x)$ is determined as the absolute value of the smallest root of \cite[Eq.~53]{Slepian1978} with the index $k$ in the formulation specialized to the subset $\cK_2$,
 yielding, to leading order,
\begin{equation}
    E(N,x) \beta = 
    x + \cO \! \left(\frac{1}{\log N }\right) 
    \qquad N\to \infty.
\end{equation} 
Using the identity $|\Gamma(\frac{1}{2} +\j y)|^2 = \pi \cosh^{-1}(\pi y)$ \cite[Eq.~6.1.30]{AbramowitzStegun} and the limit $y=-E \beta/2 \sim x/2$ into \eqref{e2},
asymptotically yields  
\begin{equation} \label{e2_asym}
c_2(N,x) \sim (-1)^{\left\lfloor\frac{N_0}{2} + \frac{x}{2 \pi} \log(N) \right\rfloor}    
\sqrt{\frac{3 \pi^{2} e^{\pi x/2}}{\cosh(\frac{\pi x}{2}) (1+e^{\pi x})}} \sqrt{\frac{N}{\log N}}
\qquad N\to\infty.
\end{equation}
{\color{blue}
Plugging \eqref{e2_asym} into \eqref{Xitildek_logN} and reversing the variable change, 
\begin{align} \label{Uk_logN} 
    \Xi_k(N)  & \sim   (-1)^{\left\lfloor\frac{N_0}{2} + \frac{x}{2 \pi} \log(N) \right\rfloor}    
e^{-\j (\beta/2) N \sfu}  \sqrt{\frac{3 \pi^{2} e^{\pi x/2}}{\cosh(\frac{\pi x}{2}) (1+e^{\pi x})}} \sqrt{\frac{N}{\log N}}  \\&
\hspace{2cm}   \nonumber
\cdot   \Phi\!\left(\frac{1}{2}- \frac{\j x}{2}, 1; -\j \beta N \sfu \right) = \Xi^{(2)}(N,x)
    \qquad N\to \infty.
\end{align}
As $N\to\infty$ with $\sfu=o(N^{-1})$, the argument of $\Phi$ is $o(1)$, and applying the same reasoning as in $\cK_1$ asymptotically yields}
\begin{align}   \label{U_allpass}
   |\Xi^{(2)}_\endf(N,x)|^{2}   \sim
      \frac{3\pi^2 e^{\pi x/2}}{\cosh(\frac{\pi x}{2}) (1+e^{\pi x})} \frac{N}{\log N} \qquad N\to \infty.
\end{align}
The eigenvalues exhibit a smooth transition from $1$ to $0$ as the parameter $x$ spans its entire domain, asymptotically giving {\color{blue} \cite[Eq.~60]{Slepian1978}}
    \begin{equation} \label{mu_2}
   \mu_k(N) \sim 
       (1+ e^{\pi x})^{-1} {\color{blue} = \mu^{(2)}(N,x)} \qquad N\to \infty.
\end{equation}

\item Case {\unboldmath $-1 < A^\prime < B < 1 \; \mathrm{or} \; \cK_3 = \{k = \lfloor N_0 (1+x)\rfloor, 0 < x < \frac{1}{2\sfd}-1 \}$}

Capitalizing on spectral symmetry, let us introduce the auxiliary variables \cite[Eqs.~56-57]{Slepian1978}
\begin{equation} \label{auxiliary_variables_app}
\sfff^\prime = \frac{1}{2}-\sfff  \qquad\quad \sfd^\prime = \frac{1}{2} - \sfd \qquad\quad k^\prime = N-1-k,
\end{equation}
such that $A^\prime = \cos(2\pi \sfd^\prime) = -A$. 
{\color{blue}(Note that $\sfff^\prime$ here denotes the symmetry-transformed frequency, rather than the beamforming direction as used elsewhere.)}
{\color{blue}In turn, define $\sfu^{\prime} = \cos(2\pi \sfff^\prime) - A^\prime = - \sfu$.}
From $k\in \cK_3$, 
\begin{align} \label{ksim_I4}
    k^\prime   
     \sim 2\sfd^\prime N (1-x^\prime) \qquad\quad x^\prime \sim (\sfd/\sfd^\prime) \, x,
\end{align}
where $0 < x < \sfd^\prime/\sfd$ ensuring $ 0<x^\prime <1$.
Applying the substitution $s = \cos(2\pi \sfff^\prime) = -t$ and expressing $k$ in terms of $k^\prime$ via the asymptotic identity $k + k^\prime \sim N$, the constraint in \eqref{con1} specializes for $k\in \cK_3$ to
\begin{equation} \label{L1_B_I4}
    B(x^\prime) = \left\{A^\prime < B< 1 : \int_{-1}^{-B} \sqrt{\frac{s+B}{(s+A^\prime)(1-s^2)}} \, ds  
    \sim 2 \pi \sfd^\prime (1-x^\prime) \right\}.
\end{equation}
Due to symmetry, the positive integrals in \eqref{Li_Slepian} morph into
\begin{equation} \label{Li_Slepian_kprime} 
\begin{aligned}
       L_1^\prime(B) & = \int_{-1}^{-B} R^\prime(B,s) \, ds \hspace{1.5cm} 
       L_2^\prime(B) = \int_{-1}^{-B} S^\prime(B,s) \, ds  \\
   L_3^\prime(B) & = \int_{-B}^{-A^\prime} R^\prime(B,s) \, ds \hspace{1.5cm} 
   L_4^\prime(B) = \int_{-B}^{-A^\prime} S^\prime(B,s) \, ds,
    \end{aligned}
\end{equation}
where 
 \begin{align}  \label{RS_prime}
              R^\prime(B,s)   = \left|\frac{s+B}{(s+A^\prime)(1-s^2)}\right|^{1/2}  \qquad
    S^\prime(B,s)  = \left|(s+B)(s+A^\prime)(1-s^2)\right|^{-1/2}.
\end{align}
Because each integral is an implicit function of $x^\prime$ via $B$ through the constraint  \eqref{L1_B_I4}, we define $L_n^{\prime}(x^\prime) = L_n^\prime(B(x^\prime))$ for $n=1,2,3,$ and $4$.
{\color{blue}The mapping $x^\prime \mapsto B$ and the numerical integration of $L_n^\prime$ are detailed in Appendices~\ref{app:existence_uniqueness} and \ref{app:numerical}, respectively.
Finally, note that while the main text drops the primed notation ($x = x^{\prime}$, $L_n = L_n^{\prime}$) for readability, this Appendix retains them to avoid confusion with the unprimed counterparts in $\cK_1$.}

{\color{blue}Also by symmetry, the eigenfunctions $U_k(N,\sfd,\sfff)$ and $U_{k^\prime}(N,\sfd^\prime,\sfff^\prime)$ coincide up to a sign depending on $k$.}
Then, the corresponding asymptotic behavior is governed by \cite[Eq.~42]{Slepian1978}
\begin{equation} \label{Xikd3}
    \Xi_k(N) \sim  c_{3}(N,x^\prime) \, J_0\!\left(N \sqrt{\frac{B+A^\prime}{1-(A^\prime)^2} \sfu^\prime}\right)  {\color{blue}= \Xi^{(3)}_\endf(N,x^{\prime})} \qquad N\to \infty,
\end{equation}
where $J_0(\cdot)$ is the Bessel function of the first kind, and $c_{3}$ follows from $c_{1}$ in \eqref{d4k} upon replacing the $L_n$'s by their counterparts $L_n^\prime$'s, and replacing $C$ in \eqref{C_eq_k} by 
\begin{equation}
    C(N,x^\prime) = \frac{4}{L_2^\prime(x^\prime)} \left[\frac{N}{2} L_1^\prime(x^\prime) + \left(2 + (-1)^{\lfloor 2 d^\prime N (1-x^\prime)\rfloor}\right) \frac{\pi}{4} \right]_{\text{rem} \, 2\pi}.
\end{equation}
{\color{blue}As $N\to\infty$ with $\sfu^\prime=o(N^{-2})$, the argument of $J_0$
is $o(1)$, so that the same reasoning as in the subset $\cK_1$ applies. Hence, asymptotically,}
\begin{align} \label{U_highpass}
 \left| \Xi^{(3)}_\endf(N,x^\prime) \right|^2  \sim   \frac{\pi^2 e^{-C(N,x^\prime) L_4^\prime(x^\prime)/2}}{L_2^\prime(x^\prime) \sqrt{1-(A^\prime)^2}} \, N  \, e^{- L_3^\prime(x^\prime) N} 
    \qquad  N\to \infty.
\end{align} 
By symmetry, $\mu_k(N,\sfd) = 1-\mu_{k^\prime}(N,\sfd^\prime)$, implying from \eqref{mu_1} that the eigenvalues decay exponentially fast with increasing $N$, \cite[Eq.~63]{Slepian1978}
    \begin{equation}  \label{mu_3}
  \mu_k(N) \sim 
       e^{- C(N,x^\prime) L_4^\prime(x^\prime)/2} \, e^{- L_3^\prime(x^\prime) N} {\color{blue} = \mu^{(3)}(N,x^\prime)}  \qquad N\to \infty.
\end{equation}
\end{itemize}

\subsection{Removal of Singularities in the $L$ Integrals} \label{app:numerical}

{\color{blue}Let us regularize the endpoint singularities of the integral family $L_n^\prime$ in \eqref{Li_Slepian_kprime}, given by} 
\begin{equation}
L_n^\prime(B) = \int_{s_\text{L}}^{s_\text{U}} \ell_n^\prime(B,s) \, ds    \qquad -1 \le s_\text{L} < s_\text{U} < 1,
\end{equation}
where the kernel $\ell_n^\prime(B,s)$ is either $R^\prime(B,s)$ or $S^\prime(B,s)$, {\color{blue}according to the index $n$.}
{\color{blue}(The regularization of the $L_n$'s in \eqref{Li_Slepian} follows similarly.)}
The kernels exhibit endpoint singularities originating from simple poles. Denoting by
 $\cP_n(\ell_n^\prime)$ the set of simple poles {\color{blue}associated with a constituent function $\ell_n^\prime$}, these are 
\begin{align}
    \cP_n(R^\prime) =
    \begin{cases}
        \{-1\} & n=1 \\
        \{-A^\prime\} & n=3
    \end{cases}
\qquad
    \cP_n(S^\prime) =
    \begin{cases}
        \{ -B, -1\} & n=2 \\
        \{-B, -A^\prime\} & n=4.
    \end{cases}
\end{align}
To resolve these singularities, the change of variables $\phi \mapsto s$ is applied:
\begin{equation}  \label{transformation_trigon} 
    s = s_\text{L} + (s_\text{U} - s_\text{L}) \, \sin^2(\phi) \qquad \quad ds = 2(s_\text{U} - s_\text{L}) \sin(\phi) \cos(\phi) \, d\phi,
\end{equation}
where the differential cancels the singularities, yielding the regularized integrals in \eqref{Li_x3prime}.

{\color{blue}To derive the implicit constraint in \eqref{B_d_main}, \eqref{transformation_trigon} is specialized to $s_\text{L}=-1$ and $s_\text{U} = -B$. The change of variables and its differential become
\begin{equation}  \label{transformation_trigon_L1} 
    s = -1 + (1-B) \, \sin^2(\phi) \qquad \quad ds = 2(1-B) \sin(\phi) \cos(\phi) \, d\phi.
\end{equation}
Substituting these into the radical of \eqref{L1_B_I4} and replacing the auxiliary variables in \eqref{auxiliary_variables_app}, $\sfd^{\prime}=1/2-\sfd$ ($A^{\prime}=-A$),  yields
\begin{equation}
2(1-B) \int_{0}^{\pi/2} \frac{\cos^2(\phi) \, d\phi}{\left| \left(A - \psi_1(\phi) \right) \left(1 - \psi_1(\phi)\right) \right|^{1/2}}  \sim  \pi (1-2\sfd)  (1- x^{\prime}),
\end{equation}
where $\psi_1(\phi) = - 1 + (1-B) \sin^2(\phi)$ and $x^{\prime}\in (0,1)$. Dropping the primed notation by defining $x = x^\prime$ returns \eqref{B_d_main}.}


\subsection{Existence and Uniqueness of the Mapping $x^\prime \mapsto B$ in \eqref{L1_B_I4}}
\label{app:existence_uniqueness}

{\color{blue}Let
\begin{equation} \label{cL1B}
\cL_1^{\prime}(B) = \int_{-1}^{-B} \sqrt{\frac{s+B}{(s+A^\prime)(1-s^2)}} \, ds.
\end{equation}
The asymptotic constraint in \eqref{L1_B_I4} is then compactly rewritten as
\begin{equation}  \label{cL1B_constraint}
\cL_1^{\prime}(B) \sim 2 \pi \sfd^\prime (1-x^\prime)
\end{equation}
for all $x^\prime\in (0,1)$ and $A^{\prime} < B < 1$. 
Next, we establish the existence and uniqueness of the mapping $x^\prime \mapsto B$ defined implicitly by \eqref{cL1B_constraint}. Figure~\ref{fig:B_d} illustrates this implicit relation, corroborating the theoretical results.}


To prove existence, we show that the mapping $x^\prime \mapsto B$ is surjective.
Because $s \in  (-1,-B)$, both $(s+B)$ and $(s+A^{\prime})$ are negative, while $(1-s^{2})$ is positive.  Then, the integrand of \eqref{cL1B} is positive and real-valued, making $\cL_1^{\prime}(B)$ a continuous function of $B$.
Moreover, evaluating the boundary limits of  $\cL_1^{\prime}(B)$ via the transformation in \eqref{transformation_trigon}:
\begin{align}  \label{L1prime_B1}
   \lim_{B\to 1}  \cL_1^{\prime}(B) & = \int_0^{\pi/2} \frac{0}{2 \sqrt{1 + A}} \, d\phi = 0
\\ \label{L1prime_B1_Aprime}
    \lim_{B\to A^\prime}  \cL_1^{\prime}(B)
    &  = \int_{-1}^{-A^\prime} \frac{ds}{\sqrt{1-s^2}}
    = \arccos(A^\prime)
    = 2\pi \sfd^\prime.
\end{align}
These integrals ensure that $x^\prime$ fully spans the domain of $B$ through the limits $x^\prime \to 1$ and $x^\prime \to 0$ applied to \eqref{cL1B_constraint}. Combined with the continuity of the integral proves the existence of at least one solution $B$ for every $x^{\prime}$.


To prove uniqueness, we show that the mapping $x^\prime \mapsto B$ is also injective. 
Applying the Leibniz's integral rule to differentiate $\cL_1^{\prime}(B)$ with respect to $B$, and since the integrand vanishes at the upper limit $s=-B$,
\begin{align}
    \frac{d \cL_1^{\prime}(B)}{dB} 
    & =  \int_{-1}^{-B} \left|\frac{1}{(s+A^\prime)(1-s^2)}\right|^{1/2} \frac{s+B}{2 |s+B|^{3/2}} \, ds.
\end{align}
Because $s<-B$ throughout the integration interval, the term $s+B$ is negative whereas the remaining terms are positive, implying that $\frac{d}{dB} \cL_1^{\prime}(B) < 0$ for $A^{\prime} < B < 1$.
Further differentiating the constraint in \eqref{cL1B_constraint} with respect to $x^{\prime}$ using the chain rule yields 
\begin{equation} \label{derivative_B_vs_x}
  \frac{dB}{d x^\prime} \sim - 2 \pi \sfd^\prime \left(\frac{d \cL_1^{\prime}(B)}{dB} \right)^{-1} > 0.
\end{equation}
Since $B(x^\prime)$ is monotonically increasing, the inverse mapping is injective, ensuring that the solution $B$ is unique for every $x^{\prime}$, consistent with Fig.~\ref{fig:B_d}. 

{\color{blue}\subsection{Monotonicity and Supremum of $L_3^\prime(\sfd,x^\prime)$}
\label{app:L3_monotonicity}

Combining \eqref{RS_prime} and \eqref{Li_Slepian_kprime}, the integral $L_3^\prime(x^\prime) = L_3^\prime(\sfd,x^\prime)$ for fixed $\sfd$ rewrites as
\begin{equation} \label{cL3B}
L_3^\prime(x^\prime) = \int_{-B}^{A} \sqrt{\frac{s+B}{(A-s)(1-s^2)}} \, ds.
\end{equation}
To study the integral behavior with respect to $x^\prime(B)$, we differentiate \eqref{cL3B} via the chain rule:
\begin{equation} \label{derivative_B_vs_x_L3}
   \frac{d L_3^\prime(x^\prime)}{d x^\prime}  =  \frac{dB}{d x^\prime} \, \frac{d L_3^\prime(B)}{dB}.
\end{equation}
Since the mapping $x^\prime \mapsto B$ is monotonically increasing as per \eqref{derivative_B_vs_x}, the behavior of $L_3^\prime(x^\prime)$ is governed by its derivative with respect to $B$.
Applying Leibniz's integral rule to \eqref{cL3B}, and noting that the boundary term at $s = -B$ vanishes, we obtain
\begin{align} 
    \frac{d}{dB} L_3^\prime(B) & = \int_{-B}^{A} \frac{d}{dB} \left(\sqrt{\frac{s+B}{(A-s)(1-s^2)}}\right) \, ds \\ \label{dL3_dB}
    & = \int_{-B}^{A}  \frac{ds}{2 \sqrt{(s+B)(A-s)(1-s^2)}}.
\end{align}
Because all factors $(s+B)$, $(A-s)$, and $(1-s^2)$ are positive for any $s \in (-B, A)$ with $B\in (-A, 1)$, the integrand of \eqref{dL3_dB} is positive, making $L_3^\prime(x^\prime)$ a monotonically increasing function of $x^\prime\in (0,1)$ by virtue of \eqref{derivative_B_vs_x_L3}.

Since $x^\prime \mapsto B$ is strictly increasing mapping onto the open interval $B\in(-A,1)$, we have
\begin{align}
  \sup_{x^\prime \in (0,1)} L_3^\prime(x^\prime) 
  & = \sup_{B \in (-A,1)} L_3^\prime(B) 
   = \lim_{B \to 1^-} L_3^\prime(B) \\
  & =  \int_{-1}^{A} \frac{1}{\sqrt{(A-s)(1-s)}} \, ds \\
  & = \ln\left(\frac{2 \sqrt{s^2 -(1+A)s + A} + 2 s -(1+A)}{4 A - (1+A)^2} \right)\bigg|_{-1}^A \\
  & = \ln\left(\frac{1}{1-A} \right) - \ln\left(\frac{3+A - 2 \sqrt{ 2(1+A)}}{(1-A)^2} \right) \\
  & = \ln\left(\frac{1-A}{3+A - 2 \sqrt{ 2(1+A)}} \right),
\end{align}
after using \cite[Eq.~2.261]{IntegralsBook}
\begin{equation}
    \int \frac{ds}{\sqrt{R(s)}} = \frac{1}{\sqrt{c}} \ln\left(\frac{2 \sqrt{c R(s)} + 2c s + b}{\Delta} \right) \qquad c>0,
\end{equation}
where $R(s) = A + b s + c s^2$ and $\Delta = 4Ac - b^2$ with $A = \cos(2\pi \sfd)$, $b=-(1+A)$, and $c=1$.}

\section{} \label{app:endfire_supergain_wide_apertures}




By applying \eqref{map_u_w} and combining \eqref{eig_lossy} with \eqref{coupling_prolate_identity}, the endfire supergain in~\eqref{max_directivity_current_fprime} can be expressed as a sum of positive terms over the index set $\cK = \{0, \ldots, N-1\}$, namely
\begin{equation}      \label{S_star_endfire_app_losses}
\gamma_\endf(N,\sfd,\rho) 
 = \frac{1}{N} \, \sum_{k \in \cK}  r_k^{(i)}(N,\sfd,\rho),
\end{equation}
where the individual contributions are defined by
\begin{equation} \label{f_k_i}
    r_k^{(i)}(N,\sfd,\rho) = \frac{|\Xi_k(N,\sfd,0)|^2}{(2\sfd)^{-1} \mu_k(N,\sfd) + \rho}.
\end{equation}
Define a non-disjoint cover of the set $\cK$ by $\bigcup_{i=1}^3 \cK_i = \cK$. These subsets satisfy $\cK_1 \cap \cK_3 = \emptyset$, while $\cK_2$ overlaps with both $\cK_1$ and $\cK_3$.
Let $S_N^{(i)}(\sfd,\rho)$ be the partial sum over each subset:
\begin{equation} \label{sum_Ki}
    S_N^{(i)}(\sfd,\rho) = \frac{1}{N} \, \sum_{k \in \cK_i}  r_k^{(i)}(N,\sfd,\rho).
\end{equation}
As every $r_k^{(i)}$ is positive, the partial subset overlap implies that a summation of $S_N^{(i)}(\sfd,\rho)$ overcounts certain indices, providing an upper bound on the supergain. Conversely, considering only the region $\cK_3$ in isolation yields a lower bound.
Altogether
\begin{equation} \label{inequality_superdirgain}
\underline{\gamma}_\endf(N,\sfd,\rho)  < \gamma_\endf(N,\sfd,\rho)  < \overline{\gamma}_\endf(N,\sfd,\rho) ,    
\end{equation}
where the nonachievable bounds are
\begin{align}
 \underline{\gamma}_\endf(N,\sfd,\rho) 
    = S_N^{(3)}(\sfd,\rho) 
    \qquad
\text{and} \qquad
\overline{\gamma}_\endf(N,\sfd,\rho) 
     = \sum_{i=1}^3 S_N^{(i)}(\sfd,\rho).
    \end{align}
As $N\to\infty$, the summands within each subset $\cK_i$ are asymptotically described by the continuous functions $\Xi^{(i)}_\endf(N,\sfd,x)$ and $ \mu^{(i)}(N,\sfd,x)$, replacing the dependence on the discrete index $k\in \cK_i$ by the continuous variables $x$.
Invoking the asymptotic results from Appendix~\ref{app:dpswf_Nlarge},  the ratio in \eqref{f_k_i} converges to the continuous function 
\begin{equation} \label{f_i_largeN}
\begin{split}
    r^{(i)}(N,\sfd,\rho,x) & = \frac{|\Xi^{(i)}_\endf(N,\sfd,x)|^2}{\mu^{(i)}(N,\sfd,x)/(2\sfd) + \rho}    
    \\ & \hspace{1cm}  
    \sim \begin{cases} 
        \begin{aligned}
            &\frac{\pi e^{- C(N,x) L_4(x)/2}}{L_2(x) \sinc(2 \sfd)}
            \\
            &  \cdot 
             \frac{N \, e^{- L_3(x) N}}{ 1-e^{- C(N,x) L_4(x)/2} e^{- L_3(x) N} + 2 \sfd \rho } 
        \end{aligned} &  i=1 \\[6ex]
        \displaystyle 
        \frac{e^{\pi x}}{1+e^{\pi x}} \frac{12\pi^2 \sfd}{1 + 2 \sfd \rho (1+ e^{\pi x})} \, \frac{N}{\log N}  &  i=2 \\[3ex]
        \begin{aligned}
        & \frac{\pi N}{L_2^\prime(x^\prime) \sinc(2 \sfd)} 
            \\
            & \, \cdot 
            \frac{1}{1 + 2 \sfd \rho \, e^{C(N,x^\prime) L_4^\prime(x^\prime)/2} \, e^{ L_3^\prime(x^\prime) N}}
        \end{aligned}   &  i=3
    \end{cases}
\end{split}
\end{equation}
where $\sqrt{1-A^2} = \sin(2\pi \sfd)$ from \eqref{A_cos_def} and $2\cosh(x)=e^{x}+e^{-x}$ were used.

To assess the behavior of the sums $S_N^{(i)}(\sfd,\rho)$ as $N\to \infty$, each is converted to a Riemann sum and the resulting expression is replaced by the convergent integral \cite[Sec.~6.7]{ODEBook}.  
For each subset $\cK_i$, the discrete points $x_{k}$ are obtained by inverting the map $x \mapsto k$ in Appendix~\ref{app:dpswf_Nlarge} while a step size $\Delta_{x,k} = x_{k+1}-x_{k}$ is naturally revealed by the normalization of the supergain factor in \eqref{supergain_def_def}.

\begin{itemize}
\item
Subset $\cK_1$: By mapping $1 - x_{k} = \frac{k}{2\sfd N}$, the step size becomes $\Delta_{x,k} = -\frac{1}{2\sfd N} = \Delta_{x}$.
Evaluating the endpoints $k=0$ and $k=\lfloor 2\sfd N \rfloor$ maps the limits to $x_{0} = 1$ and $x_{\lfloor 2\sfd N \rfloor} = 0$ (up to a vanishing error due to round-off). 
Since the differential vanishes asymptotically as $N\to \infty$, the sum \eqref{sum_Ki} for $\cK_1$ converges to
\begin{align}
    \lim_{N\to\infty} S_N^{(1)}(\sfd,\rho) & = \lim_{N\to\infty} \frac{1}{N} \sum_{k \in \cK_1}  r^{(1)}(N,\sfd,\rho,x_{k}) \\
    & = 2\sfd \lim_{\Delta_{x}\to 0} \sum_{k \in \cK_1}  r^{(1)}(N,\sfd,\rho,x_{k}) \, \Delta_{x} \\
    & = 2\sfd \int_0^1 r^{(1)}(N,\sfd,\rho,x) \, dx,
\end{align}
where the minus sign in the differential cancels out by swapping integration limits.

\item Subset $\cK_2$: Defining $x_{k} 
    = \frac{\pi}{\log N} (k - 2\sfd N)$, the step size becomes $\Delta_{x,k} = \frac{\pi}{\log N} = \Delta_{x}$.
The endpoints $k=0$ and $k=N$ map to 
\begin{align}
x_{0} = -\frac{2\pi \sfd N}{\log N}
\qquad \text{and} \qquad
x_{N} = \frac{\pi N}{\log N} (1 - 2\sfd).
\end{align}
While the differential vanishes more slowly than in $\cK_1$, the discrete set becomes more dense as $N\to \infty$.
The sum in \eqref{sum_Ki} for $\cK_2$ behaves as
\begin{align}
    \lim_{N\to\infty} S_N^{(2)}(\sfd,\rho) & = \lim_{N\to\infty} \frac{1}{N} \sum_{k \in \cK_2}  r^{(2)}(N,\sfd,\rho,x_{k}) \\
    & = \frac{1}{\pi} \lim_{\Delta_{x}\to 0} \sum_{k \in \cK_2}  \left(\frac{\log N}{N} r^{(2)}(N,\sfd,\rho,x_{k}) \right) \, \Delta_{x} \\
    & = \frac{1}{\pi} \int_{-\infty}^{\infty} \left(r^{(2)}(N,\sfd,\rho,x) \frac{\log N}{N}\right) \, dx,
\end{align}
where the integration limits tend to $\pm \infty$ as $N\to\infty$. 
Notice how the multiplication factor suggests that the contribution vanishes, but this effect is compensated by the integrand $r^{(2)}(N,\sfd,\rho,x)$ being of order $\cO(N/\log N )$ as per \eqref{U_allpass}, ensuring the sum remains nonzero asymptotically.
Particularly, replacing $r^{(2)}(N,\sfd,\rho,x)$ by \eqref{f_i_largeN} yields
\begin{align}   \label{GGG}
\lim_{N\to\infty} S_N^{(2)}(\sfd,\rho) 
& = 
12\pi \sfd \int_{-\infty}^{\infty} \frac{ e^{\pi x}}{(1+e^{\pi x}) (1+ 2\sfd \rho + 2\sfd \rho e^{\pi x})}  \, dx \\ & =
12 \sfd \int_{0}^{\infty} \frac{1}{(1+u) (1+ 2\sfd \rho + 2\sfd \rho u)}  \, du \\ & =
12 \sfd \int_{0}^{\infty} \left(\frac{1}{1+u} {\color{blue}-} \frac{2\sfd \rho}{1 + 2\sfd \rho + 2\sfd \rho u} \right)  \, du \\& = 
{\color{blue} 12 \sfd \ln \! \left( \frac{1+ \sfu}{1 + 2\sfd \rho + 2\sfd \rho u}  \right)\Big|_{0}^\infty }
\\& = 
12 \sfd \ln \! \left(1+\frac{1}{2\sfd \rho} \right)
 \qquad\quad N\to \infty,
\end{align}
after applying the substitution $u=e^{\pi x}$ and using the partial fraction decomposition.
The subset $\cK_2$ presents a subtle integrability issue in the lossless case ($\rho = 0$) as the integrand remains flat as $x\to \infty$. 
Assuming a small loss, however, integrability is restored, 
compounded by the exponential decay of the integrand as $x \to -\infty$.

\item Subset $\cK_3$:
Leveraging the spectral symmetry in \eqref{auxiliary_variables_app}, let us define $1 - x_{k^\prime}^\prime
    = \frac{k^\prime}{2\sfd^\prime N} $ such that {\color{blue}$\Delta_{x,k}^\prime = - (2\sfd^\prime N)^{-1} = \Delta_{x}^\prime$,} 
expressed in terms of the auxiliary variables $k^\prime \in \cK_1$ and $\sfd^\prime$ given in \eqref{ksim_I4}.
The endpoints $k^\prime=0$ and $k^\prime =\lfloor 2\sfd^\prime N \rfloor$ map to  $x_{0}^\prime = 1$ and $x_{\lfloor 2\sfd^\prime N \rfloor}^\prime = 0$ (up to a round-off error).
Akin to $\cK_1$, the sum in \eqref{sum_Ki} for $\cK_3$ converges to
\begin{align}
    \lim_{N\to\infty} S_N^{(3)}(\sfd,\rho) & = \lim_{N\to\infty} \frac{1}{N} \sum_{k \in \cK_3}  r^{(3)}(N,\sfd,\rho,x_{k}) \\
    & = \lim_{N\to\infty} \frac{1}{N} \sum_{k^\prime \in \cK_1}  r^{(3)}(N,\sfd,\rho,x_{k^\prime}^\prime) \\
    & = 2\sfd^\prime \lim_{\Delta_{x}^\prime \to 0} \sum_{k^\prime \in \cK_1}  r^{(3)}(N,\sfd,\rho,x_{k^\prime}^\prime) \, \Delta_{x}^\prime \\
    & = 2\sfd^\prime \int_0^1 r^{(3)}(N,\sfd,\rho,x^\prime) \, dx^\prime,
\end{align}
where $r^{(3)}(\cdot)$ inherently accounts for symmetry transformations (see Appendix~\ref{app:dpswf_Nlarge}).
\end{itemize}

The upper bound encompasses all three index subsets,
\begin{align}  \notag  
\overline{\gamma}_\endf(N,\sfd,\rho) & 
\sim 
2\sfd \int_0^1 r^{(1)}(N,\sfd,\rho,x) \, dx
+
\frac{1}{\pi} \int_{-\infty}^{\infty} \left(r^{(2)}(N,\sfd,\rho,x) \frac{\log N}{N}\right) \, dx \\ \label{pippo} &  \hspace{2cm} +
2\sfd^\prime \int_0^1 r^{(3)}(N,\sfd,\rho,x^\prime) \, dx^\prime \qquad N\to \infty,
\end{align}
whereas the lower bound retains only the final term,
\begin{equation} \label{pippo_lower}
\underline{\gamma}_\endf(N,\sfd,\rho)  
\sim  2\sfd^\prime \int_0^1 r^{(3)}(N,\sfd,\rho,x^\prime) \, dx^\prime \qquad N\to \infty.
\end{equation}

\section{} \label{app:superdirectivity_largeN_subsetK3}

By specializing {\color{blue}\eqref{inequality_superdirgain}} 
 to the lossless case ($\rho=0$), the endfire superdirectivity factor becomes
\begin{equation} \label{inequality_superdirectivity}
\underline{\gamma}_{0,\endf}(N,\sfd) < \gamma_{0,\endf}(N,\sfd) < \overline{\gamma}_{0,\endf}(N,\sfd),    
\end{equation}
{\color{blue}where upper and lower bounds are established by the aggregate or individual contributions. From \eqref{pippo} and \eqref{pippo_lower}, defining for $i=1,2$, and $3$,
\begin{align}
r_0^{(i)}(N,\sfd,x) = r^{(i)}(N,\sfd,0,x) ,
\end{align}
we have asymptotically}
\begin{align}  \notag  
\overline{\gamma}_{0,\endf}(N,\sfd) & 
\sim 
2\sfd \int_0^1 r_0^{(1)}(N,\sfd,x) \, dx
+
\frac{1}{\pi} \int_{-\infty}^{\infty} \left(r_0^{(2)}(N,\sfd,x) \frac{\log(N)}{N}\right) \, dx \\ \label{pippo_D} & \quad
+
2\sfd^\prime \int_0^1 r_0^{(3)}(N,\sfd,x^\prime) \, dx^\prime \qquad N\to \infty,
\end{align}
and
\begin{equation}  \label{pippo_D_lower}
\underline{\gamma}_{0,\endf}(N,\sfd)  
\sim 
2\sfd^\prime \int_0^1 r_0^{(3)}(N,\sfd,x^\prime) \, dx^\prime \qquad N\to \infty.
\end{equation}
In the absence of losses, the scaled eigenvalues $\mu^{(i)}/(2\sfd)$ dominate the denominator of the asymptotic ratios $r_0^{(i)}(N,\sfd,0,x)$ in \eqref{f_i_largeN}, yielding 
\begin{equation} \label{ratios_asymp}
\begin{split}
    r_0^{(i)}(N,\sfd,x) &=  \frac{2\sfd \, |\Xi^{(i)}_\endf(N,\sfd,x)|^2}{\mu^{(i)}(N,\sfd,x)} 
     \\ & \hspace{1cm}   
     \sim
     \begin{cases} 
        \begin{aligned}
            &\frac{\pi}{L_2(\sfd,x) \sinc(2 \sfd)} \\
            &\quad \cdot \frac{N}{e^{ C(N,\sfd,x) L_4(\sfd,x)/2} \, e^{ L_3(\sfd,x) N}-1}
        \end{aligned} &  i=1 \\[6ex]
        \displaystyle \frac{12\pi^2 \sfd e^{\pi x}}{1 + e^{\pi x}} \, \frac{N}{\log N} &  i=2 \\[3ex]
        \displaystyle \frac{\pi}{L_2^\prime(\sfd,x^\prime) \sinc(2 \sfd)} \, N &  i=3
    \end{cases}
\end{split}
\end{equation}
where, recall, the $L_n$'s integrals are positive and $C = \cO(1)$.

Analyzing the individual contributions in \eqref{pippo_D} reveals that the superdirectivity is asymptotically governed by the subset $\cK_3$, which provides the dominant $\cO(N)$ scaling as $N\to \infty$. 
The contributions from the remaining subsets are sublinear:
the subset $\cK_1$ exhibits an exponential decay as $\cO(N e^{- L_3 N})$,  while the contribution from $\cK_2$ converges to $\cO(1)$.

Because the gaps between the bounds---stemming from subsets $\cK_1$ and $\cK_2$---are sublinear, i.e., $o(N)$, their contributions normalized by $N$ vanish as $N\to \infty$. 
By the squeeze theorem, the bounds in \eqref{inequality_superdirectivity} are asymptotically tight, 
with the scaling factor derived as
\begin{align} \label{tau_d}
  \tau(\sfd)  
  = \lim_{N\to\infty} \frac{\underline{\gamma}_{0,\endf}(N,\sfd)}{N} =  
\frac{\pi (1-2\sfd)}{\sinc(2 \sfd)} \int_0^1 \frac{1}{L_2^\prime(\sfd,x^\prime)} \, dx^\prime
\end{align} 
after using $2\sfd^\prime=1-2\sfd$.
{\color{blue}The derived expression is reported in \eqref{tau_d_lossless_factor} where, recall, the primed notation was dropped by defining $x = x^\prime$.} 

As $\sfd \to 1/2$, $A^\prime = -A$ approaches unity in light of \eqref{A_cos_def}. Consequently, $B(\sfd,x)$, bounded within $A^\prime < B < 1$, is squeezed toward $1$ for all $x^{\prime}$, as shown in Fig. \ref{fig:B_d}. 
Evaluating the integral $L_2^\prime(\sfd,x^\prime)$ in \eqref{L2_prime} at $B(\sfd,x)=1$ and $\sfd=1/2$ causes the integral to diverge (its reciprocal $1/L_2^\prime$ thus vanishes), allowing for the interchange of the limit and the integral in \eqref{tau_d}:
\begin{align} \label{tau_d_half_final}
\lim_{\sfd\to 1/2}  \tau(\sfd)  
  & = 
  \int_0^1 \lim_{\sfd\to 1/2} \frac{1}{L_2^\prime(\sfd,x^\prime)} \, dx^\prime 
   = \int_0^1 0 \, dx^\prime = 0.
\end{align} 
As $\sfd \to 0$, the scaling factor $\tau(\sfd)$ remains {\color{blue}bounded and nonzero. Expectedly, because $L_2^\prime$ is monotonically decreasing (its reciprocal is increasing) with $x^\prime$, the dominant contribution is concentrated in the vicinity of $x^\prime = 1$ (see also Appendix~\ref{app:dpswf_Nlarge}).} 

\section{} \label{app:supergain_condition}


{\color{blue}To sustain the supergain regime in the large-$N$ asymptotic expansion \eqref{supergain_endfire_lossy_KK3}, the loss factor must depend on
$N$, $\sfd$, and $x^\prime$, and satisfy $\rho = \rho(N,\sfd,x^\prime)$ with 
\begin{equation} \label{condition_supergain}
    \rho(N,\sfd,x^\prime)  \ll  E_1(N,\sfd,x^\prime) \, E_2(N,\sfd,x^\prime),
    \end{equation}
where 
\begin{equation} \label{expexp}
    E_1(N,\sfd,x^\prime) = 
    e^{-\ln(2\sfd)-C(N,\sfd,x^\prime) L_4^\prime(\sfd,x^\prime)/2} \qquad \quad
    E_2(N,\sfd,x^\prime) = e^{-L_3^\prime(\sfd,x^\prime) N}
\end{equation}
with $C(N,\sfd,x^\prime)$ as defined in \eqref{phase_discont_main} and  $L_n^\prime(\sfd,x^\prime) > 0$ denoting the integrals given by \eqref{Li_x3prime}. 
Next, we examine how \eqref{condition_supergain}, obtained for $N\to \infty$,  behaves when subsequently taking $\sfd\to0$ for any fixed $x^\prime\in (0,1)$.}

 {\color{blue}Because $0 \le C \le 8\pi/L_2^\prime$, the first exponential is bounded by 
\begin{equation} \label{eheh_aux}
e^{-\ln(2\sfd)- 4 \pi L_4^\prime(\sfd,x^\prime)/L_2^\prime(\sfd,x^\prime)}  \le  E_1(N,\sfd,x^\prime) \le (2\sfd)^{-1}.
\end{equation}
We start evaluating $ L_2^\prime(0,x^\prime) = \lim_{\sfd\to 0}  L_2^\prime(\sfd,x^\prime)$. 
To justify interchanging of the limit and the integral, the uniform boundedness of the integrand must be established.
From \eqref{L2_prime}, we have that (since $B < 1$ in $\cK_3$ for all admissible $x^\prime$)
\begin{equation}
-\psi_1(B,\phi) = 1-(1-B)\sin^2(\phi) \in [B,1].
\end{equation}
Consequently, the integrand is bounded uniformly for $\phi\in[0,\pi/2]$ by
\begin{equation}
\left| \left(A - \psi_1(B,\phi)  \right) \left(1 - \psi_1(B,\phi)  \right) \right|^{-1/2}
\le \frac{1}{\sqrt{(A+B)(1+B)}}.
\end{equation}
Since $A+B$ and $1+B$ remain positive $\forall x^{\prime} > 0$ (see Fig.~\ref{fig:B_d}), the right-hand side is bounded by a constant and is integrable over $[0,\pi/2]$. 
(The singularity at $x^\prime = 0$ is immaterial, as the behavior within $\cK_3$ is governed by $x^\prime$ near unity.)
By the dominated convergence theorem, the limit $\sfd\to 0$ and the integral can be swapped. Setting $A=1$ as per \eqref{A_cos_def},}
\begin{align} 
    L_2^\prime(0,x^\prime) & = 2 \int_{0}^{\pi/2} \left| \left(1 - \psi_1(B(0,x^\prime),\phi)  \right) \right|^{-1} \, d\phi \\ \label{L2L2}
    & =  \int_{0}^{\pi/2} \frac{d\phi}{1 - \frac{1 - B(0,x^\prime)}{2}  \sin^{2}(\phi)}.
  \end{align}
{\color{blue}Letting $a = 1 - (1 - B)/2 = (1+B)/2$ and applying the change of variables $t = \tan(\phi)$, such that $\sin^2(\phi) = t^2/(1+t^2)$ and $dt = (1+t^2) d\phi$, the denominator of \eqref{L2L2} becomes 
\begin{equation}
    1 - \frac{1 - B(0,x^\prime)}{2}  \sin^{2}(\phi)  = 1 - (1 - a)\sin^{2}(\phi) = \frac{1+at^2}{1+t^2}.
\end{equation}
Because the the factors $1+t^2$ cancel with the differential, \eqref{L2L2} becomes
\begin{align} 
    L_2^\prime(0,x^\prime) & = \int_{0}^\infty \frac{dt}{1 + a t^2} \\
    & = \frac{1}{\sqrt{a}} \int_{0}^\infty \frac{dz}{1 + z^2} \\ \label{L2_prime_dzero}
    & = \frac{\pi}{2 \sqrt{a}} 
     = \frac{\pi}{\sqrt{2}} \frac{1}{\sqrt{1 + B(0,x^\prime)}},
    \end{align} 
    where $\frac{d}{dz} \arctan(z) = 1/(1 + z^2)$ was used.
A similar domination argument applied to $L_4^\prime(0,x^\prime) = \lim_{\sfd \to 0}L_4^\prime(\sfd,x^\prime)$  fails due to an asymptotic singularity of the constituent integrand at $\phi = \pi/2$. Precisely, from \eqref{L4_prime}, we have (since $-1 < -A < B(\sfd,x^\prime) < 1$ in $\cK_3$ for all $x^\prime$) that
\begin{equation}
\psi_2(B(\sfd,x^\prime),\phi) = - B(\sfd,x^\prime) + (A+B(\sfd,x^\prime)) \sin^2(\phi) \in [-B(\sfd,x^\prime),A],
\end{equation}
whereby, as $\sfd \to 0$ ($A  = \cos(2\pi\sfd) \to 1$ in  \eqref{A_cos_def}),
\begin{equation} \label{singular_behavior}
\left| 1 - \psi_2^{2}(B,\phi) \right|^{-1/2} \le \frac{1}{\sqrt{1 - A^2}} \to \infty.
\end{equation}
Hence, the dominated convergence theorem cannot be applied here. }
Instead, the integral is first evaluated and then the limit $\sfd \to 0$ taken.
{\color{blue}Since $B(\sfd,x^\prime) \in (-1,1)$ for all admissible $x^\prime$ (see Fig.~\ref{fig:B_d}), its dependence on $\sfd$ is omitted. }
The kernel of $L_4^\prime(\sfd,x^\prime)$ in \eqref{L4_prime} factors as
\begin{align} \label{L4_prime_aux}
    L_4^\prime(\sfd,x^\prime) & = \int_{0}^{\pi/2} \frac{2}{\alpha(\sfd,x^\prime) \sqrt{ (k_1(\sfd,x^\prime) - \sin^2(\phi) ) (k_2(\sfd,x^\prime) + \sin^2(\phi) )} } \, d\phi 
\end{align}
with positive (in light of $-1 < -A < B <1$) parameters
\begin{equation} \label{parameterski}
    \alpha(\sfd,x^\prime) = A+B \qquad k_1(\sfd,x^\prime) = \frac{1+B}{A+B} \qquad k_2(\sfd,x^\prime) = \frac{1-B}{A+B}.
\end{equation}
Applying the change of variables $s=\sin^2(\phi) $, such that $d\phi = \frac{ds}{2\sqrt{s(1-s)}}$, \eqref{L4_prime_aux} rewrites as
\begin{equation}
    L_4^\prime(\sfd,x^\prime)  = \frac{1}{\alpha(\sfd,x^\prime)} \int_{0}^{1} \frac{ds}{\sqrt{s (1-s) (k_1(\sfd,x^\prime) - s) (k_2(\sfd,x^\prime) + s)} }.
    \end{equation}
   {\color{blue} Define the roots of the quartic polynomial,
    \begin{equation} \label{roots}
    a(\sfd,x^\prime) = k_1 \left(= \frac{1+B}{A+B}\right), \quad b =1, \quad c =0, \quad \text{and} \quad d(\sfd,x^\prime) = -k_2 \left( = \frac{B-1}{A+B}\right).
    \end{equation}
The result of the integration depends on the ordering of these roots \cite[Eq.~3.147]{IntegralsBook}.
Because the sorting into $a>b>c>d$ is fixed for every admissible $B(\sfd,x^\prime)$, substituting the parameters into  \cite[Eq.~3.147.4]{IntegralsBook} and using \cite[Eq.~8.112.1]{IntegralsBook}, leads to (omitting algebraic steps and removing the dependence of every variable on $(\sfd,x^\prime)$) 
\begin{align}
L_4^\prime(\sfd,x^\prime) &  = \frac{1}{\alpha} \frac{2}{\sqrt{k_1 (1+k_2)}} K\left(\sqrt{\frac{k_1 + k_2}{k_1 (1 + k_2)}}\right) \\ \label{Kcomplete}
& = \frac{2}{\sqrt{(1+ A)(1+B)} } \, K \! \left(\sqrt{\frac{2 (A+B)}{(1+A)(1+B)}} \right),
\end{align}
where $K(\cdot)$ is the complete elliptic integral of the first kind. 
Expanding $A = 1 - 2\pi^2 \sfd^2 + \mathcal{O}(\sfd^4)$ for $\sfd\to 0$, the multiplicative factor in \eqref{Kcomplete} becomes, to leading order,
\begin{equation} \label{mult_factor}
    \frac{2}{\sqrt{(1+ A)(1+B)} } \sim \sqrt{\frac{2}{1+B}} \qquad \sfd\to 0,
\end{equation}
whereas the argument of the integral $z = \sqrt{\frac{2 (A+B)}{(1+A)(1+B)}}$ satisfies
\begin{equation} \label{argument_int}
1 - z^2 
= \frac{(1-A)(1-B)}{(1+A)(1+B)} \sim \pi^2 \sfd^2 \frac{1-B}{1+B}.
\end{equation}
Substituting this into the asymptotic expansion \cite{Carlson1985}
\begin{equation} \label{Kz_expansion}
    K(z) \sim \ln \left(\frac{4}{\sqrt{1-z^2}}\right) \qquad z \to 1
\end{equation}
while also using \eqref{mult_factor} leads to, retaining only the dominant term,} 
\begin{align} \label{L4_prime_dzero}
    L_4^\prime(\sfd,x^\prime) 
    & \sim - \sqrt{\frac{2}{1+ B(\sfd,x^\prime)}} \,  \ln(\sfd) \qquad \sfd\to 0,
\end{align}
which is uniformly bounded $\forall x^\prime >0$ ($B>-1$). 
{\color{blue}Combining \eqref{L2_prime_dzero} and \eqref{L4_prime_dzero}, $L_4^\prime/L_2^\prime \sim - \frac{2}{\pi} \, \ln(\sfd)$ for $\sfd\to 0$.
Consequently, the lower bound in \eqref{eheh_aux} asymptotically becomes
\begin{equation} \label{eheh}
E_1(N,\sfd,x^\prime) \ge e^{-\ln(\sfd)- 4 \pi L_4^\prime(\sfd,x^\prime)/L_2^\prime(\sfd,x^\prime)} \sim  \sfd^7 \qquad \sfd\to 0,
\end{equation}
pointwise for any fixed $x^\prime \in (0,1)$.

{\color{blue}The term $E_2(\sfd,x^\prime)$ depends on $L_3(\sfd,x^\prime)$ in \eqref{L3_prime}, which requires evaluating the limit $L_3^\prime(0,x^\prime) = \lim_{\sfd\to0} L_3^\prime(\sfd,x^\prime)$.
Due to the shared singularity with $L_4^\prime(\sfd,x^\prime)$ in 
\eqref{singular_behavior} the dominated convergence theorem cannot be applied. 
Using the same factorization as for \eqref{L4_prime_aux}, the integral is transformed  as}
\begin{align}
    L_3^\prime(\sfd,x^\prime) & = \int_{0}^{\pi/2} \frac{2 \sin^2 \phi }{\sqrt{(k_1(\sfd,x^\prime) - \sin^2(\phi)) (k_2(\sfd,x^\prime) + \sin^2 \phi )} } \, d\phi \\ \label{L3333}
    & = \int_{0}^{1} \frac{s}{\sqrt{s (1-s) (k_1(\sfd,x^\prime) - s) (k_2(\sfd,x^\prime) + s)} } \, ds,
\end{align}
where $k_1(\sfd,x^\prime)$ and $k_2(\sfd,x^\prime)$ are defined in \eqref{parameterski}.
{\color{blue}Specializing \cite[Eq.~3.148.4]{IntegralsBook} to \eqref{L3333}, using the roots in \eqref{roots}, we have (omitting algebraic steps and removing the dependence of every variable on $(\sfd,x^\prime)$)  
\begin{align} 
L_3^\prime(\sfd,x^\prime) 
& = k_2 \Biggl[ \Pi\left(\frac{1}{1+k_2}, \sqrt{\frac{k_1 + k_2}{k_1 (1 + k_2)}}\right)  - K\left(\sqrt{\frac{k_1 + k_2}{k_1 (1 + k_2)}}\right) \Biggr]
 \\ \nonumber
&= \frac{1-B}{A+B} \Biggl[ \Pi\left(\frac{A+B}{1+A}, \sqrt{\frac{2 (A+B)}{(1+A)(1+B)}}\right) \\   \label{L3PiK}
&\hspace{4cm} - K\left( \sqrt{\frac{2 (A+B)}{(1+A)(1+B)}} \right) \Biggr],
\end{align}
where $\Pi(\cdot)$ is the complete elliptic integral of the third kind. 
Expanding for $\sfd\to 0$ according to \eqref{argument_int} and using that into the asymptotics of $\Pi(n,z)$ \cite{Karp2007},
\begin{equation} \label{Piz_expansion}
\Pi(n,z) \sim \frac{1}{1-n} \ln \! \left(\frac{4}{\sqrt{1-z^2}}\right) 
\qquad z\to 1, \quad \left(n = \frac{A+B}{1+A}\right).
\end{equation} 
Combining the expansion \eqref{Piz_expansion} to that of $K(z)$ for $z\to 1$ in \eqref{Kz_expansion} yields, to leading order,
\begin{align} 
\Pi(n,z) - K(z) & \sim \frac{n}{1-n} \ln\left(\frac{4}{\sqrt{1-z^2}}\right)   \\ \label{PiK}
& = \frac{A+B}{1-B} \left[ - \ln(\sfd) + \ln \! \left(\frac{4}{\pi} \sqrt{\frac{1+B}{1-B}}\right)  \right] \qquad \sfd\to 0.
\end{align} 
Substituting \eqref{PiK} into \eqref{L3PiK} and simplifying retaining only the dominant term gives, pointwise for any fixed $x^\prime \in (0,1)$,
\begin{align}
L_3^\prime(\sfd,x^\prime) 
& \sim - \ln(\sfd) \qquad \sfd\to 0.
\end{align}
Plugged into \eqref{condition_supergain}, the second exponential term is asymptotically governed by
\begin{equation} \label{ohoh}
   e^{-L_3^\prime(\sfd,x^\prime) N} \sim \sfd^{N}
   \qquad \sfd\to 0.
\end{equation}
Multiplying \eqref{eheh} and \eqref{ohoh}, a sufficient condition for the supergain regime is 
\begin{equation}
    \rho(N,\sfd,x^\prime) = o(\sfd^{N+7}) \qquad \sfd \to 0, \quad N\to \infty,
\end{equation}
pointwise for any fixed $x^\prime \in (0,1)$, whose dependence is only implicit and it is therefore omitted in \eqref{supergain_regime_condition_d}.
Here, the limit $N\to\infty$ was taken first and then $\sfd\to0$ was subsequently applied to the resulting expression.}

{\color{blue}\section{} \label{app:beamwidth_N}

From the mapping $\sfff \mapsto \sfu$ in \eqref{map_u_w} and the implicit dependence of $\sfu$ on $N$ (see Appendix~\ref{app:largeN_analysis}), we have, for any $\alpha>0$,
\begin{equation} \label{u_asymptotic}
\sfu(N) + A = \cos(2\pi\sfff(N)) = A + o(N^{-\alpha}) \qquad N\to \infty,
\end{equation}
where $A = \cos(2\pi \sfd)$ is defined in \eqref{A_cos_def} and $d \in (0,1/2)$ is kept fixed.
Applying the inverse cosine function to both sides of \eqref{u_asymptotic} and dividing by $2\pi$ yields
 \begin{equation} 
|\sfff(N)| = \frac{1}{2\pi} \arccos \! \left( A + o(N^{-\alpha}) \right) \qquad N\to \infty.
\end{equation}
Since $A \in (-1, 1)$, $\arccos(x)$ is continuously differentiable at $x = A$. Its first-order expansion about $A$ is given by
\begin{equation}
\arccos(A + t) = \arccos(A) -\frac{t}{\sqrt{1-A^2}} + o(t) \qquad t\to 0.
\end{equation}
Evaluating this expansion at $t(N) = o(N^{-\alpha})$ gives, as $N \to \infty$,
 \begin{align} 
|\sfff(N)| & = \frac{1}{2\pi}  \arccos(A) + o(N^{-\alpha})  + o\big(o(N^{-\alpha})\big)  \\
& =  \sfd  + o(N^{-\alpha}) + o(N^{-\alpha}) \\ \label{sfffN}
& = \sfd + o(N^{-\alpha})  \qquad\quad N\to \infty,
\end{align}
where $\arccos(A) = 2\pi \sfd$ and $o(o(g)) = o(g)$ were used.
From \eqref{f_discrete}, substituting $\sfff(N) = \sfd \sin(\theta(N))$ into \eqref{sfffN} and dividing both sides by $\sfd > 0$ yields
\begin{equation}  \label{sin_theta_N}
|\sin(\theta(N))|  = 1 + o(N^{-\alpha})  \qquad N\to \infty.
\end{equation}
Since $\sin(\theta) \to 1$ asymptotically, $\theta(N)$ can be parameterized as 
\begin{equation} \label{theta_N}
 \theta(N) = \pm \frac{\pi}{2} + x(N) \qquad  x(N) \to 0
\end{equation}
with increasing $N$.
Expanding \eqref{sin_theta_N} around $x=0$ gives
\begin{equation} \label{sin_x_N_expansion}
\left|\sin\left(\pm \frac{\pi}{2} + x\right)\right| = \cos(x) = 1 - \frac{x^2}{2} + o(x^2) \qquad x\to 0.
\end{equation}
Equating \eqref{sin_x_N_expansion} with \eqref{sin_theta_N} and subtracting $1$ from both sides results in
\begin{equation} \label{x_asymptotic_relation}
- \frac{x^2(N)}{2} + o(x^2(N)) = o(N^{-\alpha}) \qquad N\to \infty.
\end{equation}
Factoring out $x^2(N)$,  \eqref{x_asymptotic_relation} simplifies to $x^2(N)\left(-\frac{1}{2} + o(1)\right) = o(N^{-\alpha})$, which implies $x^2(N) \cO(1) = o(N^{-\alpha})$, and equivalently
\begin{equation} \label{xNasymp}
x(N) = o\big(N^{-\alpha/2}\big) \qquad  N \to \infty.
\end{equation}
Plugging \eqref{xNasymp} into \eqref{theta_N} finally gives
\begin{equation} \label{theta_N_final}
 \theta(N) = \pm \frac{\pi}{2} + o\big(N^{-\alpha/2}\big) \qquad  N\to \infty.
\end{equation}
Specializing \eqref{theta_N_final} to $\alpha=1$ and $\alpha=2$, corresponding to the requirements in the subsets $\cK_2$ and $\cK_3$ (see Appendix~\ref{app:largeN_analysis}), leads to \eqref{loss_theta} and \eqref{supergain_theta}, respectively.}}
\end{appendices}

\bibliography{refs}

\end{document}